\documentclass{aa}  
\usepackage{graphicx}
\usepackage{txfonts}
\usepackage[colorlinks=true,citecolor=blue]{hyperref}
\usepackage{newtxtext,newtxmath}
\usepackage[T1]{fontenc}
\usepackage{amsmath}
\usepackage{amssymb}
\usepackage{color}
\usepackage{hyperref}
\usepackage{pdflscape}

\usepackage{multirow}

\newcommand{\Msun}{\ensuremath{\mathrm{M}_\odot}}
\newcommand{\Msunpyr}{\ensuremath{\Msun~\mathrm{yr^{-1}}}}
\newcommand{\kmps}{\ensuremath{\mathrm{km~s^{-1}}}}

\def\SIIHa{[{\ion{S}{ii}}]/H$\alpha$}

\def\OIIIHb{[{\ion{O}{iii}}]/H$\beta$}
\def\OIII5007Hb{[{\ion{O}{iii}}] $\lambda5007$/H$\beta$}

\def\OH{$12+\log({\rm O/H})$}
\def\NIIHa{[\ion{N}{ii}]/H$\alpha$}
\def\OIIId{[{\ion{O}{iii}}] $\lambda$5007}

\def\NIIu{[{\ion{N}{ii}}] $\lambda$6548}
\def\NIId{[{\ion{N}{ii}}] $\lambda$6583}
\def\NII{[{\ion{N}{ii}}]}

\def\SII{[{\ion{S}{ii}}]}
\newcommand{\Ha}{\rm{H}$\alpha$ $\lambda$6563}
\newcommand{\Had}{\rm{H}$\alpha$}
\newcommand{\Hb}{\rm{H}$\beta$ $\lambda$4861}

\graphicspath{{./}{fig/}}

\begin{document} 
\title{Environmental dependence of Type~IIn supernova properties}
\titlerunning{Type IIn SN properties and environments}
\authorrunning{Moriya, Galbany, et al.}
\author{
Takashi J. Moriya\inst{1,2}\thanks{E-mail: takashi.moriya@nao.ac.jp (TJM)},
Llu\'is Galbany\inst{3,4}\thanks{E-mail: lgalbany@ice.csic.es (LG)}, 
Cristina Jim\'enez-Palau\inst{3,4},
Joseph P. Anderson\inst{5,6},
Hanindyo Kuncarayakti\inst{7,8},
Sebasti\'an F. S\'anchez\inst{9},
Joseph D. Lyman\inst{10},
Thallis Pessi\inst{11,5},
Jose L. Prieto\inst{11,6},
Christopher S. Kochanek\inst{12,13},
Subo Dong\inst{14},
Ping Chen\inst{15}
}
\institute{
National Astronomical Observatory of Japan, National Institutes of Natural Sciences, 2-21-1 Osawa, Mitaka, Tokyo 181-8588, Japan \and
School of Physics and Astronomy, Faculty of Science, Monash University, Clayton, Victoria 3800, Australia \and
Institute of Space Sciences (ICE, CSIC), Campus UAB, Carrer de Can Magrans, s/n, E-08193 Barcelona, Spain \and
Institut d’Estudis Espacials de Catalunya (IEEC), E-08034 Barcelona, Spain \and
European Southern Observatory, Alonso de C\'ordova 3107, Casilla 19, Santiago, Chile \and
Millennium Institute of Astrophysics MAS, Nuncio Monsenor Sotero Sanz 100, Off. 104, Providencia, Santiago, Chile \and
Tuorla Observatory, Department of Physics and Astronomy, FI-20014 University of Turku, Finland \and
Finnish Centre for Astronomy with ESO (FINCA), FI-20014 University of Turku, Finland \and
Instituto de Astronom\'ia, Universidad Nacional Aut\'onoma de M\'exico, A.P. 70-264, 04510 M\'exico, D.F., Mexico \and
Department of Physics, University of Warwick, Coventry CV4 7AL, UK \and
N\'ucleo de Astronom\'ia de la Facultad de Ingenier\'ia y Ciencias, Universidad Diego Portales, Av. Ej \'ercito 441, Santiago, Chile \and
Department of Astronomy, The Ohio State University, 140 W. 18th Ave., Columbus, OH, 43210, USA \and
Center for Cosmology and Astroparticle Physics (CCAPP), The Ohio State University, 191 W. Woodruff Ave., Columbus, OH, 43210, USA \and
Kavli Institute for Astronomy and Astrophysics, Peking University, Yi He Yuan Road 5, Hai Dian District, Beijing 100871, China \and
Department of Particle Physics and Astrophysics, Weizmann Institute of Science, 234 Herzl St, 7610001 Rehovot, Israel 
}
\date{Received 19 April 2023; accepted 15 June 2023}

\abstract%{}{}{}{}
{
Type~IIn supernovae occur when stellar explosions are surrounded by dense hydrogen-rich circumstellar matter. The dense circumstellar matter is likely formed by extreme mass loss from their progenitors shortly before they explode. The nature of Type~IIn supernova progenitors and the mass-loss mechanism forming the dense circumstellar matter are still unknown. In this work, we investigate if there are any correlations between Type~IIn supernova properties and their local environments.
We use Type~IIn supernovae with well-observed light-curves and host-galaxy integral field spectroscopic data so that we can estimate both supernova and environmental properties. We find that Type~IIn supernovae with a higher peak luminosity tend to occur in environments with lower metallicity and/or younger stellar populations. 
The circumstellar matter density around Type~IIn supernovae is not significantly correlated with metallicity, so the mass-loss mechanism forming the dense circumstellar matter around Type~IIn supernovae might be insensitive to metallicity.
}
\keywords{supernovae: general -- stars: massive -- stars: mass-loss}
\maketitle

%%%%%%%%%%%%%%%%%%%%%%%%%%%%%%%%%%%%%%%%%%%%%%%%%%
%%%%%%%%%%%%%%%%%%%%%%%%%%%%%%%%%%%%%%%%%%%%%%%%%%
%%%%%%%%%%%%%%%%%%%%%%%%%%%%%%%%%%%%%%%%%%%%%%%%%%

\begin{table*}
\caption{The SN IIn sample and its properties.}
\label{tab:samples}      
\centering          
\begin{tabular}{lccccccc} 
\hline\hline       
\multicolumn{1}{c}{Name} & Redshift & Distance Modulus & Rise time & Peak mag.\tablefootmark{a} & $A_*$ & Mass-loss rate\tablefootmark{b} & Reference\tablefootmark{c} \\
& & mag &  days & mag & & \Msunpyr & \\ 
\hline                    
SN~1997bs  &$0.00145$&$29.7$\tablefootmark{d} & $11.0$ & $-13.0$ & $ 2200$ & $2.2\times10^{-3}$ &  (1) \\
SN~1998S   &$0.00298$&$30.9$\tablefootmark{e} & $18.1$ & $-15.9$ & $ 5800$ & $5.8\times10^{-3}$ &  (2) \\
SN~2005cl  &$0.0258$ &$34.7$\tablefootmark{f} & $26.6$ & $-18.1$ & $12000$ & $1.2\times10^{-2}$ &  (3) \\
SN~2005db  &$0.0151$ &$33.8$\tablefootmark{e} & $25.1$ & $-16.9$ & $ 9100$ & $9.1\times10^{-3}$ &  (3) \\
SN~2005kd  &$0.0150$ &$34.3$\tablefootmark{e} & $20.7$ & $-20.3$ & $15000$ & $1.5\times10^{-2}$ &  (4) \\
SN~2007cm  &$0.0160$ &$34.5$\tablefootmark{g} & $17.8$ & $-17.7$ & $ 8200$ & $8.2\times10^{-3}$ &  (4) \\
SN~2008B   &$0.0188$ &$33.3$\tablefootmark{h} & $15.3$ & $-17.0$ & $ 6300$ & $6.3\times10^{-3}$ &  (4) \\
SN~2015Z   &$0.0289$ &$35.5$\tablefootmark{g} & $23.0$ & $-18.3$ & $11000$ & $1.1\times10^{-2}$ &  (5) \\
ASASSN-15ab&$0.0178$ &$34.5$\tablefootmark{i} & $14.2$ & $-19.2$ & $ 9300$ & $9.3\times10^{-3}$ &  (6) \\
SN~2016bdu &$0.0170$ &$34.4$\tablefootmark{i} & $11.5$ & $-18.1$ & $ 6200$ & $6.2\times10^{-3}$ &  (7) \\
SN~2016iaf &$0.0619$ &$37.3$\tablefootmark{i} & $47.3$ & $-20.1$ & $22000$ & $2.2\times10^{-2}$ &  (8) \\
ASASSN-16bw&$0.0100$ &$33.3$\tablefootmark{i} & $18.9$ & $-17.8$ & $ 8800$ & $8.8\times10^{-3}$ &  (9) \\
ASASSN-16in&$0.0161$ &$34.3$\tablefootmark{i} & $22.6$ & $-18.1$ & $11000$ & $1.1\times10^{-2}$ & (10) \\
ASASSN-16jt&$0.0108$ &$33.2$\tablefootmark{i} & $13.2$ & $-18.4$ & $ 8500$ & $8.5\times10^{-3}$ & (11) \\
SN~2017bzm &$0.0452$ &$36.6$\tablefootmark{i} & $21.5$ & $-19.2$ & $13000$ & $1.3\times10^{-2}$ & (12) \\
SN~2017cin &$0.0328$ &$35.9$\tablefootmark{i} & $21.0$ & $-18.2$ & $10000$ & $1.0\times10^{-2}$ & (13) \\
SN~2017fav &$0.0456$ &$36.6$\tablefootmark{i} & $ 8.0$ & $-18.4$ & $ 5000$ & $5.0\times10^{-3}$ & (14) \\
SN~2017ggv &$0.0264$ &$35.4$\tablefootmark{i} & $24.0$ & $-18.4$ & $12000$ & $1.2\times10^{-2}$ & (15) \\
SN~2017ghw &$0.0762$ &$37.7$\tablefootmark{i} & $48.8$ & $-18.9$ & $23000$ & $2.3\times10^{-2}$ & (16) \\
SN~2017hcc &$0.0169$ &$34.4$\tablefootmark{i} & $59.5$ & $-20.6$ & $38000$ & $3.8\times10^{-2}$ & (17) \\
SN~2021fpn &$0.0424$ &$36.4$\tablefootmark{i} & $48.4$ & $-17.8$ & $18000$ & $1.8\times10^{-2}$ & (18) \\
\hline                  
\end{tabular}
\tablefoot{
\tablefoottext{a}{Host galaxy extinction is not taken into account.}
\tablefoottext{b}{A wind velocity of 100~\kmps\ is assumed.}
\tablefoottext{c}{(1) \citet{vandyk2000}, (2) \citet{poon2011}, (3) \citet{kiewe2012}, (4) \citet{hicken2017}, (5) \citet{nyholm2020}, (6) \citet{dong2015,shappee2015}, (7) \citet{pastorello2018}, (8) \citet{2016TNSTR.909....1T,2016TNSCR.932....1T}, (9) \citet{2016ATel.8703....1B,2016ATel.8727....1E}, (10) \citet{2016ATel.9344....1B,2016TNSCR.544....1R}, (11) \citet{2016ATel.9439....1B,2016ATel.9445....1B}, (12) \citet{2017TNSTR.284....1T,2017TNSCR.293....1B}, (13) \citet{2017TNSTR.336....1T,2017TNSCR.354....1O}, (14) \citet{2017TNSTR.709....1T,2017TNSCR.745....1P}, (15) \citet{2017TNSTR.903....1X,2017TNSCR.937....1L}, (16) \citet{2017TNSTR.923....1M,2017TNSCR.937....1L}, (17) \citet{kumar2019,smith2020,chandra2022,moran2023}, (18) \citet{2021TNSTR.771....1T,2021TNSAN.118....1P}}
\tablefoottext{d}{\citet{willick2001}}
\tablefoottext{e}{\citet{sorce2014}}
\tablefoottext{f}{\citet{springob2014}}
\tablefoottext{g}{\citet{theureau2007}}
\tablefoottext{h}{\citet{tully2013}}
\tablefoottext{i}{DM from redshift.}
}
\end{table*}

\section{Introduction}
Type~IIn supernovae (SNe~IIn) occur when stars explode within a dense hydrogen-rich circumstellar matter (CSM, \citealt{schlegel1990,filippenko1997}). The dense CSM is created by strong mass loss from the progenitors with typical mass-loss rate estimates of more than $10^{-4}~\Msunpyr$ \citep[e.g.,][]{taddia2013,kiewe2012,moriya2014,ofek2014}. Such mass-loss rates are much higher than those measured for typical stars \citep[e.g.,][]{smith2014}, and the progenitors and mass-loss mechanisms of SNe~IIn are still not well understood. It is suggested that the high mass-loss rates are similar to those of massive ($\gtrsim 25~\Msun$) luminous blue variable stars (LBVs, e.g., \citealt{weis2020}). Indeed, the progenitor of the Type~IIn SN~2005gl is consistent with a massive LBV \citep[][]{gal-yam2009}. On the other hand, the progenitor of the Type~IIn SN~2008S was relatively low mass ($\simeq 10~\Msun$, \citealt{prieto2008}). These events suggest that progenitors and mass-loss mechanisms of SNe~IIn are diverse. In addition, SNe~Ia are sometimes hidden below dense hydrogen-rich CSM and observed as SNe~IIn \citep[``SN~Ia-CSM,'' e.g.,][]{sharma2023}.

The local environments where SNe explode provide rich information on their progenitors (see \citealt{anderson2015} for a review). For example, SNe~II and Ibc but not SNe~Ia preferentially occur in star-forming environments, which indicates that SNe~II and Ibc are associated with massive star explosions \citep[e.g.,][]{li2011,galbany2014}. There have been several studies of the local environments of SNe~IIn. \citet{habergham2014} found that their locations are not necessarily associated with the most actively star-forming regions in their host galaxies and some may not be associated with very massive progenitors. A later study by \citet{ransome2022} found that 60\% of SNe~IIn originate from actively star-forming regions and could be linked to very massive progenitors such as LBVs. The remaining 40\% were not correlated with ongoing star formation and could have relatively low-mass progenitors \citep[see also][]{kuncarayakti2018}. Similarly, \citet{Gal18a} estimated the age distributions of SN~IIn progenitors based on spectra of their surroundings and found that they may have a bimodal age distribution with one peak at $0-20~\mathrm{Myr}$ and the other at $100-300~\mathrm{Myr}$. These studies suggest that SN~IIn progenitors are a mixture of massive ($\gtrsim 25~\Msun$ like LBVs) and low-mass ($\simeq 10~\Msun$ like the progenitor of SN~2008S) stars.
Other local properties may also provide information on their nature.
For example, \citet{taddia2015} investigated the relationship between SN~IIn progenitor mass-loss rates and their local metallicity. They found that the progenitors of SNe~IIn may have higher mass-loss rates in higher metallicity environments.

In this work, we explore the environmental dependence of SN~IIn properties by using SNe~IIn with integral-field spectroscopy (IFS) of their host galaxies.
The IFS data allow us to estimate not only the metallicity but also environmental parameters such as the local star-formation rates (SFRs). 
We introduce our SN~IIn samples in Section~\ref{sec:definition}. We estimate the local environmental parameters of the SN~IIn explosion sites in Section~\ref{sec:local_environments} and estimate the SN~IIn properties in Section~\ref{sec:density_estimate}. We investigate possible correlations between the environmental and SN properties in Section~\ref{sec:correlations} and discuss them in Section~\ref{sec:discussion}. We conclude this paper in Section~\ref{sec:conclusions}. We adopt a $\Lambda$CDM cosmology with $H_0 = 68.3~\mathrm{km~s^{-1}~Mpc^{-1}}$, $\Omega_M = 0.28$, and $\Omega_\Lambda = 0.72$ \citep[][]{hinshaw2013}.

\begin{table*}
\caption{Local metallicity measurements of our SN IIn sample. Numbers in parentheses are the standard deviation.}
\label{tab:metallicity}      
\centering          
\begin{tabular}{lccc} 
\hline\hline       
\multicolumn{1}{c}{Name} & \multicolumn{1}{c}{$12+\log \mathrm{(O/H)_{N2}}$} & \multicolumn{1}{c}{$12+\log \mathrm{(O/H)_{O3N2}}$} & \multicolumn{1}{c}{$12+\log \mathrm{(O/H)_{D16}}$}  \\
 &  &  &  \\
\hline                    
SN~1997bs  & $8.699(0.014)$ & $8.568(0.025)$ & $8.697(0.022)$ \\
SN~1998S   & $8.577(0.003)$ & $8.560(0.004)$ & $8.720(0.004)$ \\
SN~2005cl  & $8.617(0.008)$ & $8.582(0.019)$ & $8.833(0.025)$ \\
SN~2005db  & $8.550(0.004)$ & $8.639(0.001)$ & $8.856(0.005)$ \\
SN~2005kd  & $8.299(0.002)$ & $8.285(0.001)$ & $8.141(0.004)$ \\
SN~2007cm  & $8.639(0.020)$ & $8.500(0.005)$ & $8.716(0.020)$ \\
SN~2008B   & $8.561(0.019)$ & $8.532(0.005)$ & $8.612(0.022)$ \\
SN~2015Z   & $8.571(0.002)$ & $8.459(0.001)$ & $8.607(0.002)$ \\
ASASSN-15ab& $8.332(0.002)$ & $8.350(0.004)$ & $8.376(0.013)$ \\
SN~2016bdu & $8.429(0.259)$ & $8.316(0.114)$ & $8.250(0.001)$ \\
SN~2016iaf & $8.105(0.121)$ & $8.453(0.099)$ & $7.484(0.807)$ \\
ASASSN-16bw& $8.612(0.041)$ & $8.424(0.035)$ & $8.567(0.051)$ \\
ASASSN-16in& $8.585(0.060)$ & $8.395(0.033)$ & $8.500(0.089)$ \\
ASASSN-16jt& $8.529(0.028)$ & $8.456(0.011)$ & $8.603(0.037)$ \\
SN~2017bzm & $8.200(0.020)$ & $8.185(0.013)$ & $8.108(0.071)$ \\
SN~2017cin & $8.372(0.005)$ & $8.381(0.003)$ & $8.373(0.011)$ \\
SN~2017fav & $8.429(0.019)$ & $8.392(0.009)$ & $8.412(0.036)$ \\
SN~2017ggv & $8.505(0.004)$ & $8.512(0.005)$ & $8.555(0.006)$ \\
SN~2017ghw & $8.561(0.188)$ & $8.631(0.077)$ & $8.642(0.264)$ \\
SN~2017hcc & $8.247(0.117)$ & $8.232(0.065)$ & $8.058(0.331)$ \\
SN~2021fpn & $8.756(0.010)$ & $8.334(0.005)$ & $8.728(0.011)$ \\
\hline                  
\end{tabular}
%\tablefoot{
%}
\end{table*}

\begin{table*}
\caption{Local environmental properties of our SN IIn sample. Numbers in parentheses are the standard deviation.}
\label{tab:numbers}      
\centering          
\begin{tabular}{lrrrrcc} 
\hline\hline       
\multicolumn{1}{c}{Name} &  \multicolumn{1}{c}{$\log\Sigma_\mathrm{SFR}$} & \multicolumn{1}{c}{EW(H$\alpha$)}& \multicolumn{1}{c}{$\log (\mathrm{sSFR})$} & \multicolumn{1}{c}{$\langle\log t_{*,L}\rangle$} & $E(B-V)$ & $A_{V*}$ \\
 & \multicolumn{1}{c}{$\mathrm{\Msun~yr^{-1}~kpc^{-2}}$} & \multicolumn{1}{c}{\AA} & \multicolumn{1}{c}{$\mathrm{yr^{-1}}$} & \multicolumn{1}{c}{years} & \multicolumn{1}{c}{mag} & \multicolumn{1}{c}{mag} \\
\hline                    
SN~1997bs  & $-4.174(0.006)$ & $  8.66(0.09)$ & $-10.359(0.694)$ & $8.47(0.66)$ & $0.24(0.05)$ & $1.25$ \\
SN~1998S   & $-1.614(0.001)$ & $ 39.99(0.08)$ & $-11.097(0.169)$ & $8.80(1.21)$ & $0.43(0.07)$ & $0.77$ \\
SN~2005cl  & $-2.376(0.003)$ & $ 22.07(0.16)$ & $ -9.246(0.001)$ & $9.00(1.34)$ & $0.25(0.04)$ & $0.41$ \\
SN~2005db  & $-1.711(0.001)$ & $ 51.29(0.16)$ & $ -9.371(0.001)$ & $8.17(1.62)$ & $0.38(0.05)$ & $0.13$ \\
SN~2005kd  & $-1.920(0.001)$ & $ 52.67(0.12)$ & $ -9.063(0.001)$ & $8.05(0.84)$ & $0.20(0.02)$ & $0.12$ \\
SN~2007cm  & $-3.354(0.008)$ & $ 12.13(0.17)$ & $-11.471(0.001)$ & $9.28(1.53)$ & $0.04(0.01)$ & $0.32$ \\
SN~2008B   & $-2.946(0.007)$ & $ 37.40(0.90)$ & $ -8.060(0.001)$ & $7.93(1.50)$ & $0.10(0.01)$ & $0.00$ \\
SN~2015Z   & $-2.386(0.001)$ & $ 39.68(0.29)$ & $ -9.242(0.001)$ & $8.08(1.38)$ & $0.22(0.02)$ & $0.00$ \\
ASASSN-15ab& $-3.347(0.001)$ & $313.14(1.23)$ & $ -7.622(0.001)$ & $6.40(0.71)$ & $0.13(0.01)$ & $0.00$ \\
SN~2016bdu & $-4.021(0.081)$ & $  4.40(0.75)$ & $-10.964(0.016)$ & $8.14(2.66)$ & $0.04(0.01)$ & $0.20$ \\
SN~2016iaf & $-4.577(0.015)$ & $ 35.08(0.74)$ & $ -8.242(0.001)$ & $6.50(0.01)$ & $0.06(0.01)$ & $0.49$ \\
ASASSN-16bw& $-4.286(0.016)$ & $ 33.16(0.88)$ & $ -8.743(0.017)$ & $7.05(0.28)$ & $0.36(0.03)$ & $1.57$ \\
ASASSN-16in& $-4.736(0.022)$ & $ 29.82(1.24)$ & $ -8.352(0.004)$ & $6.48(0.37)$ & $0.00(0.01)$ & $0.00$ \\
ASASSN-16jt& $-4.041(0.010)$ & $ 24.37(0.26)$ & $-10.467(0.008)$ & $7.73(1.89)$ & $0.15(0.02)$ & $1.23$ \\
SN~2017bzm & $-3.395(0.005)$ & $133.56(1.03)$ & $ -8.662(0.001)$ & $7.37(1.42)$ & $0.04(0.01)$ & $0.78$ \\
SN~2017cin & $-3.191(0.002)$ & $112.16(0.32)$ & $ -9.228(0.001)$ & $8.32(1.56)$ & $0.23(0.02)$ & $0.07$ \\
SN~2017fav & $-2.547(0.006)$ & $ 53.97(0.59)$ & $ -9.246(0.001)$ & $7.72(1.55)$ & $0.18(0.01)$ & $0.19$ \\
SN~2017ggv & $-3.360(0.001)$ & $ 34.10(0.07)$ & $ -9.916(0.001)$ & $8.30(1.20)$ & $0.25(0.03)$ & $1.02$ \\
SN~2017ghw & $-4.817(0.066)$ & $  7.20(0.70)$ & $-10.305(0.001)$ & $7.14(1.76)$ & $0.00(0.01)$ & $0.00$ \\
SN~2017hcc & $-4.192(0.000)$ & $ 13.44(0.59)$ & $ -9.322(0.004)$ & $6.86(1.12)$ & $0.20(0.03)$ & $0.16$ \\
SN~2021fpn & $-4.150(0.004)$ & $187.74(1.45)$ & $ -8.589(0.001)$ & $8.23(1.60)$ & $0.20(0.02)$ & $0.00$ \\
\hline                  
\end{tabular}
%\tablefoot{
%}
\end{table*}

\section{Sample definition}\label{sec:definition}

We constructed our sample using all galaxies observed with IFS from the PISCO, AMUSING, and MaNGA surveys to host a Type~IIn SN.

The PMAS/PPak Integral field Supernova hosts COmpilation (PISCO; \citealt{Gal18a}) is a compilation of IFS observations of more than 400 SN host galaxies obtained with the Potsdam Multi Aperture Spectograph (PMAS; \citealt{2005PASP..117..620R}) on the 3.5m telescope of the Centro Astronomico Hispano-Aleman (CAHA) at the Calar Alto Observatory. The observations were obtained in PPak mode \citep{2004AN....325..151V,2006PASP..118..129K}.
About a third of the objects were observed by the CALIFA survey \citep{2016A&A...594A..36S}.
Each observation consists of a 3D datacube with a 100\% covering factor within a hexagonal field-of-view (FoV) of $\sim$1.3 arcmin$^2$, with 1"$\times$1" spatial pixels (spaxel) and a spectral resolution of $\sim$6~\AA\ over the wavelength range 3750$-$7300 \AA, providing $\sim$4000 spectra per object.

The All-weather MUse Supernova Integral-field of Nearby Galaxies (AMUSING; \citealt{Gal16a}; \citealt{2020AJ....159..167L}; Galbany et al. in prep.) survey has been running for 11 semesters (P95-P106), and has compiled observations for more than 800 nearby SN host galaxies with the Multi-Unit Spectroscopic Explorer (MUSE; \citealt{2014Msngr.157...13B}), located at the Nasmyth B focus of Yepun, the VLT UT4 telescope at Cerro Paranal Observatory. 
MUSE is composed of 24 identical IFS. Wide Field Mode (WFM) samples a nearly contiguous 1 arcmin$^2$ FoV with spaxels of 0.2 $\times$ 0.2 arcsec, and over a wavelength range of 4650-9300~\AA\ with a mean resolution of R$\sim$3000. Each 3D cube consists of $\sim$100,000 spectra per pointing.

The Mapping Nearby Galaxies at APO (MaNGA; \citealt{2015ApJ...798....7B}) 
was part of Sloan Digital Sky Survey (SDSS) IV (\citealt{2017AJ....154...28B}) and 
obtained IFS data of $\sim$10,000 nearby galaxies using 17 units of different hexagonal FoVs ranging from 12 to 32 arcsec in diameter at the 2.5m SDSS telescope at the Apache Point Observatory, in New Mexico.
The square spaxels are of 0.5 arcsec across, with a spectral resoulution of R$\sim$2000 over a wavelength range of 3600-10000 \AA.

After a thorough search of these three datasets, we compiled an initial sample of 66 SN~IIn host galaxies where the SN location was within the FoV. 
Next, we performed a thorough search for public light-curves of the 66 SNe~IIn in the literature. For those objects that exploded in 2016 or after, we also used the ATLAS forced photometry service\footnote{\href{https://fallingstar-data.com/forcedphot/queue/}{https://fallingstar-data.com/forcedphot/queue/}} to obtain light curves. For those objects that exploded in 2018 or later, we utilized the ZTF forced photometry service\footnote{\href{https://ztfweb.ipac.caltech.edu/cgi-bin/requestForcedPhotometry.cgi}{https://ztfweb.ipac.caltech.edu/cgi-bin/requestForcedPhotometry.cgi}}.
From the 24 SNe~IIn with publicly available data, only 17 had light-curves with enough quality and sampling during the rise to reliably determine the peak magnitude and rise time from explosion (see Section~\ref{sec:density_estimate}).
In addition, we obtained light curves with good sampling from All-Sky Automated Survey for SuperNovae (ASAS-SN, \citealt{shappee2014,kochanek2017}) and follow-up observations with the Las Cumbres Observatory Global Telescope network (LCOGT) for four additional SNe~IIn (ASASSN-15ab, ASASSN-16bw, ASASSN-16in, and ASASSN-16jt). The LCOGT photometry was performed according to the procedures described in \citet{chen2022}.
The final 21 SNe~IIn in our sample are listed in Table \ref{tab:samples}.

\section{Local environments}\label{sec:local_environments}

The final sample of 21 SNe~IIn is composed of 13 host galaxies observed with MUSE, five with PMAS and three with MaNGA. Synthetic $r$-band images created from the IFS cubes are displayed in Figure \ref{fig:mosaic1}.

We followed a similar procedure for all 3 IFS instruments. 
We extracted a rest-frame 2.7~arcsec diameter aperture spectrum for each SN position, corresponding to the worst spatial resolution in all the cubes.
We analyzed the spectra as in \cite{galbany2014,2016A&A...591A..48G}.
We fit single stellar population (SSP) synthesis models to remove the underlying stellar continuum from the ionized gas-phase emission using {\tt STARLIGHT} \citep{2005MNRAS.358..363C,2009RMxAC..35..127C}.
{\tt STARLIGHT} determines the fractional contribution of different SSP models to the spectrum, accounting for dust extinction as a foreground screen. 
We use three parameters from the SSP fit: the stellar mass ($M_*$), the average light-weighted stellar population age ($t_{*,L}$), and the extinction derived from the stellar component ($A_{V*}$).

The best fit continuum model is then subtracted from each observed spectrum to leave the ionized gas-phase emission.
Figure~\ref{fig:envspec1} shows the aperture spectra, the best SSP fits, and their resulting gas-phase emission line spectra for all 21 SN~IIn environments.
We fit the emission lines needed to estimate oxygen abundances using several different methods. 
This included fitting Gaussian profiles to the Balmer \Ha\ and \Hb\ lines, and the \OIIId, \NIId, \SII~$\lambda\lambda$6716,31 lines.
The \Had\ and \NII\ lines were fit simultaneously with \NIIu\ as three Gaussian profiles with fixed positions and similar width, but free amplitudes. 
In seven cases of relatively recent SNe (SN~2016bdu, SN~2016iaf, 
ASASSN-16bw, 
ASASSN-16in, 
ASASSN-16jt, 
SN~2017ghw, SN~2017hcc; see Figure \ref{fig:envspec1}) it was necessary to include a fourth component  to account for a broad underlying \Had~emission coming from the CSM interaction (see also Mart\'inez-Rodr\'iguez in prep.).

The flux of the emission lines was corrected for dust extinction along the line of sight using the color excess ($E(B-V)$) estimate from the H$\alpha$/H$\beta$ Balmer line flux ratios assuming the Case~B recombination intrinsic ratio $I(\mathrm{H}\alpha)/I(\mathrm{H}\beta)=2.86$ for $T=10,000$ K and an electron density of $10^2$ cm$^{-3}$ \citep{2006agna.book.....O}, and a \cite{1999PASP..111...63F} extinction law.

% ESFR and sSFR.
The ongoing SFR can be directly estimated from the extinction-corrected H$\alpha$ flux following \cite{1998ApJ...498..541K}, 
\begin{equation}
    \mathrm{SFR} [\Msun~\mathrm{yr^{-1}}] = 7.9 \times 10^{-42} L(\mathrm{H}\alpha),
\end{equation}
where
\begin{equation}
L(\mathrm{H}\alpha) = 4\pi d_L^2 F(\mathrm{H}\alpha),
\end{equation}
is the extinction-corrected H$\alpha$ luminosity in units of erg s$^{-1}$ and $d_L$ is the luminosity distance to the galaxy. The SFR density ($\Sigma_\mathrm{SFR}$) is obtained by dividing the SFR by the area of the aperture in kpc$^2$, and the specific SFR (sSFR) is obtained by dividing the SFR by the stellar mass obtained from the \texttt{STARLIGHT} fit.

% HaEW
While the H$\alpha$ flux is an indicator of the ongoing SFR, the H$\alpha$ equivalent width, EW(H$\alpha$), is a measurement of how strong the emission line is compared with the stellar continuum. The stellar continuum is dominated by the contribution from old stars, which also contain most of the stellar mass. The EW(H$\alpha$) represents the fraction of young stars, and it can be thought of as an indicator of the strength of the ongoing SFR compared with the past SFR and it decreases with time if no ongoing star-formation is present. It is a reliable proxy for the age of the youngest stellar components \citep{2016A&A...593A..78K,2015A&A...574A..47S}.
To estimate EW(H$\alpha$), we divided the observed spectrum by the \texttt{STARLIGHT} fit, and repeated the weighted
nonlinear least-squares fit of the H$\alpha$ line in the normalized spectra.

%N2, O3N2, D16
The most commonly used metallicity indicator in interstellar medium (ISM) studies is the oxygen abundance, since it is the most abundant metal in the gas phase and has very strong optical nebular lines. We estimated the oxygen abundances, \OH, using three different empirical calibrations based on emission-line ratios. In particular, we used the N2 index with the \citet{marino2013} calibrations updated from \cite{2004MNRAS.348L..59P} based on the \NIIHa~ratio,
\begin{equation}
12 + \log \mathrm{(O/H)_{N2}} = 8.743 + 0.462 \times \log \frac{[{\ion{N}{ii}}]}{\rm H\alpha}, %(valid in the range −2.5 < N2 < − 0.3), 
\end{equation}
and the O3N2 index based on the difference between the logs of the \OIIIHb~and \NIIHa~ratios,
\begin{equation}
12 + \log \mathrm{(O/H)_{O3N2}} = 8.533 - 0.214 \times \log\left(\frac{[{\ion{O}{iii}}]}{\rm H\beta}\frac{\rm H\alpha}{[{\ion{N}{ii}}]}\right).
%\left( \log \frac{[{\ion{O}{iii}}]}{\rm H\beta} - \log \frac{[{\ion{N}{ii}}]}{\rm H\alpha} \right). %    .−1 < O3N2 < 1.9 
\end{equation} 
Finally, we used the sulphur-based calibrator from \cite{2016Ap&SS.361...61D} based on the \SIIHa\ and \NIIHa\ ratios,
\begin{equation}
12 + \log \mathrm{(O/H)_\mathrm{D16}} = 8.77 + y + 0.45 \times (y + 0.3),
\end{equation} 
where $y = \log [{\ion{N}{ii}}]/[{\ion{S}{ii}}] + 0.264 \times \log [{\ion{N}{ii}}]/{\rm H\alpha}$. 
All these calibrations are quite insensitive to extinction because the emission lines used for the ratio diagnostics are close in wavelength. The ratios are also little affected by differential atmospheric refraction (DAR), although DAR has been corrected for during data reduction. 
The resulting metallicities are reported in Table \ref{tab:metallicity} and the other local environmental properties are summarized in Table~\ref{tab:numbers}.

\begin{sidewaystable*}
\caption{Pearson correlation coefficients, their standard deviations, and $p$ values.}
\label{tab:correlations}      
\centering          
\begin{tabular}{lccccccc} 
\hline\hline       
 & $12+\log \mathrm{(O/H)_{N2}}$ & $12+\log \mathrm{(O/H)_{O3N2}}$ & $12+\log \mathrm{(O/H)_{D16}}$ &  $\log\Sigma_\mathrm{SFR}$ & EW(H$\alpha$) & $\log \mathrm{sSFR}$ & $\langle\log t_{*,L}\rangle$  \\
\hline    
&\multicolumn{7}{c}{No host galaxy extinction correction} \\
\multirow{2}{*}{Rise time} & $-0.17\pm0.27$  & $-0.07\pm0.26$ & $-0.22\pm0.25$  & $-0.37\pm0.17$  & $-0.01\pm0.25$  &  $0.18\pm0.19$ & $-0.25\pm0.16$  \\
 & $(p=0.49)$  & $(p=0.74)$  & $(p=0.22)$  & $(p=0.031)$  & $(p=0.91)$  &  $(p=0.34)$ & $(p=0.097)$  \\
\multirow{2}{*}{Peak mag.} & $\mathbf{0.67\pm0.08}$  &  $\mathbf{0.56\pm0.13}$ &  $\mathbf{0.60\pm0.09}$ & $0.16\pm0.26$  & $-0.18\pm0.15$  & $-0.36\pm0.15$ &  $\mathbf{0.42\pm0.10}$  \\
 & $(p=0.000011)$  &  $(p=0.0010)$ &   $(p=0.0000040)$ & $(p=0.63)$  & $(p=0.17)$  & $(p=0.027)$ &  $(p=0.00032)$  \\
\multirow{2}{*}{$\log A_\ast$} & $-0.39\pm0.20$  & $-0.31\pm0.22$ &  $-0.37\pm0.16$  & $-0.25\pm0.22$  &  $0.12\pm0.18$  &  $0.31\pm0.17$ & $-0.33\pm0.12$  \\
& $(p=0.041)$  & $(p=0.13)$  & $(p=0.034)$  & $(p=0.28)$  &  $(p=0.58)$  &  $(p=0.067)$ & $(p=0.011)$  \\
\hline
&\multicolumn{7}{c}{Host galaxy extinction correction with $E(B-V)$} \\
\multirow{2}{*}{Rise time} & $ -0.16\pm0.27$  & $-0.06\pm0.25$ & $-0.21\pm0.25$  & $-0.37\pm0.18$  & $-0.01\pm0.25$  &  $0.19\pm0.19$ & $-0.25\pm0.16$  \\
 & $(p=0.51)$  & $(p=0.78)$ & $(p=0.24)$  & $(p=0.032)$  & $(p=0.92)$  &  $(p=0.34)$ & $(p=0.10)$  \\
\multirow{2}{*}{Peak mag.} & $\mathbf{0.66\pm0.08}$  &  $\mathbf{0.54\pm0.12}$ &  $\mathbf{0.57\pm0.09}$ & $0.06\pm0.25$  & $-0.20\pm0.14$  & $-0.37\pm0.15$ &  $\mathbf{0.39\pm0.11}$  \\
 & $(p=0.000016)$  &  $(p=0.0010)$ &  $(p=0.000049)$ & $(p=0.89)$  & $(p=0.15)$  & $(p=0.028)$ &  $(p=0.0021)$  \\
\multirow{2}{*}{$\log A_\ast$} & $-0.37\pm0.20$  & $-0.28\pm0.22$  & $-0.34\pm0.17$  & $-0.20\pm0.22$  &  $0.12\pm0.19$  &  $0.31\pm0.17$ & $-0.30\pm0.12$  \\
& $(p=0.051)$  & $(p=0.17)$ & $(p=0.047)$  & $(p=0.39)$  &  $(p=0.58)$  &  $(p=0.070)$ & $(p=0.018)$  \\
\hline
&\multicolumn{7}{c}{Host galaxy extinction correction with $A_{V*}$} \\
\multirow{2}{*}{Rise time} & $-0.16\pm0.27$  & $-0.06\pm0.25$ & $-0.21\pm0.25$  & $-0.37\pm0.18$  & $-0.01\pm0.25$  &  $0.19\pm0.19$ & $-0.25\pm0.16$  \\
 & $(p=0.51)$  & $(p=0.78)$ & $(p=0.24)$  & $(p=0.032)$  & $(p=0.92)$  &  $(p=0.34)$ & $(p=0.10)$  \\
\multirow{2}{*}{Peak mag.} & $\mathbf{0.66\pm 0.09}$  &  $\mathbf{0.55\pm0.13}$ &  $\mathbf{0.60\pm0.08}$ & $0.20\pm0.25$  & $-0.13\pm0.15$  & $-0.31\pm0.17$ &  $\mathbf{0.42\pm0.10}$  \\
 & $(p=0.000013)$  &  $(p=0.0015)$ &  $(p=0.000022)$ & $(p=0.49)$  & $(p=0.34)$  & $(p=0.059)$ &  $(p=0.00035)$  \\
\multirow{2}{*}{$\log A_\ast$} & $-0.39\pm0.20$  & $-0.30\pm0.22$ & $-0.37\pm0.16$  & $-0.28\pm0.21$  &  $0.09\pm0.18$  &  $0.29\pm0.18$ & $-0.33\pm0.12$  \\
& $(p=0.043)$  & $(p=0.14)$ & $(p=0.037)$  & $(p=0.21)$  &  $(p=0.69)$  &  $(p=0.097)$ & $(p=0.011)$  \\
\hline                  
\end{tabular}
%\tablefoot{
%}
\end{sidewaystable*}

\section{SN~IIn properties}\label{sec:density_estimate}
SNe~IIn are characterized by their high CSM density. 
We assume that the CSM density is $\rho_\mathrm{CSM}=Ar^{-2}$, where $A$ is constant and $r$ is the radius. Given a mass-loss rate ($\dot{M}$) and a wind velocity ($v_\mathrm{wind}$) of the progenitor, the constant is
\begin{equation}
    A = \frac{\dot{M}}{4\pi v_\mathrm{wind}}.\label{eq:A}
\end{equation}
Following convention \citep[e.g.,][]{chevalier2006}, we define 
\begin{equation}
    A_\ast = \frac{1}{4\pi} \left(\frac{\dot{M}}{10^{-6}~\Msunpyr}\right)\left(\frac{v_\mathrm{wind}}{100~\kmps}\right)^{-1}.
\end{equation}
Assuming that shock breakout occurs inside the dense CSM, the rise time and peak luminosity can be related to the density \citep[e.g.,][]{ofek2010,ofek2014,moriya2014m}. Following \citet{moriya2014m}, %we adopt the following relation to estimate the CSM density around SNe~IIn,
\begin{equation}
    A = C_2^{-\frac{n-2}{n}}C_3^{-\frac{n-2}{4n-5}}\varepsilon^{-\frac{n-2}{4n-5}}\kappa^{-\frac{3(n-1)}{4n-5}}t_d^{\frac{3(n-1)}{4n-5}}L_p^{\frac{n-2}{4n-5}},
\end{equation}
where
\begin{align}
    C_2 =& c^{-\frac{1}{n-2}}\left[2\pi(n-4)(n-3)(n-\delta)\frac{[(3-\delta)(n-3)]^{\frac{n-5}{2}}}{[2(5-\delta)(n-5)]^{\frac{n-3}{2}}}\right]^{\frac{1}{n-2}} \nonumber \\
    &\times\left(\frac{n-2}{n-3}\right)^{\frac{n-3}{n-2}}, \\
    C_3 =& \frac{2\pi}{n-5}c^{\frac{n-5}{n(n-2)}}\left[\frac{1}{4\pi (n-\delta)}\frac{[2(5-\delta)(n-5)]^{\frac{n-3}{2}}}{[(3-\delta)(n-3)]^{\frac{n-5}{2}}}\right]^{\frac{4n-5}{n(n-2)}} \nonumber \\
    &\times \left[\frac{(n-4)(n-3)}{2}\right]^\frac{(n-1)(n-5)}{n(n-2)}\left(\frac{n-3}{n-2}\right)^{\frac{(n-5)(n-3)}{n(n-2)}},
\end{align}
$\varepsilon$ is the conversion efficiency from kinetic energy to radiation at the shock, $\kappa=0.34~\mathrm{cm^2~g^{-1}}$ is the electron scattering opacity in the CSM, $t_d$ is the rise time, $L_p$ is the peak bolometric luminosity, and $c$ is the speed of light. Here, the SN ejecta density $\rho_\mathrm{ejecta}$ is assumed to have a two-component power-law structure ($\rho_\mathrm{ejecta}\propto r^{-n}$ outside and $\rho_\mathrm{ejecta}\propto r^{-\delta}$ inside) with $n=7$ and $\delta=0$ \citep[][]{matzner1999}. We assume a constant $\varepsilon=0.3$ in our analysis (e.g., \citealt{ofek2014,moriya2014m,fransson2014}, but see also \citealt{tsuna2019}).

This formalism is applicable to bolometric light curves. However, it is difficult to estimate bolometric luminosity without extensive multi-wavelength observations and such observations are rarely available. Here, we use observed light curves in the \textit{o} filter ($5600-8200$~\AA), the \textit{R} band filter ($5500-8600$~\AA), or the \textit{r} band filter ($5600-7300$~\AA) to estimate the rise time and peak luminosity. We do not include a bolometric correction, because the bolometric correction near luminosity peak in this wavelength range is estimated to be small (e.g., around $-0.3~\mathrm{mag}$ in the $R$ band for SN~2010jl, \citealt{ofek2014b}). In the case of ASASSN-15ab and ASASSN-16in, we use \textit{V} band ($4800-6400$~\AA) light-curves that provide better constraints on the rising part of the light curve.

\begin{figure}
    \centering
    \includegraphics[width=\columnwidth]{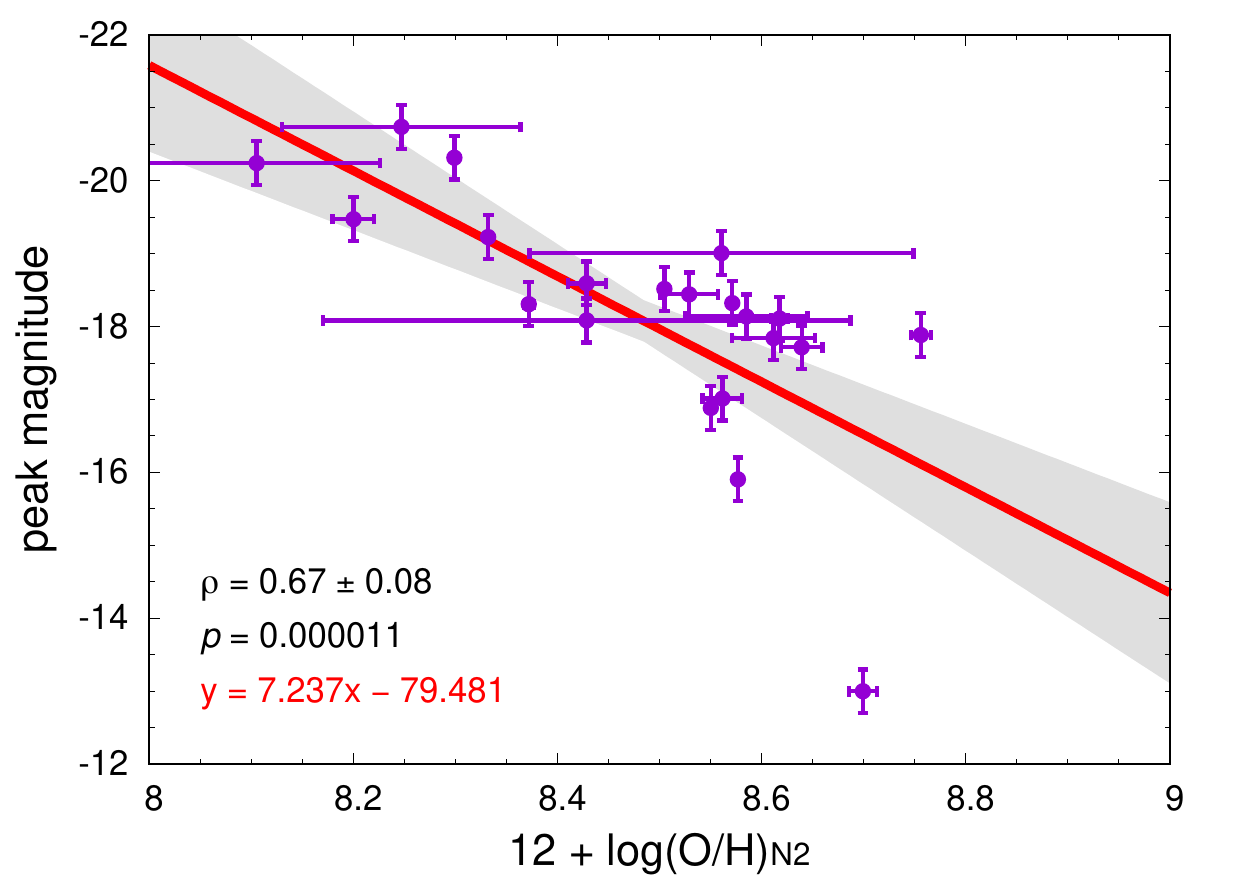} 
    \includegraphics[width=\columnwidth]{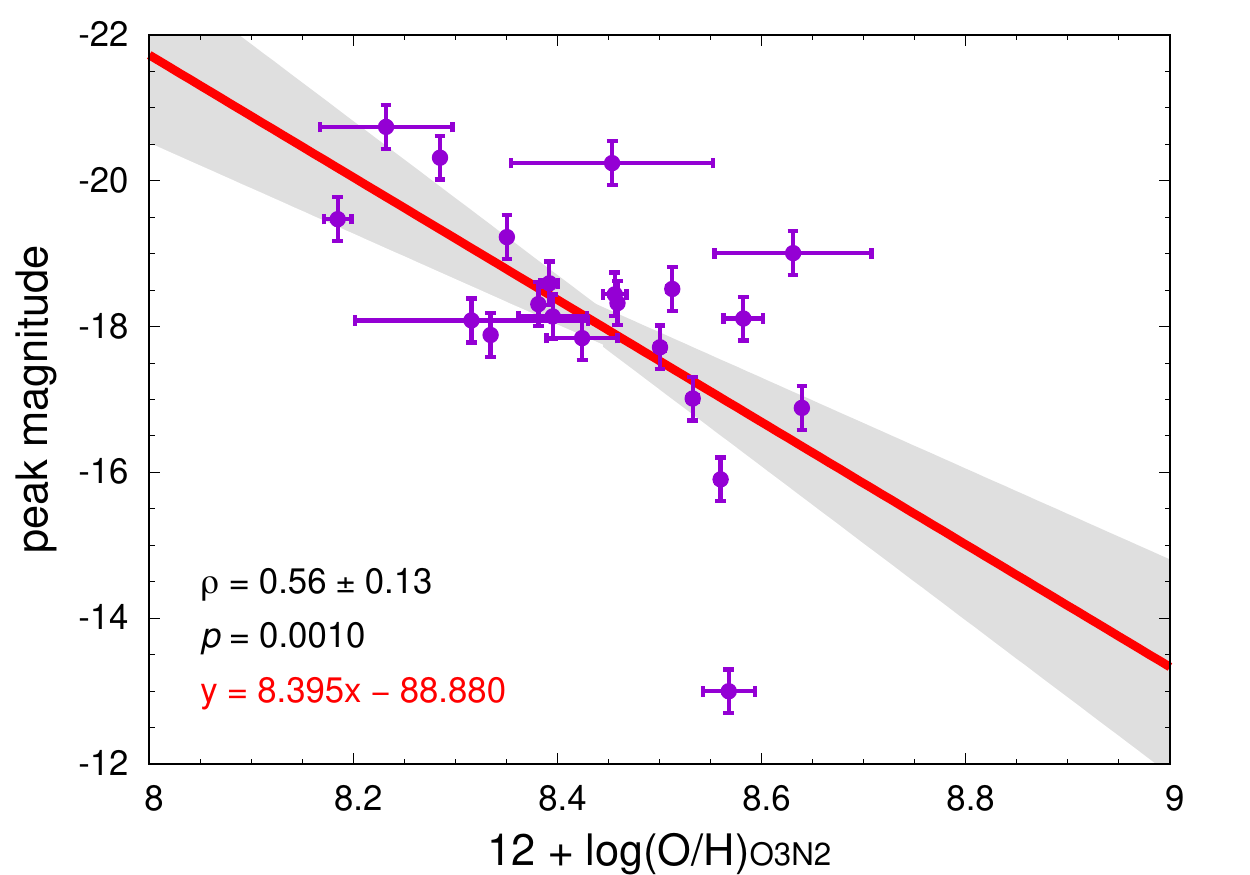} 
    \includegraphics[width=\columnwidth]{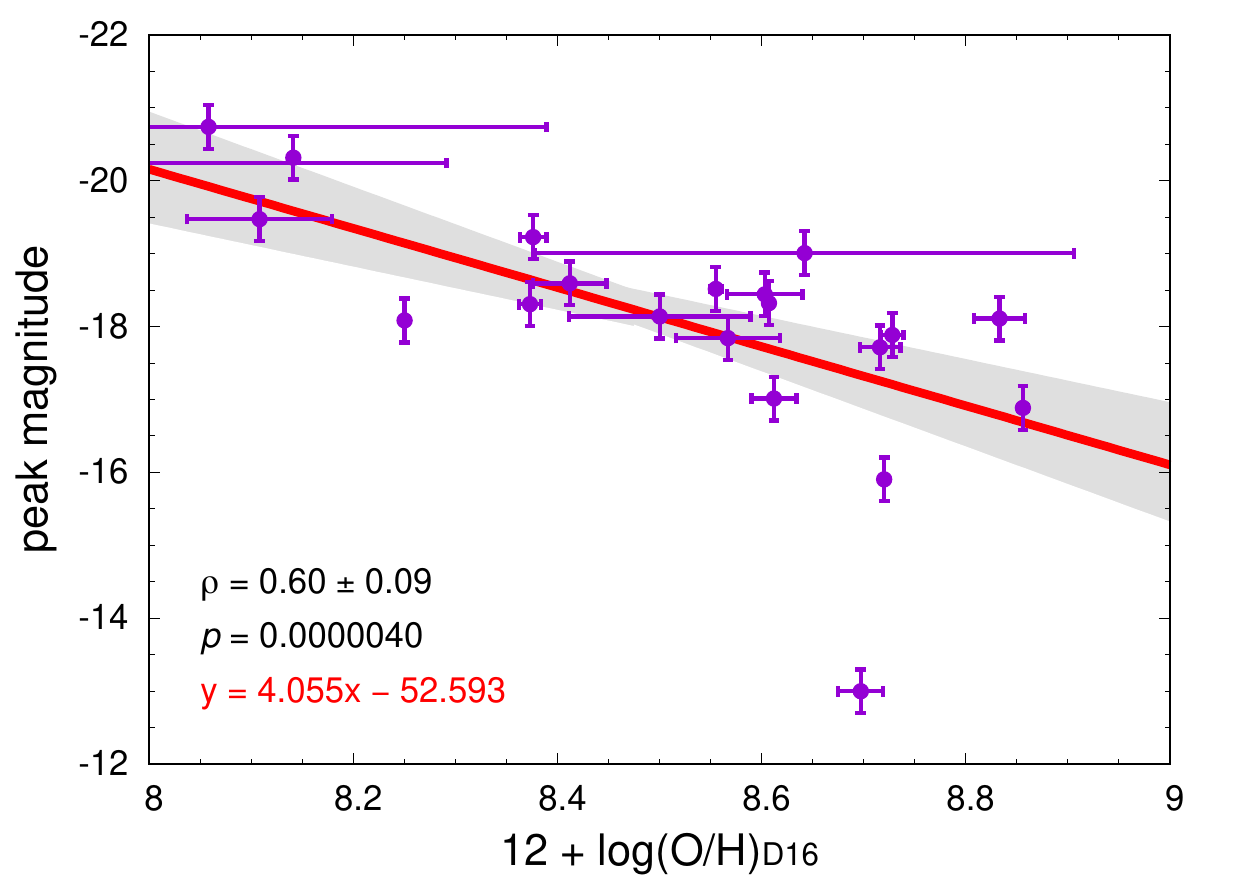} 
    \caption{
    Correlation between metallicity ($12+\log(\mathrm{O/H})$) and peak magnitude for the three different metallicity estimators (N2 at top, O3N2 at middle, and D16 at bottom). Each Pearson correlation coefficient $\rho$ is shown along with its standard deviation and $p$ value. The best linear fits are shown with the red lines and the $1\sigma$ region is indicated by the gray shades. No host extinction is applied here.
    }
    \label{fig:metallicity_peak}
\end{figure}

The light-curves are corrected for the Galactic extinction. The host galaxy extinction is uncertain. Although we estimated $E(B-V)$ and $A_{V*}$ from the host galaxy spectra, they do not necessarily represent the extinction at the exact SN location. Here, we assume three cases: no host extinction, the host extinction correction with $E(B-V)$, and the host extinction correction with $A_{V*}$. We find that our results are independent of the choice of the host galaxy extinction. We discuss the case without the host galaxy extinction in the following sections.

The rise time and peak luminosity of our SN~IIn sample are estimated using the method developed by \citet{ofek2014}. We fit the rising part of the light-curves to estimate $t_d$ and $L_p$,
\begin{equation}
    L(t) = L_p \left[1-\left(\frac{t-t_\mathrm{peak}}{t_d}\right)^2\right], \label{eq:fitting}
\end{equation}
where $t_\mathrm{peak}$ is the time of the luminosity peak. The fits are shown in Fig.~\ref{fig:lightcurve_fit}, and the estimated rise times and peak luminosities are summarized in Table~\ref{tab:samples}. Because of the uncertainties in the rise time and the peak luminosity caused by the distance uncertainties and bolometric corrections, we assume a $1\sigma$ uncertainty of 3~days and 0.3~mag in the rise time and peak luminosity, respectively.

Table~\ref{tab:samples} also includes the CSM density estimates. We also show the corresponding mass-loss rates for $v_\mathrm{wind}=100~
\kmps$. The estimated mass-loss rates with $v_\mathrm{wind}=100~\kmps$ range from $\sim 10^{-3}~\Msunpyr$ to $\sim 10^{-2}~\Msunpyr$, and they are consistent with previous studies \citep[e.g.,][]{ofek2014,moriya2014m}. In the following analysis, we assume an 0.5~dex uncertainty in CSM density estimates to account for possible systematic uncertainties as well as the uncertainties in estimating rise time and peak luminosity.

\begin{figure}
    \centering
    \includegraphics[width=\columnwidth]{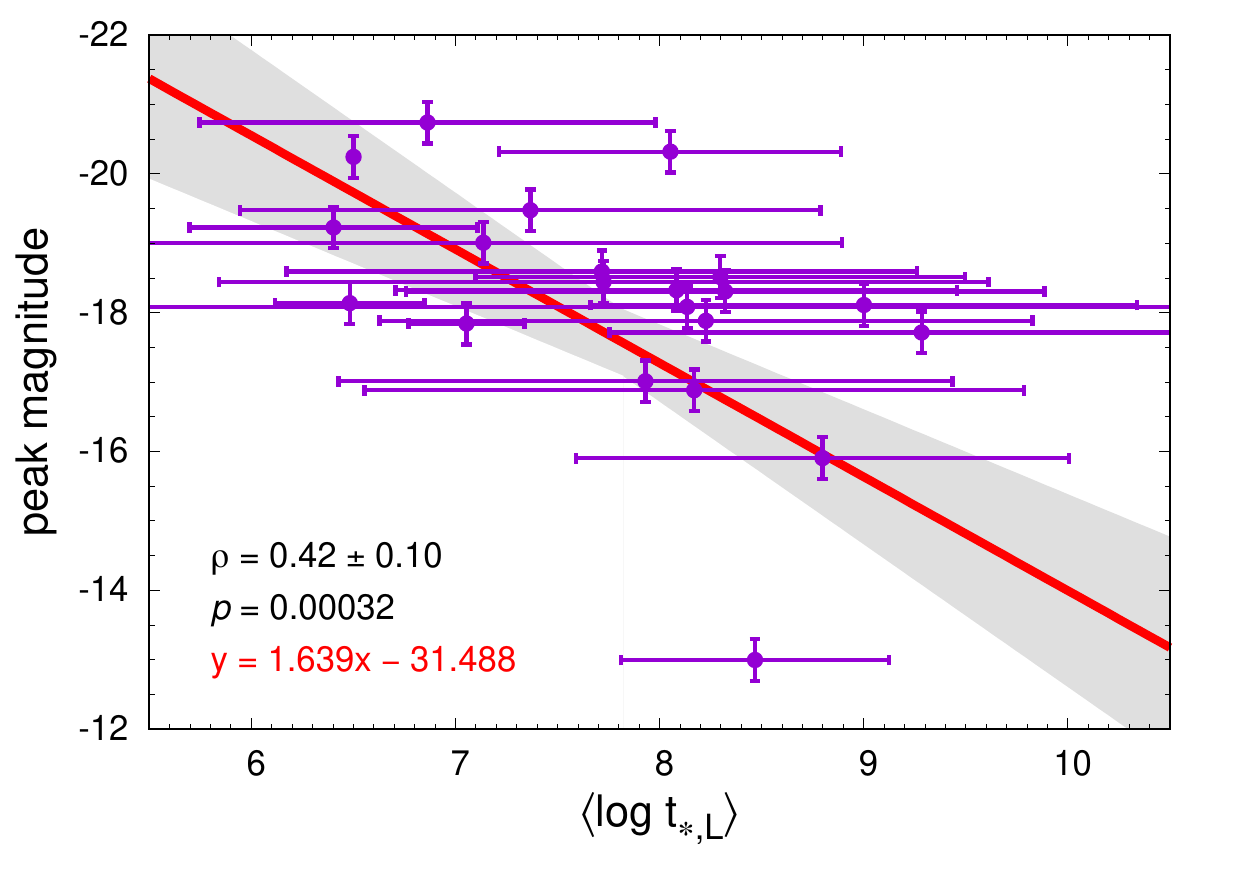} 
    \caption{
    Correlation between the average light-weighted stellar population age ($\langle\log t_{*,L}\rangle$) and the peak magnitude. See the caption of Fig.~\ref{fig:metallicity_peak} for details.
    }
    \label{fig:age_peak}
\end{figure}

\begin{figure}
    \centering
    \includegraphics[width=\columnwidth]{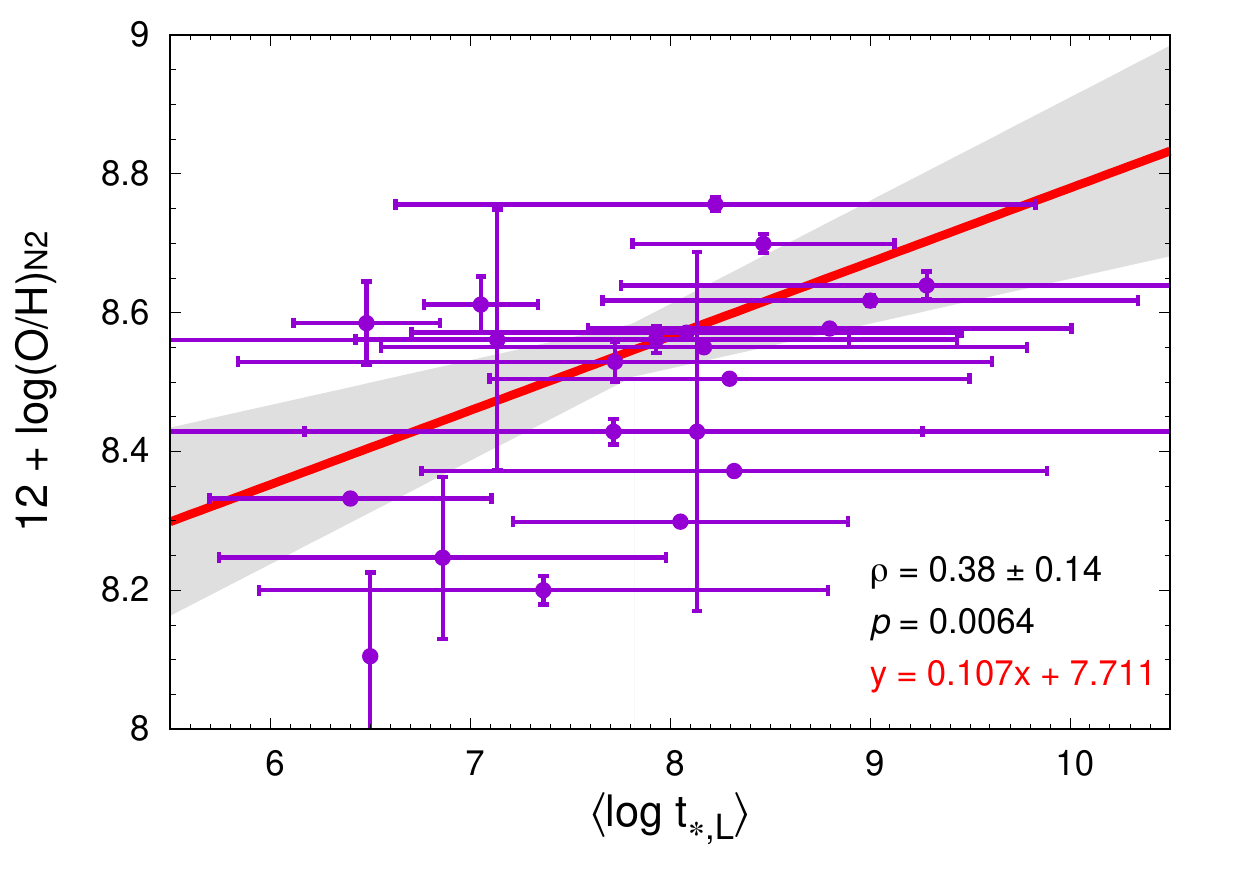} 
    \includegraphics[width=\columnwidth]{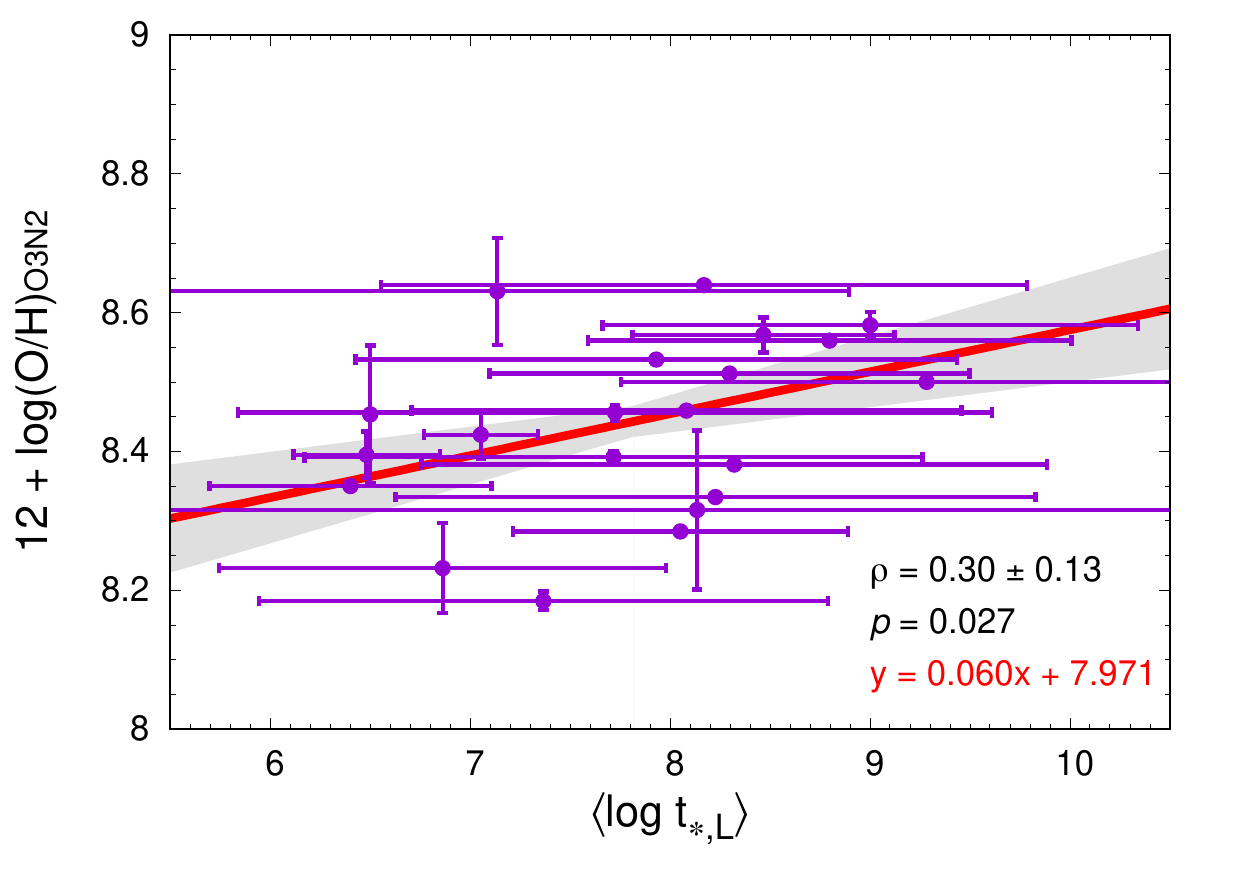} 
    \includegraphics[width=\columnwidth]{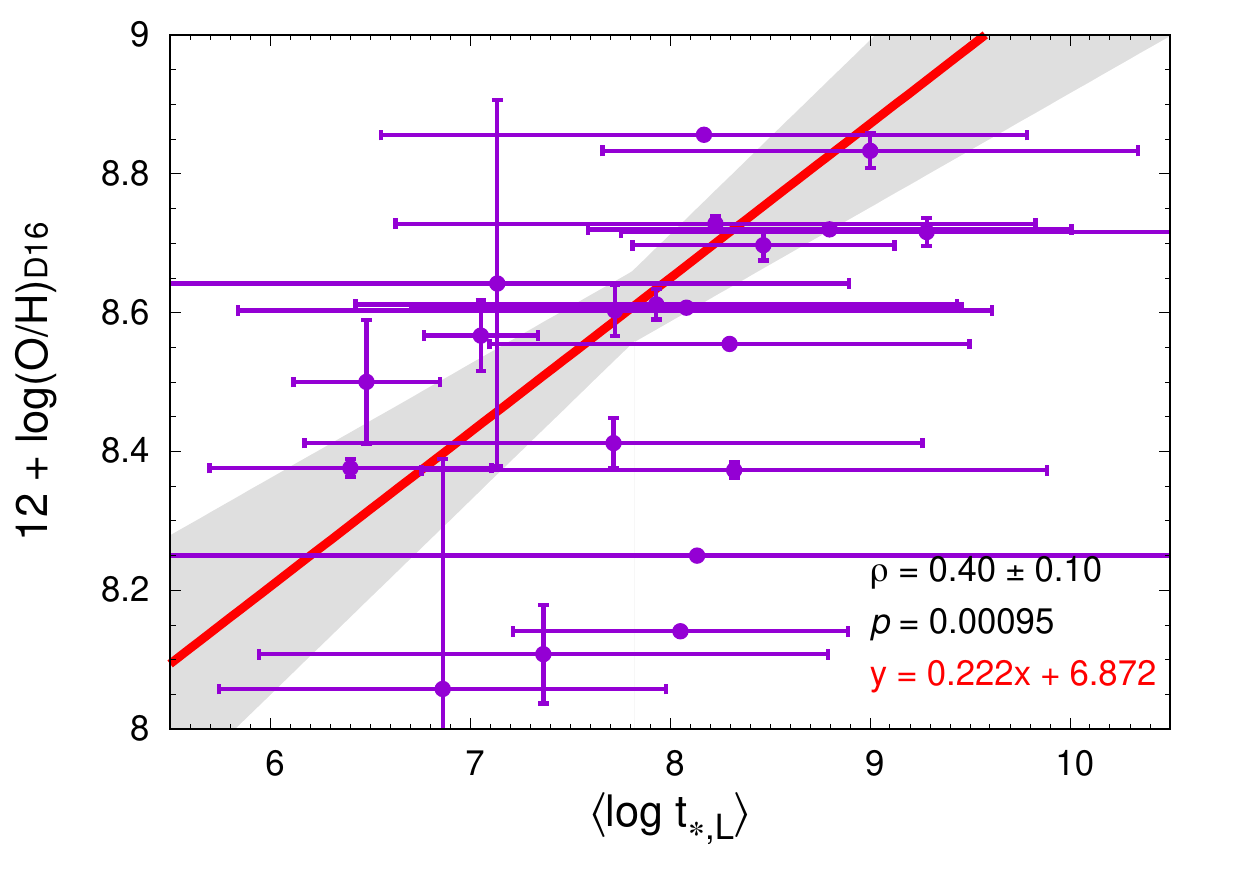}     
    \caption{
    Correlation between the average light-weighted stellar population age ($\langle\log t_{*,L}\rangle$) and the metallicity ($12+\log(\mathrm{O/H})$) for the three different metallicity estimators (N2 at top, O3N2 at middle, and D16 at bottom). See the caption of Fig.~\ref{fig:metallicity_peak} for details.
    }
    \label{fig:metallicity_age}
\end{figure}

\section{Environmental dependence}\label{sec:correlations}
Using the SN~IIn environmental properties (Section~\ref{sec:local_environments}) and SN~IIn properties (Section~\ref{sec:density_estimate}), we next investigate if there exist any correlations among them. We evaluate the Pearson correlation coefficient $\rho$ to determine the existence and strength of correlations. We employ $10^6$ bootstrapping simulations and derive the Pearson correlation coefficient, its standard deviation, and the $p$ value for each. Each bootstrapping simulation is performed by randomly selecting 21 SNe allowing multiple selections of the same SN~IIn.

Table~\ref{tab:correlations} summarizes the Pearson correlation coefficients, their standard deviations, and the $p$ values for each.
One statistically significant correlation is a positive correlation between the peak magnitude and all three metallicity indicators. This means that more luminous SNe~IIn tend to appear in lower metallicity environments. Figure~\ref{fig:metallicity_peak} illustrates the correlation. The other significant correlation is a very weak positive correlation between the peak magnitude and the average light-weighted stellar population age ($\langle\log t_{*,L}\rangle$). In other words, more luminous SNe~IIn prefer to occur in environments with younger stellar populations (Fig.~\ref{fig:age_peak}). We also found that metallicity and average light-weighted stellar population age might be weakly correlated (Fig.~\ref{fig:metallicity_age}). Thus, it is not clear if the peak luminosity correlation is driven by metallicity, stellar population age, or both. Because we found stronger correlations with metallicity, it is possible that metallicity difference is the main cause of the correlation.

\begin{figure}
    \centering
    \includegraphics[width=\columnwidth]{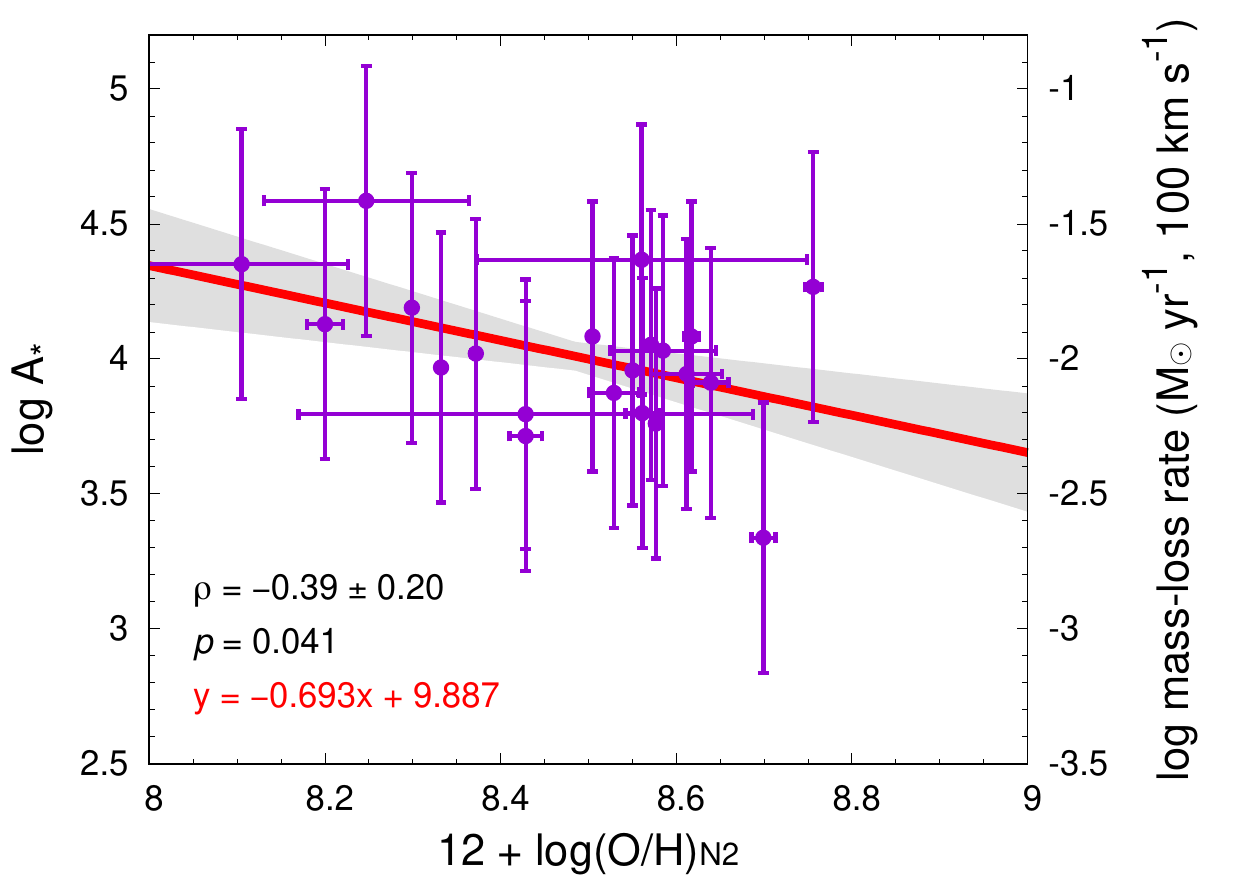} 
    \includegraphics[width=\columnwidth]{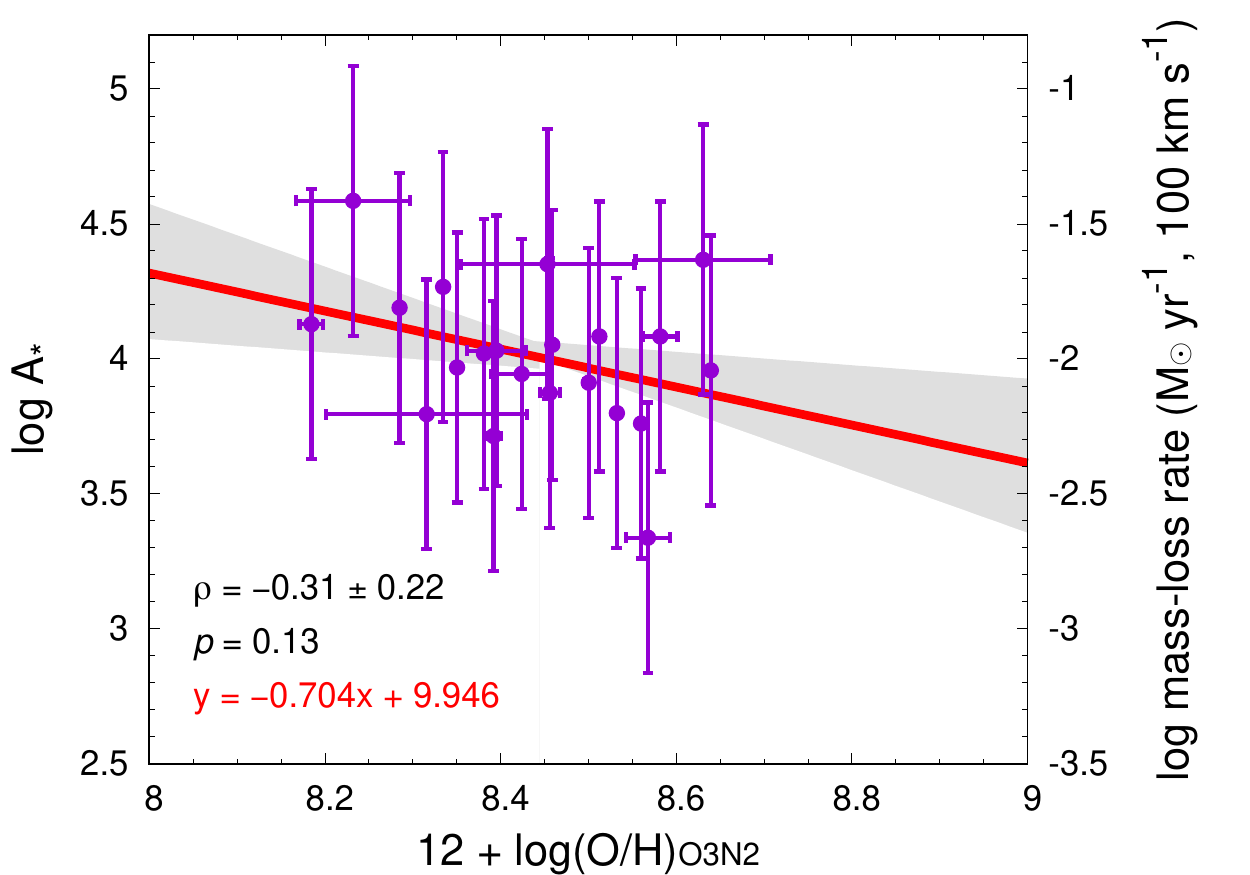} 
    \includegraphics[width=\columnwidth]{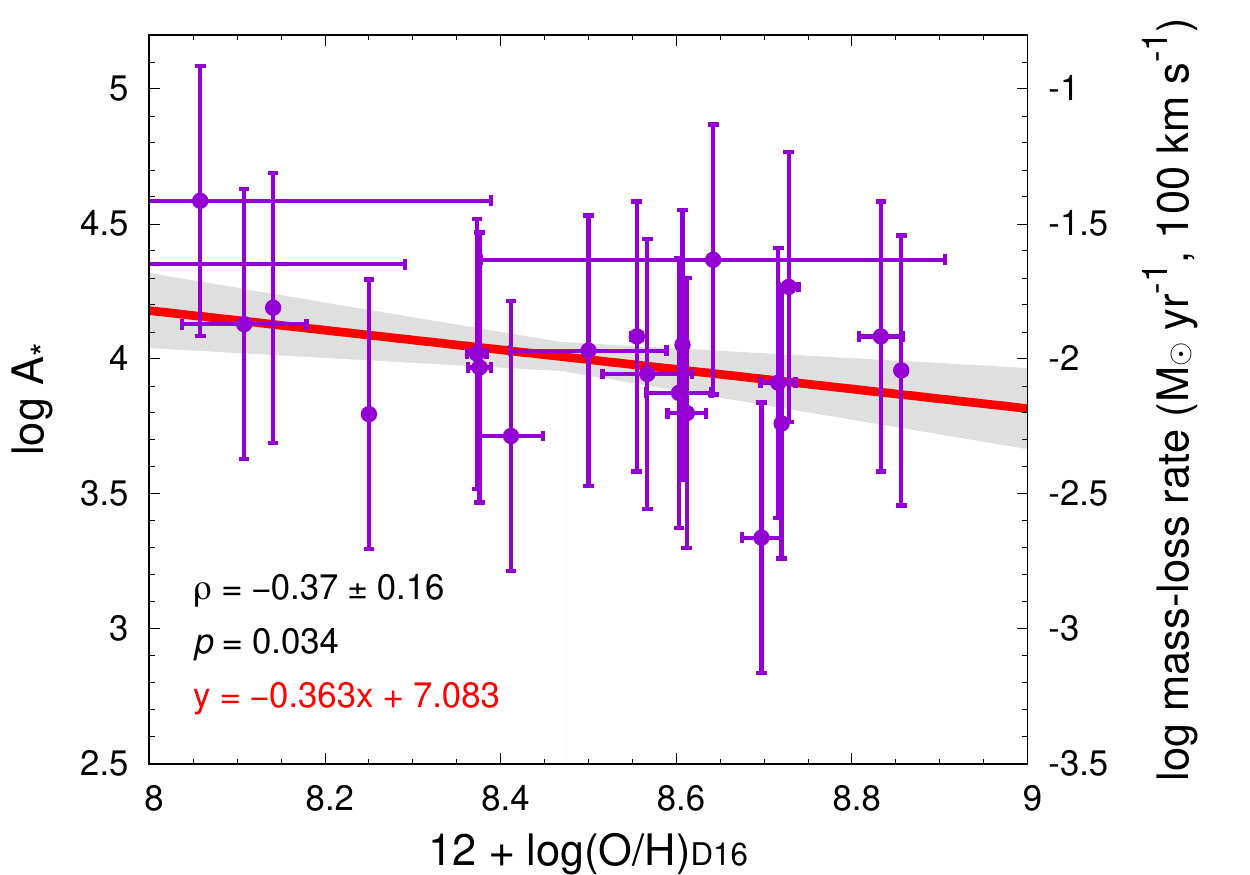} 
    \caption{
    Same as Fig.~\ref{fig:metallicity_peak}, but for CSM density ($\log A_\ast$). The right vertical axis shows the corresponding mass-loss rate for a wind velocity of 100~\kmps.
    }
    \label{fig:metallicity_dependence}
\end{figure}

\begin{figure}
    \centering
    \includegraphics[width=\columnwidth]{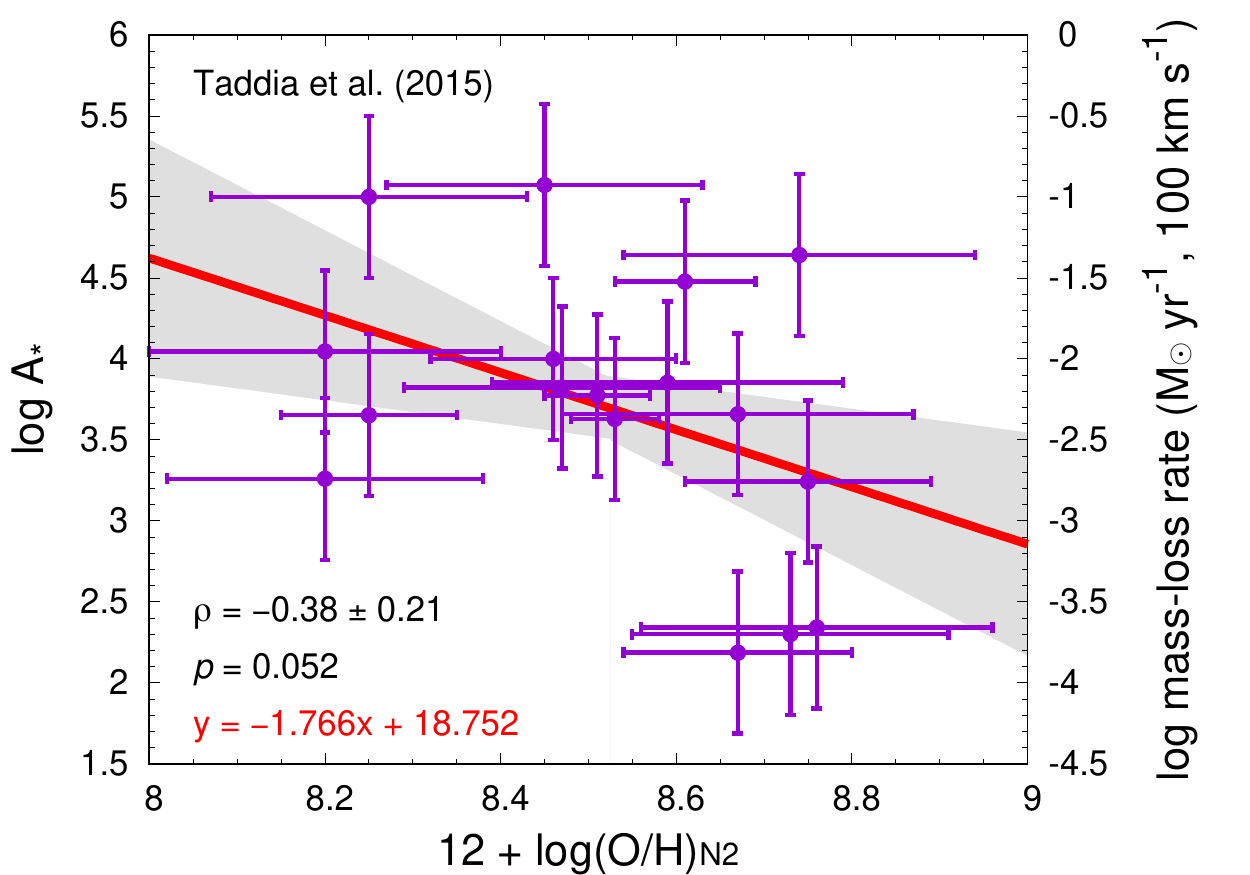} 
    \caption{
    Correlation between metallicity ($12+\log(\mathrm{O/H})_\mathrm{N2}$) and CSM density ($\log A_\ast$) for the SN~IIn sample in \citet{taddia2015}. The right vertical axis shows the corresponding mass-loss rate for a wind velocity of 100~\kmps. The best linear fitting function (red line) with the $1\sigma$ region (gray shade) is shown.
    }
    \label{fig:taddia_n2}
\end{figure}

It is worth noting that we do not find significant correlations between metallicity and CSM density (Fig.~\ref{fig:metallicity_dependence}). A very weak negative correlation between metallicity and CSM density (i.e., SNe~IIn with higher metallicity tend to have less dense CSM) may exist, but it is still statistically marginal and depends on the metallicity indicator. Interestingly, no positive correlation is likely to exist. \citet{taddia2015} previously investigated the metallicity dependence of mass-loss rates and wind velocities in SNe~IIn. They concluded that SNe~IIn from higher metallicity environments have higher mass-loss rates and wind velocities. Figure~\ref{fig:taddia_n2} shows the CSM density estimates from the SNe~IIn used in their analysis. The mass-loss rates and wind velocities in \citet{taddia2015} estimates taken from a range of sources using different methodologies, and are not necessarily estimated in a consistent way. Nonetheless, we do not find a significant correlation in the CSM density and metallicity in their sample, either. Our results show that, although mass-loss rates and wind velocities may have metallicity dependence as proposed by \citet{taddia2015}, the CSM density ($A\propto \dot{M}/v_\mathrm{wind}$) is not significantly metallicity dependent.

For the other combinations of the parameters, we do not find any statistically significant correlations. There may be other very weak correlations such as between the rise time and $\log \Sigma_\mathrm{SFR}$, between the peak magnitude and $\log\mathrm{sSFR}$, and between $\log A_\ast$ and $\langle\log t_{*,L}\rangle$. More SNe~IIn are required to determine the validity of any additional correlations.

\section{Discussion}\label{sec:discussion}
We found that there is a negative correlation between metallicity and peak luminosity of SNe~IIn in the sense that more luminous SNe~IIn are associated with lower metallicity environments. We also found a weak negative correlation between stellar population age and peak luminosity. The luminosity of SNe~IIn can be characterized by $\varepsilon E_\mathrm{kin}/t_d$, where $E_\mathrm{kin}$ is the kinetic energy in the shocked SN ejecta up to the time of the luminosity peak \citep{moriya2014m}. We found that rise time, which is related to $t_d$, does not correlate with metallicity or stellar age. The conversion efficiency $\varepsilon$ is not likely to be sensitive to metallicity and stellar population age, although it could be higher for higher metallicities because of more efficient cooling. Thus, the negative correlation could be caused by the fact that SNe~IIn tend to have higher explosion energy at lower metallicity environments and/or younger stellar populations. Because higher mass progenitors tend to have higher explosion energies \citep[e.g.,][]{martinez2022}, it may be natural to expect that SNe~IIn from younger stellar populations to have higher explosion energies. However, we do not find any correlations between EW(H$\alpha$) and peak luminosity. It is also possible that SN~IIn progenitor masses tend to be higher at lower metallicity. 

We did not find a significant correlation between metallicity and CSM density. This is interesting because some mass-loss mechanisms predict a positive correlation between mass-loss rate and metallicity. For example, in the case of hot massive stars, \citet{bjorklund2021} find that
\begin{align}
    \log\left(\frac{\dot{M}}{\Msunpyr}\right) =& -5.55 + 0.79 \log \left(\frac{Z}{Z_\odot}\right) \nonumber \\ &+ \left[2.16-0.32\log \left(\frac{Z}{Z_\odot}\right) \right]\log\left(\frac{L}{10^6 L_\odot}\right),
\end{align}
with 
$v_\mathrm{wind} \propto Z^{p(L)}$ and $p(L) = -0.41 \log \left(\frac{L}{10^6 L_\odot}\right)-0.32$.
Here, $Z$ is metallicity and $L$ is luminosity of a star. This leads to a CSM density factor scaling of
\begin{equation}
    A\propto Z^{1.11+0.09\log(L/10^6L_\odot)}L^{2.16}.
\end{equation}
For a given luminosity, the CSM density is expected to positively correlate with the metallicity. In order to have no or negative correlations between $A$ and $Z$, the SN~IIn progenitor luminosity $L$ could increase at low metallicity. Ignoring the small term $0.09\log(L/10^6L_\odot)$ and assuming $L\propto Z^\alpha$ for SN~IIn progenitors, we obtain $A\propto Z^{1.11+2.16\alpha}$. Thus, $\alpha \lesssim -0.5$ is required to have no or negative correlations between $Z$ and $A$. If the progenitor luminosity is close to the Eddington luminosity (i.e., $L\propto M$), an increase in progenitor mass by a factor of around 2 for a metallicity increase by a factor of 0.3 would produce no correlations, for example.

In the case of cool stars such as RSGs, the metallicity dependence of $\dot{M}$ is not so clear. RSG mass-loss rates have been suggested to follow a relation of $\dot{M}\propto L^{1.05}Z^{0.7}$ with $v_\mathrm{wind}\propto L^{0.35}$ \citep[][and the references therein]{mauron2011}, while \citet{goldman2017} suggested no metallicity dependence for RSG mass-loss rates ($\dot{M}\propto L^{0.9}$ with $v_\mathrm{wind}\propto ZL^{0.4}$). The two prescriptions predict quite different CSM density dependences on metallicity with $A\propto L^{0.7}Z^{0.7}$ \citep{mauron2011} or $A\propto L^{0.5}Z^{-1}$ \citep[][]{goldman2017}. In both cases, CSM density around RSGs is predicted to strongly depend on metallicity. Nonetheless, because of huge uncertainties in the metallicity dependence of RSG mass loss, it is difficult to judge from the metallicity dependence whether SN~IIn progenitors are dominated by RSGs or not. Additional investigations into the metallicity dependence of RSG mass loss are required.

Because of their high mass-loss rates, the progenitors of SNe~IIn may actually have optically-thick winds forming a dense CSM. Mass-loss rates and wind velocities from optically-thick winds are also predicted to be metallicity dependent, but their dependence may also compensate to have a metallicity-independent CSM density \citep[e.g.,][]{grafener2008,sander2020}.

It is also possible that the normal mass-loss mechanisms for hot and cool stars are irrelevant for SN~IIn progenitors. Their CSM density may be driven by a totally different mass-loss mechanism that is not strongly affected by metallicity. Precursors observed in some SNe~IIn \citep[e.g.,][]{ofek2013,strotjohann2021} may indeed indicate that their mass-loss mechanism is quite different from those of metallicity-dependent steady winds %calm stars
discussed above. For example, continuum-driven winds are not expected to have a metallicity dependence \citep[e.g.,][]{smith2006}. Further investigation of the environmental dependence of SN~IIn properties would help understanding such an unknown mass-loss mechanism in SNe~IIn.

Another possibility to explain the apparent lack of a metallicity dependence is that the CSM density actually depends on the metallicity, but we do not find it clearly because the CSM density needs to be high enough to be observed as SNe~IIn. We might be simply biased to SNe having a CSM density above a certain metallicity-independent threshold by observing SNe~IIn. In such a case, the apparent lack of the metallicity dependence would simply be an observational bias.

\section{Conclusions}\label{sec:conclusions}
Using 21 SNe~IIn with good light-curves and local IFS data, we investigated the relationship between local environments and SN properties. We found that SNe~IIn with a higher peak luminosity tend to be in environments with lower metallicities and stellar population ages. Because metallicity and stellar population age are correlated in our sample, it is unclear if metallicity, stellar population age, or both drive the correlations. The correlations may indicate that SNe~IIn have higher explosion energies in environments with lower metallicity and/or younger stellar ages.

We did not find statistically significant correlations between local metallicity and CSM density around SNe~IIn. There might be a very weak negative correlation, but no positive correlation exists. This indicates that the mass-loss mechanism triggering the formation of dense CSM around SNe~IIn could be metallicity independent. Alternatively, SN~IIn progenitor mass range may depend on metallicity. It is also possible that the lack of the metallicity dependence is an observational bias due to needing a minimum threshold CSM density to be classified as a SN~IIn.

Our study is based on 21 SNe~IIn. Some correlations are still not significant and further confirmation is required. In addition, it is possible that some bias exist in our samples. Thus, a similar study with larger numbers of SNe~IIn is encouraged. Wide-field high-cadence transient surveys are increasing the number of well-observed SNe~IIn. Follow-up observations to obtain local environment information to increase the sample size will be important in uncovering the mysterious nature of SNe~IIn.

%%%%%%%%%%%%%%%%%%%%%%%%%%%%%%%%%%%%%%%%%%%%%%%%%%
\begin{acknowledgements}
We thank the anonymous referee for thoughtful comments.
%Acknowledgement for NAOJ grant/visitor programme.
This work was supported by the NAOJ Research Coordination Committee, NINS (NAOJ-RCC-2201-0401).
TJM is supported by the Grants-in-Aid for Scientific Research of the Japan Society for the Promotion of Science (JP20H00174, JP21K13966, JP21H04997).
L.G. acknowledges financial support from the Spanish Ministerio de Ciencia e Innovaci\'on (MCIN), the Agencia Estatal de Investigaci\'on (AEI) 10.13039/501100011033, and the European Social Fund (ESF) "Investing in your future" under the 2019 Ram\'on y Cajal program RYC2019-027683-I and the PID2020-115253GA-I00 HOSTFLOWS project, from Centro Superior de Investigaciones Cient\'ificas (CSIC) under the PIE project 20215AT016, and the program Unidad de Excelencia Mar\'ia de Maeztu CEX2020-001058-M.
H.K. was funded by the Academy of Finland projects 324504 and 328898.
JDL acknowledges support from a UK Research and Innovation Fellowship (MR/T020784/1).
We acknowledge the Telescope Access Program (TAP) funded by the NAOC, CAS and the Special Fund for Astronomy from the Ministry of Finance. SD acknowledges Project number 12133005 supported by National Natural Science Foundation of China (NSFC) and the Xplorer Prize.
This work is supported by the Japan Society for the Promotion of Science Open Partnership Bilateral Joint Research Project between Japan and Chile (JPJSBP120209937, JPJSBP120239901).
This work was funded by ANID, Millennium Science Initiative, ICN12\_009.
Based on observations collected at the Centro Astron\'omico Hispano en Andaluc\'ia (CAHA) at Calar Alto, operated jointly by Junta de Andaluc\'ia and Consejo Superior de Investigaciones Cient\'ificas (IAA-CSIC).
Based on observations collected at the European Organisation for Astronomical Research in the Southern Hemisphere under ESO programmes
096.D-0296,
0100.D-0341,
0103.D-0440,
0101.D-0748, 
196.B-0578, and
1100.B-0651.
This research was partly supported by the Munich Institute for Astro-, Particle and BioPhysics (MIAPbP) which is funded by the Deutsche Forschungsgemeinschaft (DFG, German Research Foundation) under Germany´s Excellence Strategy – EXC-2094 – 390783311.
\end{acknowledgements}
%%%%%%%%%%%%%%%%%%%%%%%%%%%%%%%%%%%%%%

\appendix

\section{Figures of the SN environments}
\begin{figure*}
    \includegraphics[width=0.24\textwidth]{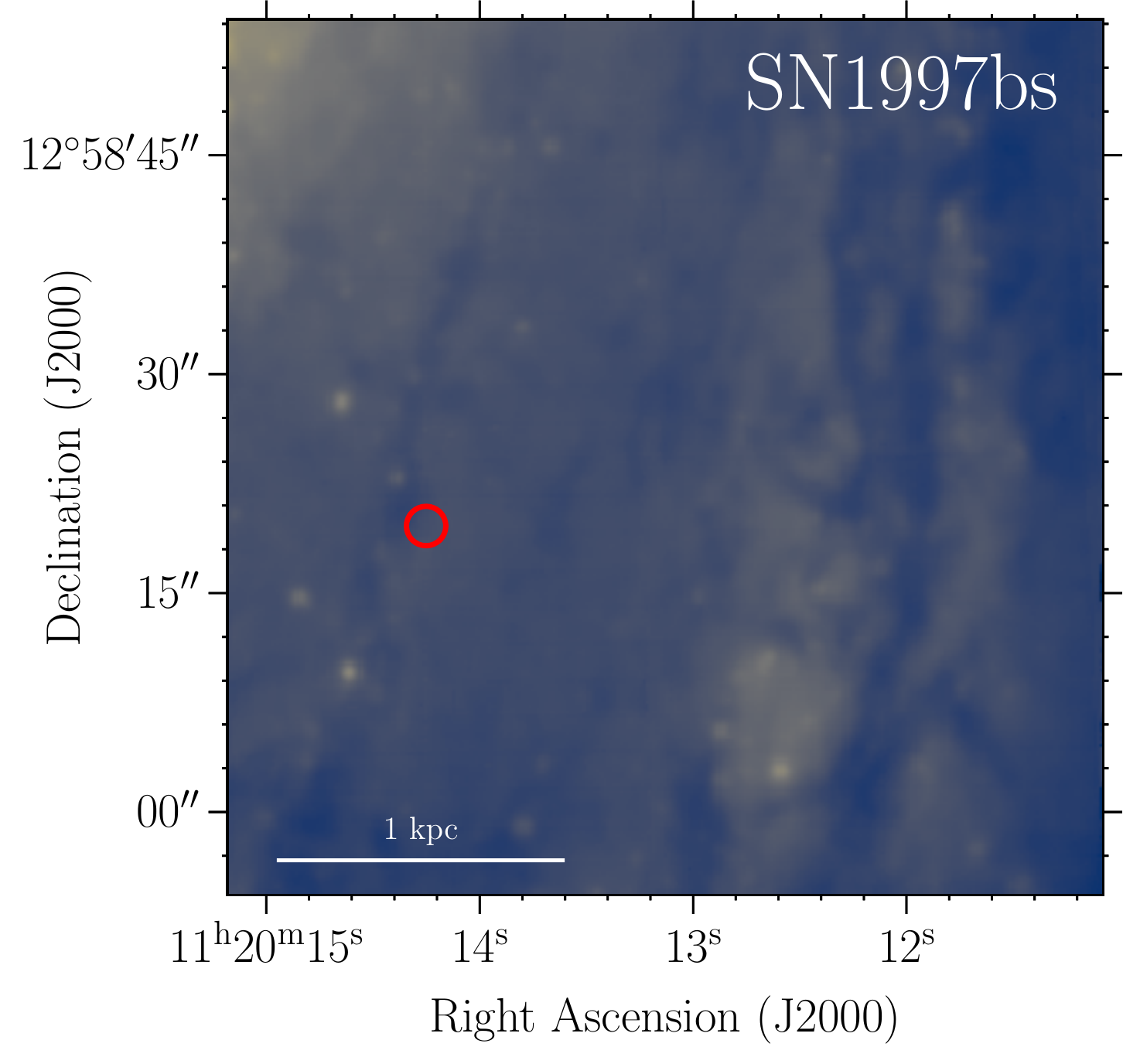}
    \includegraphics[width=0.24\textwidth]{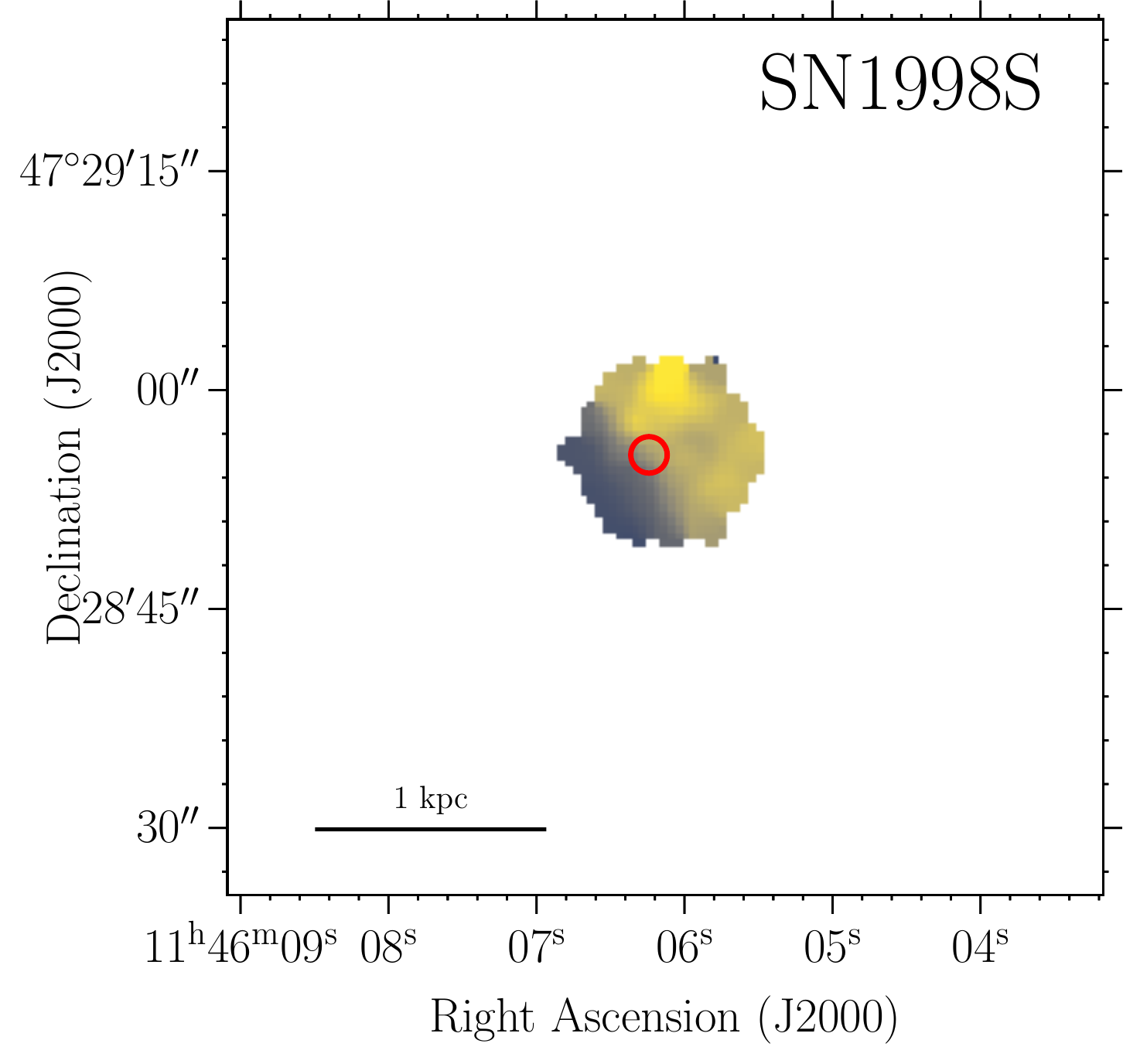}
    \includegraphics[width=0.24\textwidth]{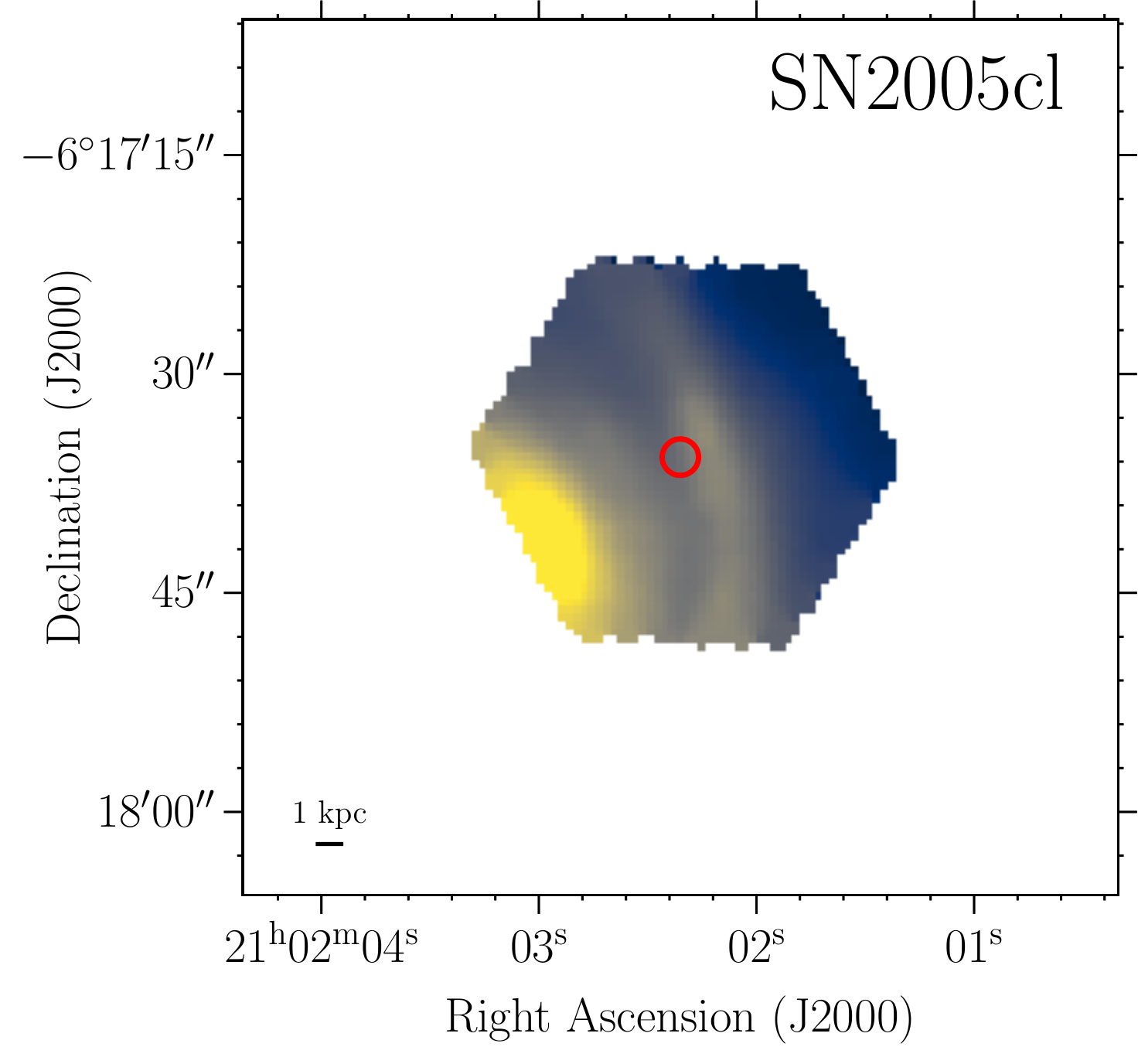}
    \includegraphics[width=0.24\textwidth]{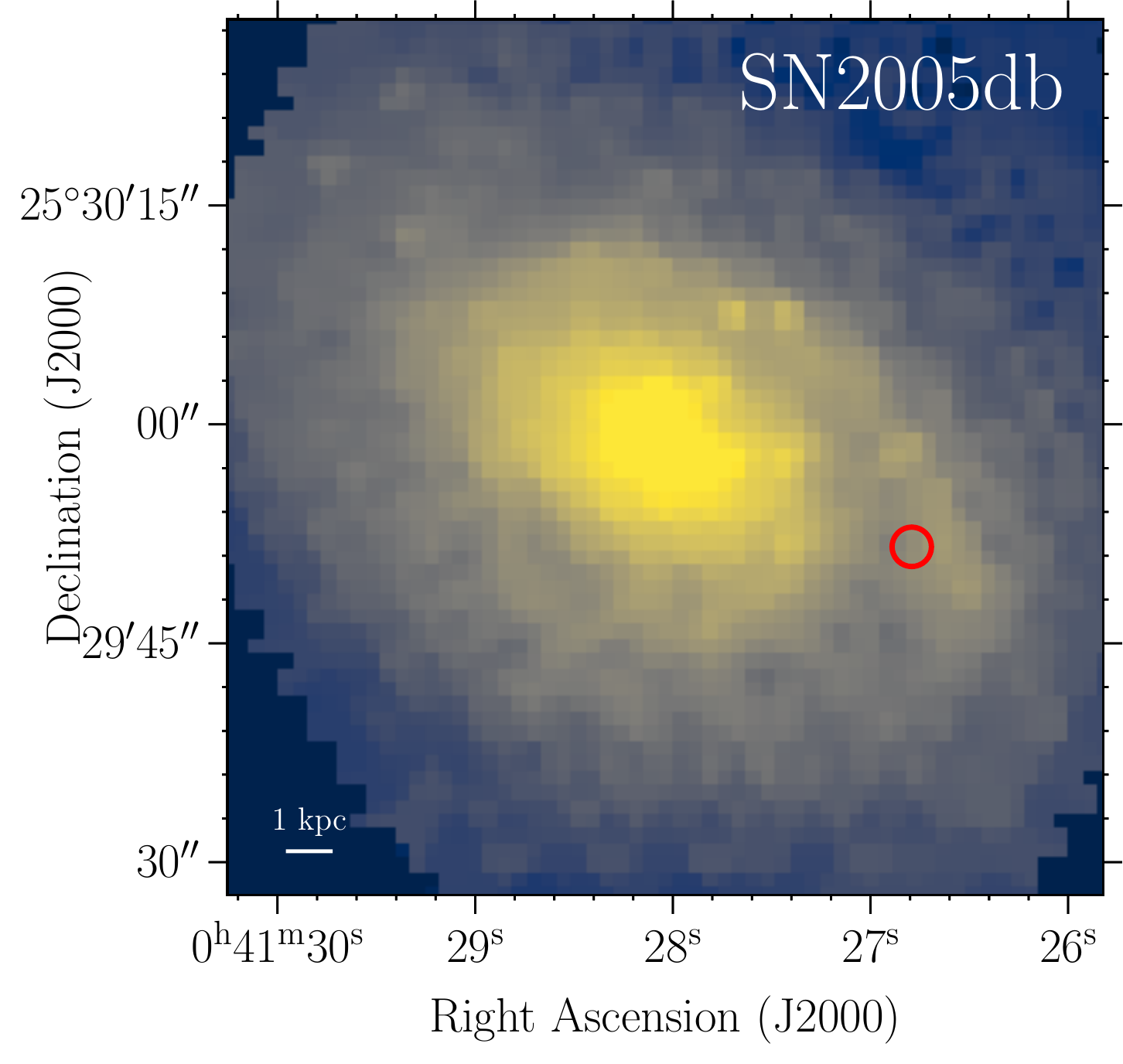}
    \includegraphics[width=0.24\textwidth]{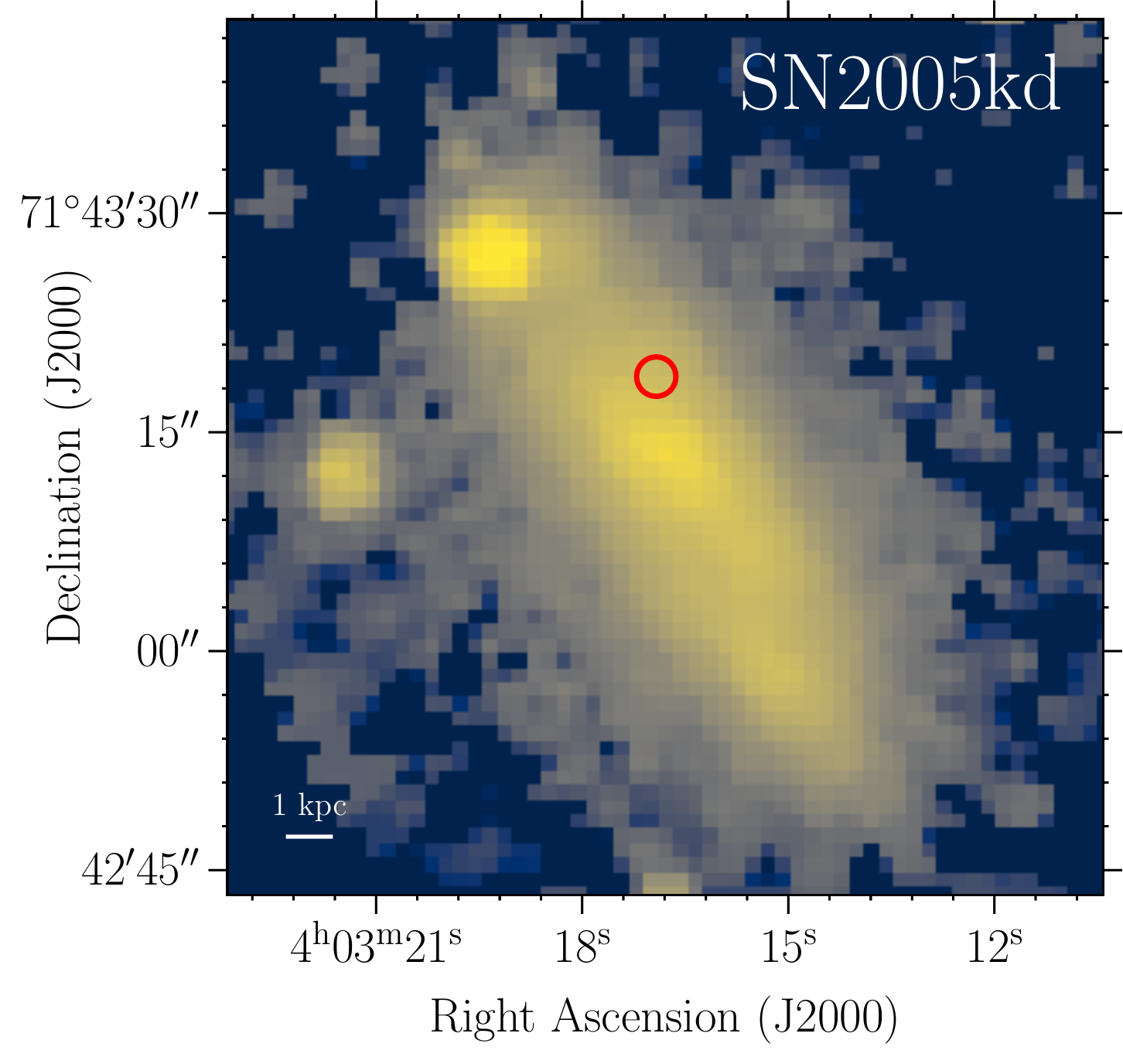}
    \includegraphics[width=0.24\textwidth]{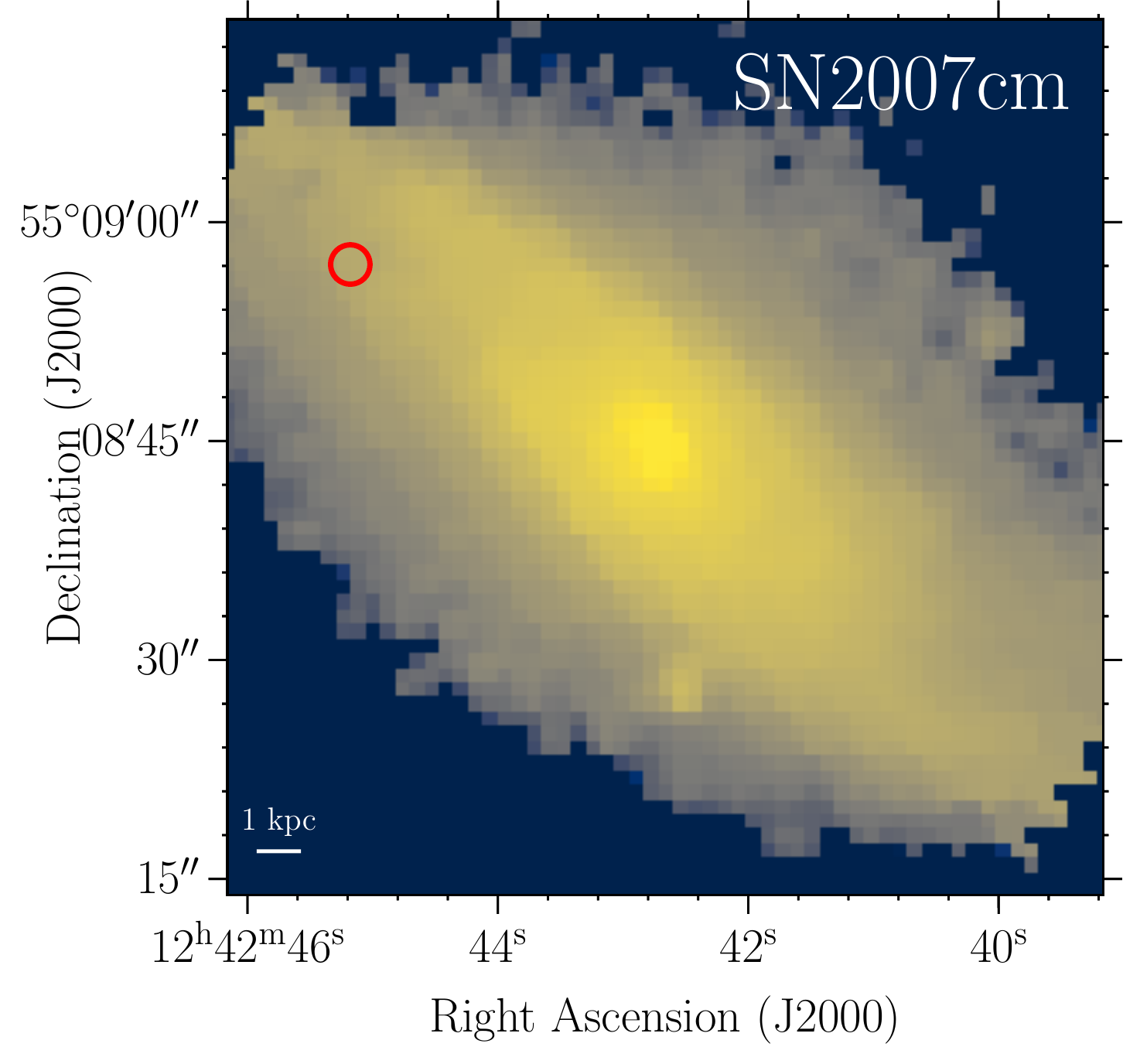}
    \includegraphics[width=0.24\textwidth]{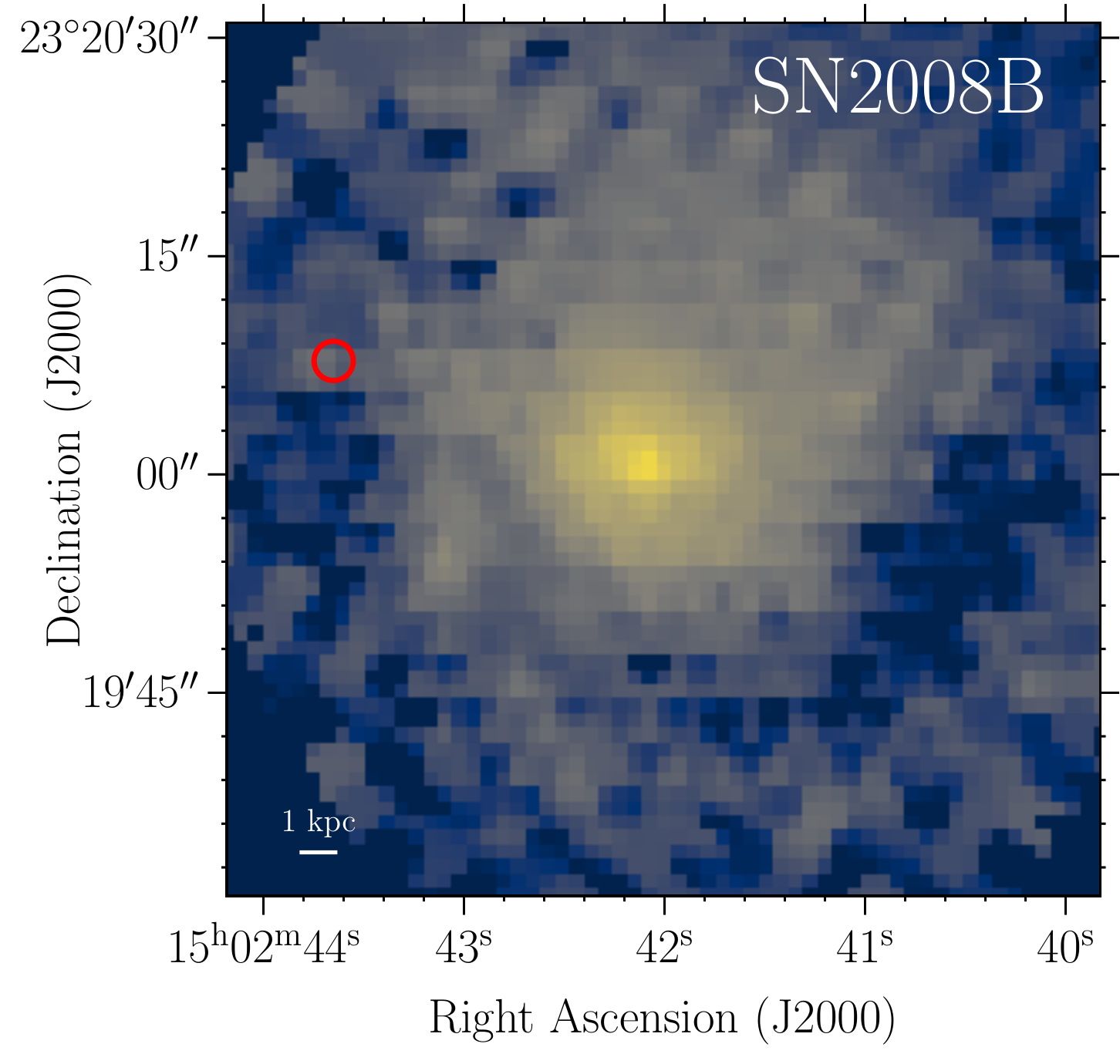}
    \includegraphics[width=0.24\textwidth]{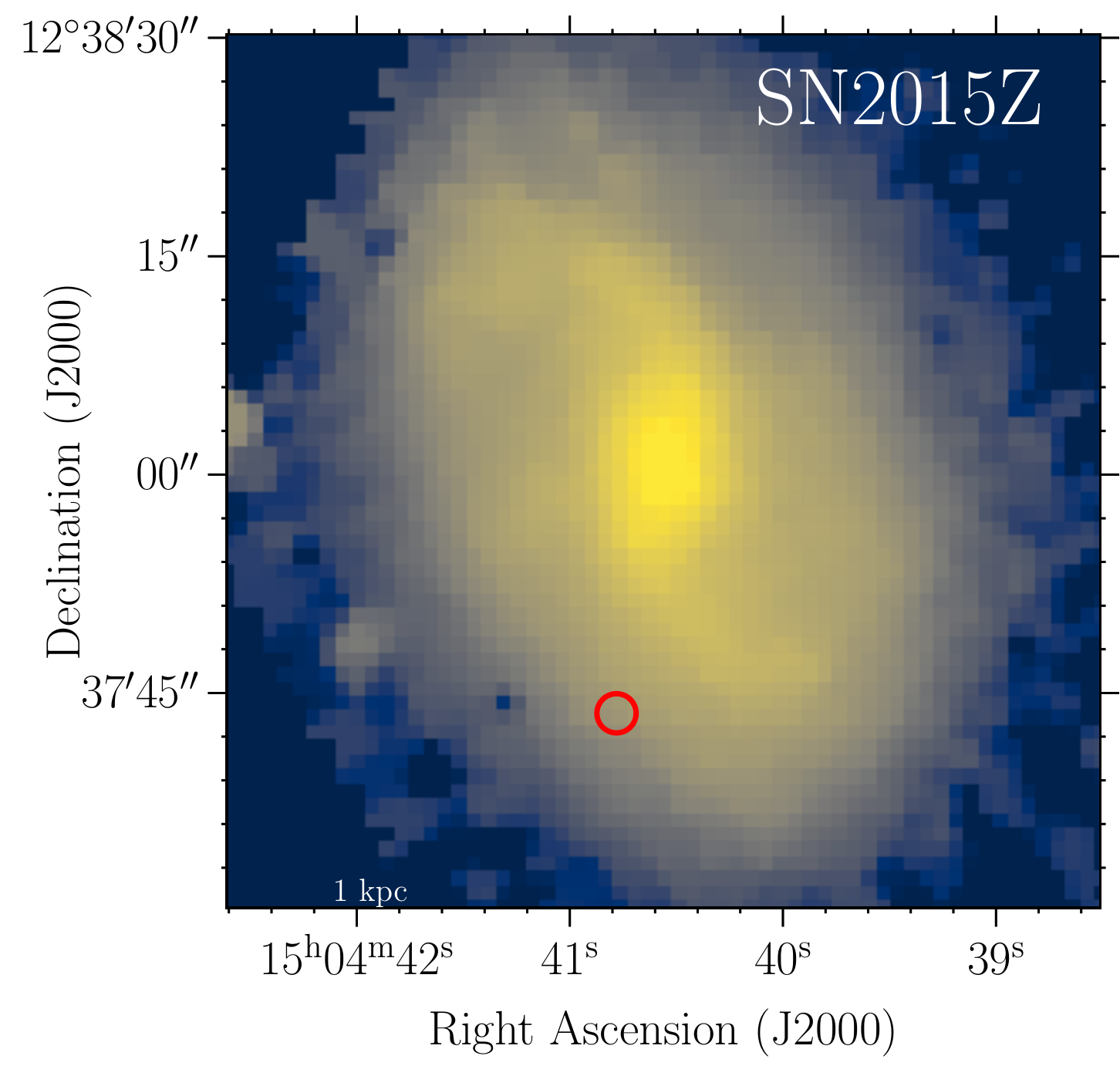}
    \includegraphics[width=0.24\textwidth]{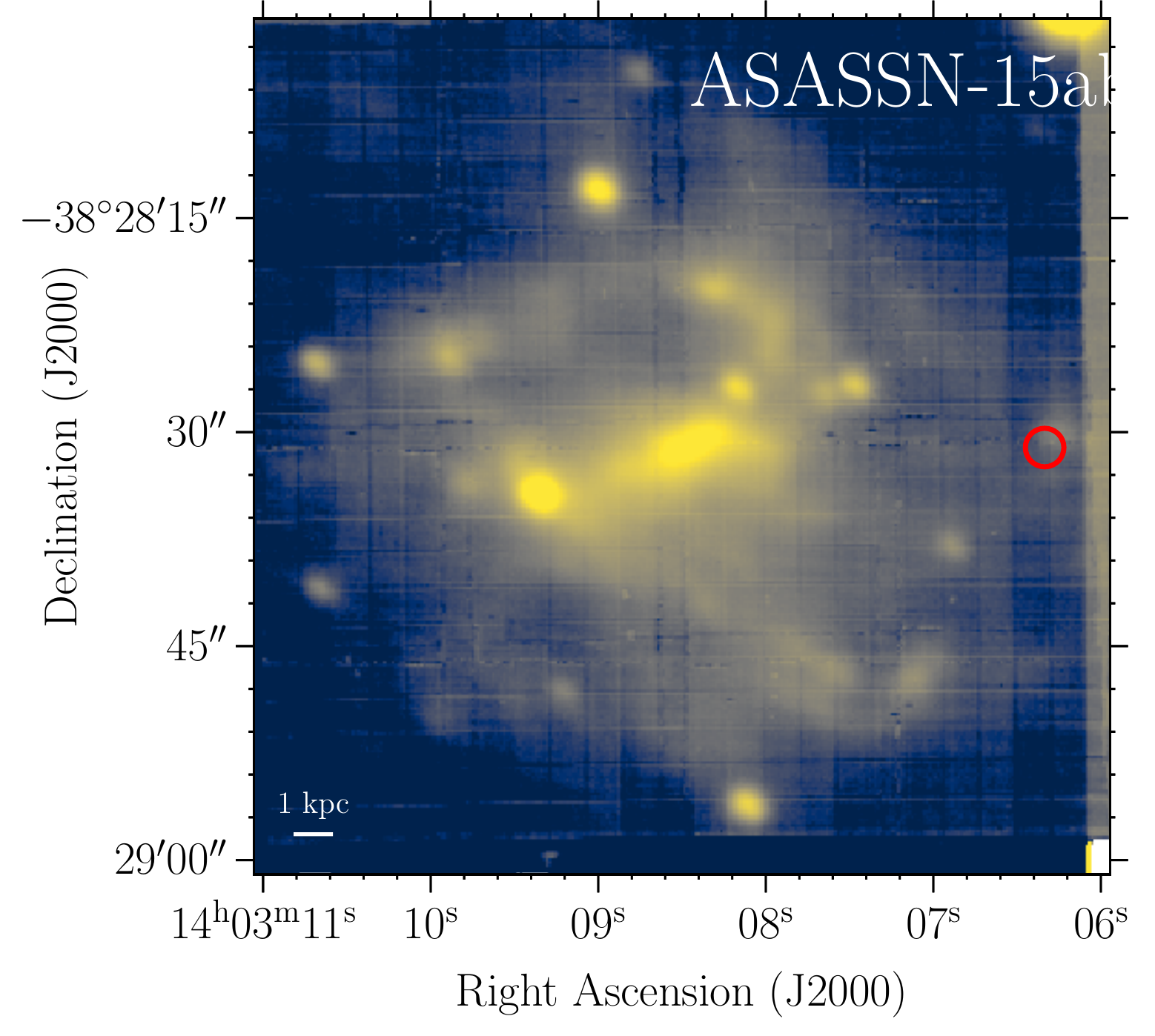}
    \includegraphics[width=0.24\textwidth]{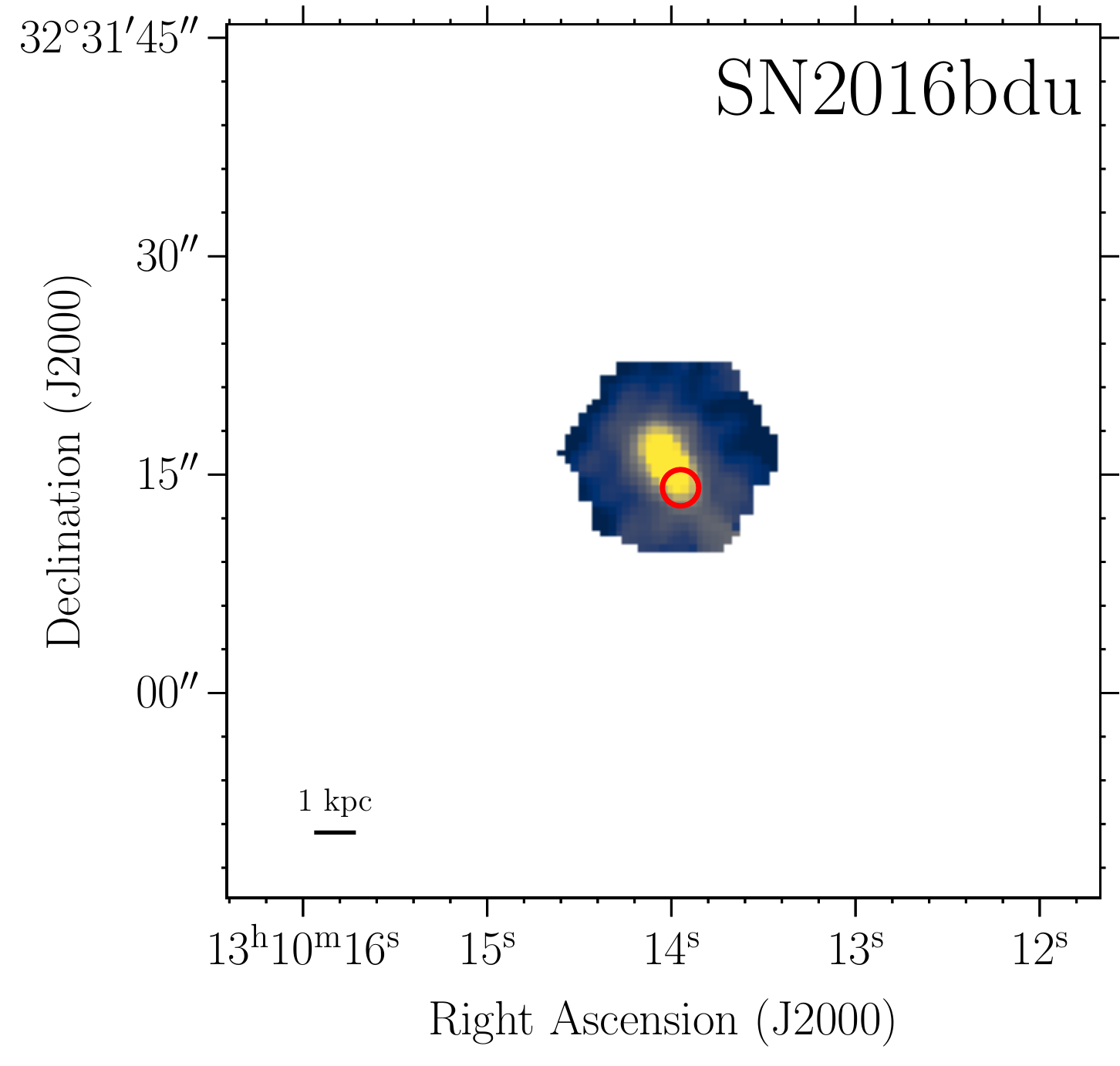}
    \includegraphics[width=0.24\textwidth]{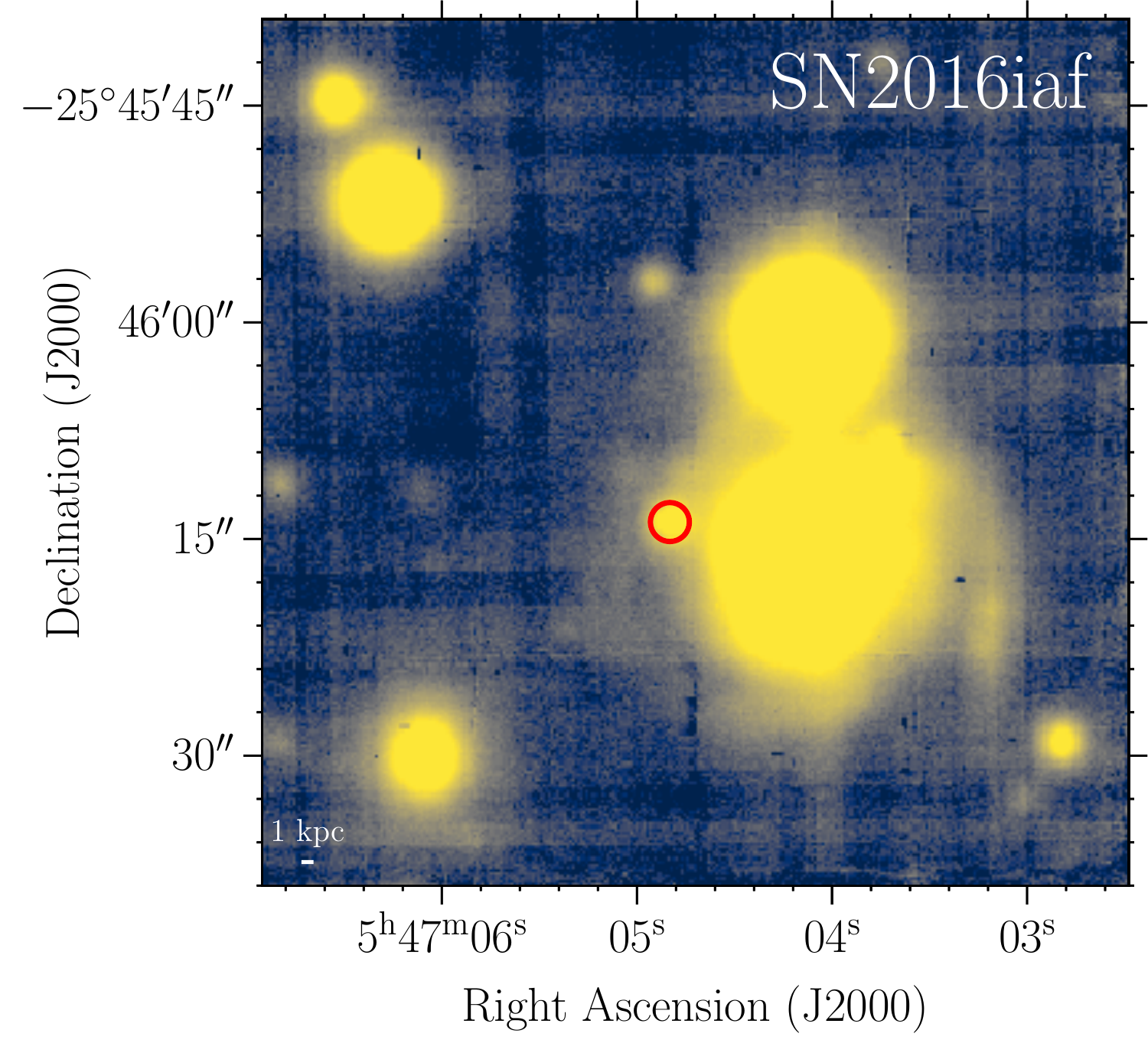}
    \includegraphics[width=0.24\textwidth]{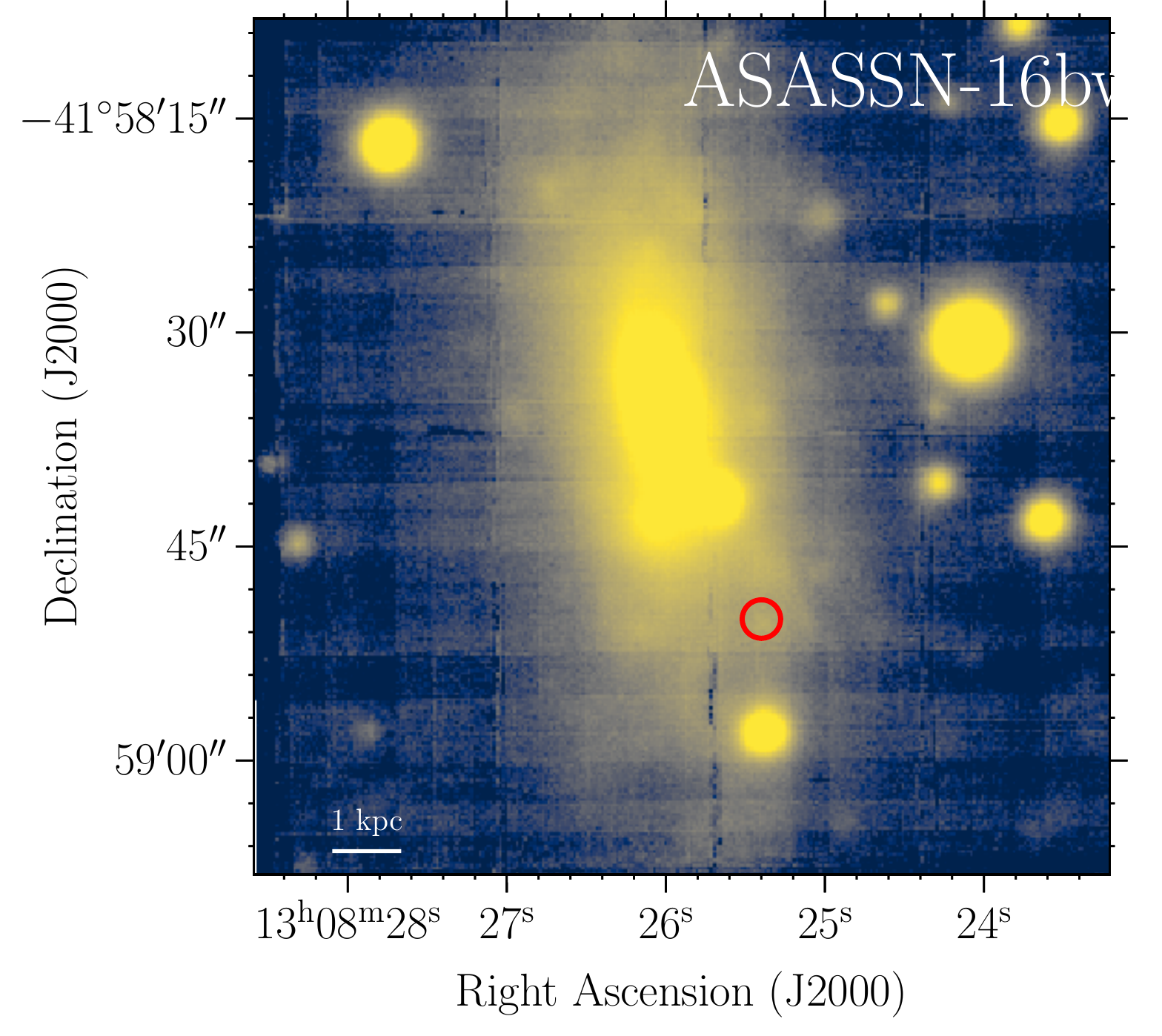}
    \includegraphics[width=0.24\textwidth]{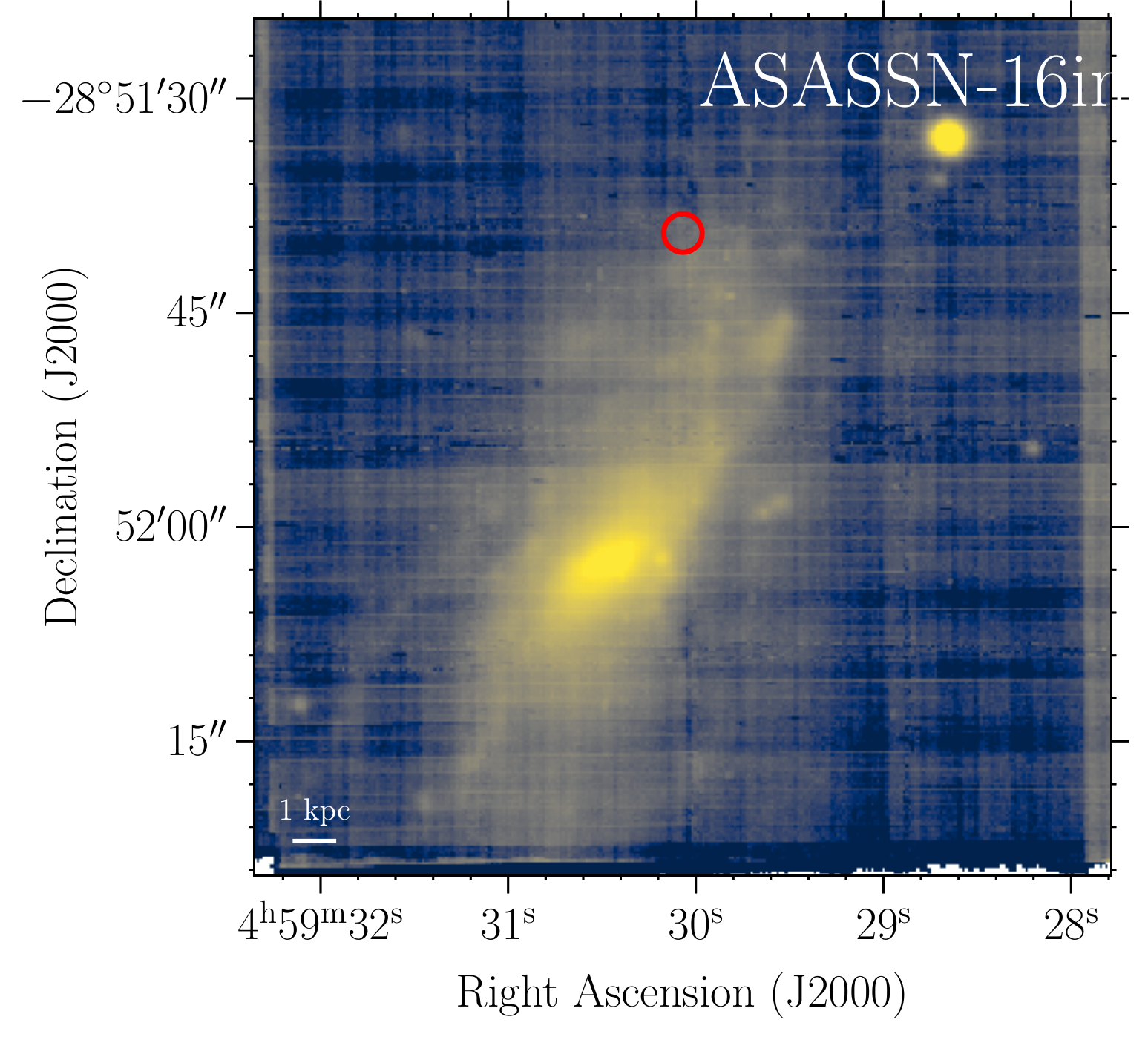}
    \includegraphics[width=0.24\textwidth]{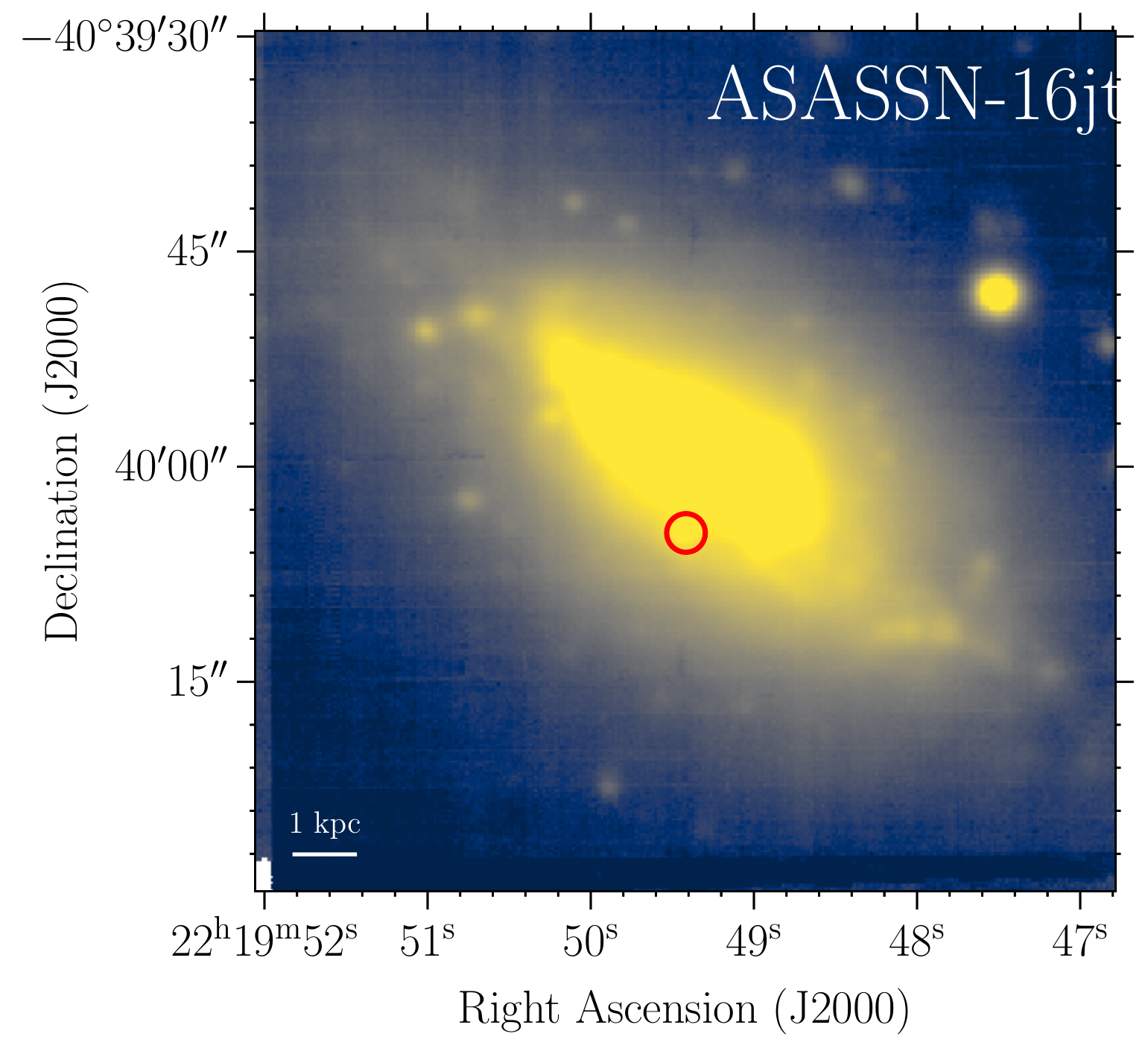}
    \includegraphics[width=0.24\textwidth]{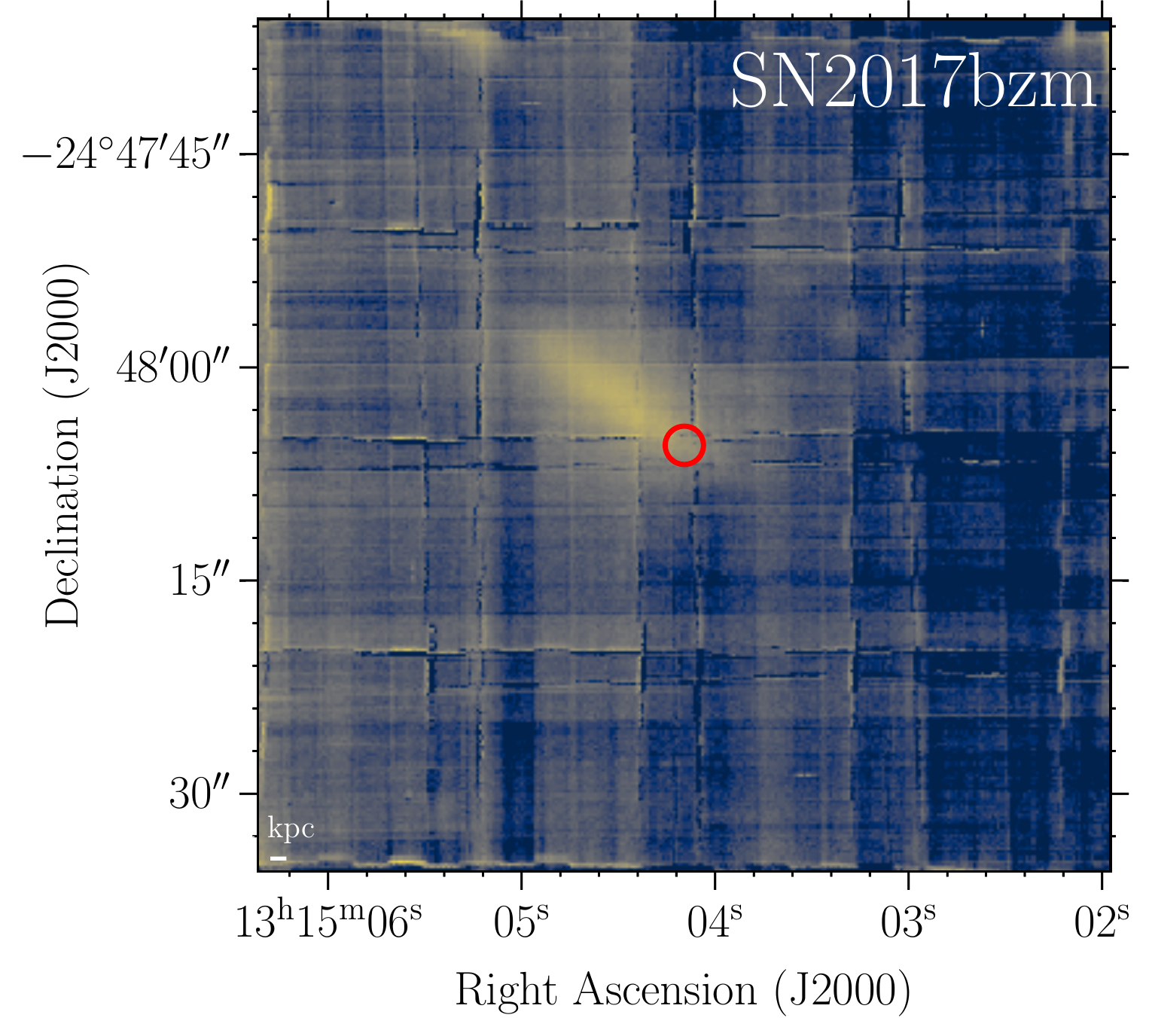}
    \includegraphics[width=0.24\textwidth]{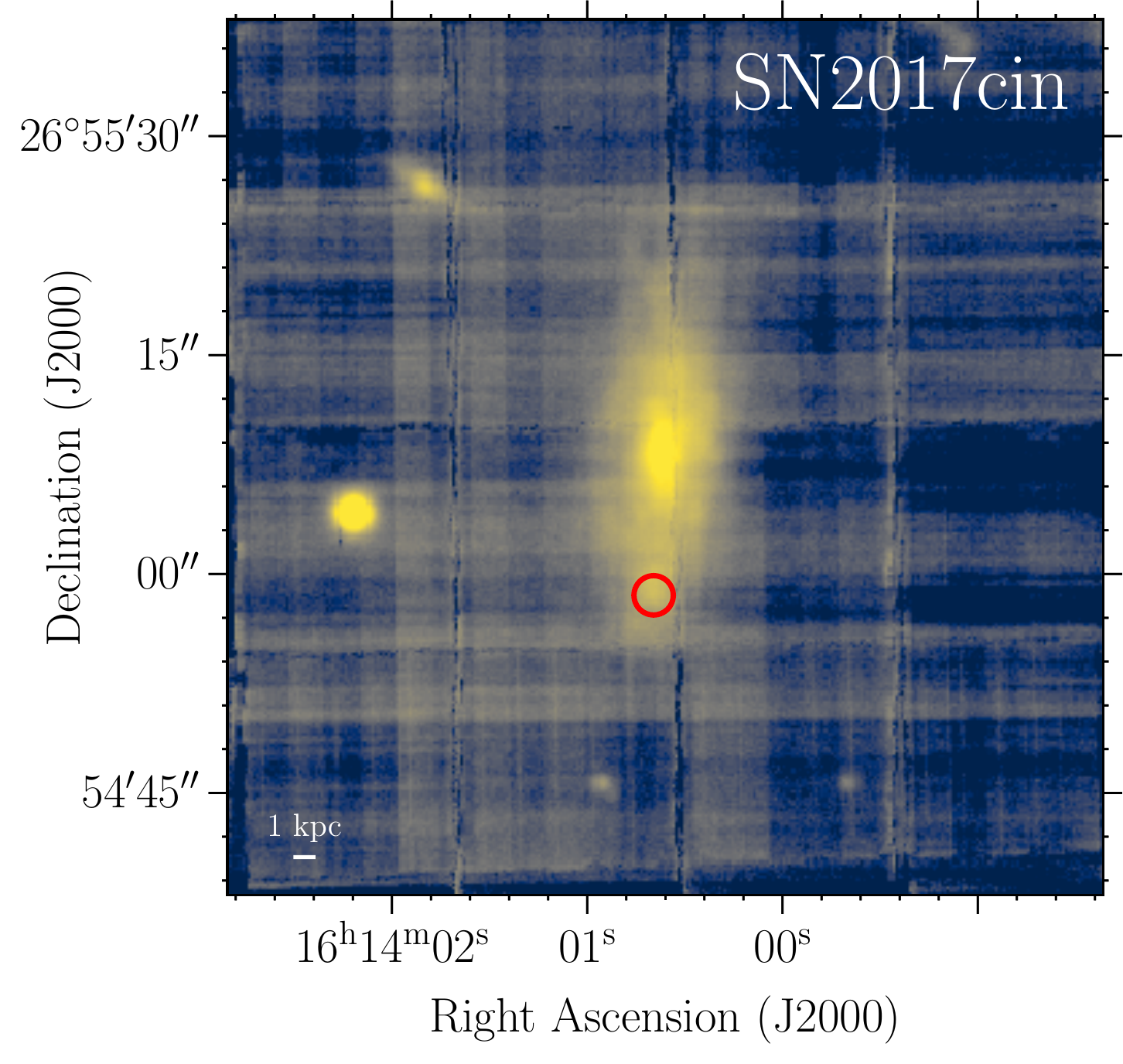}
    \includegraphics[width=0.24\textwidth]{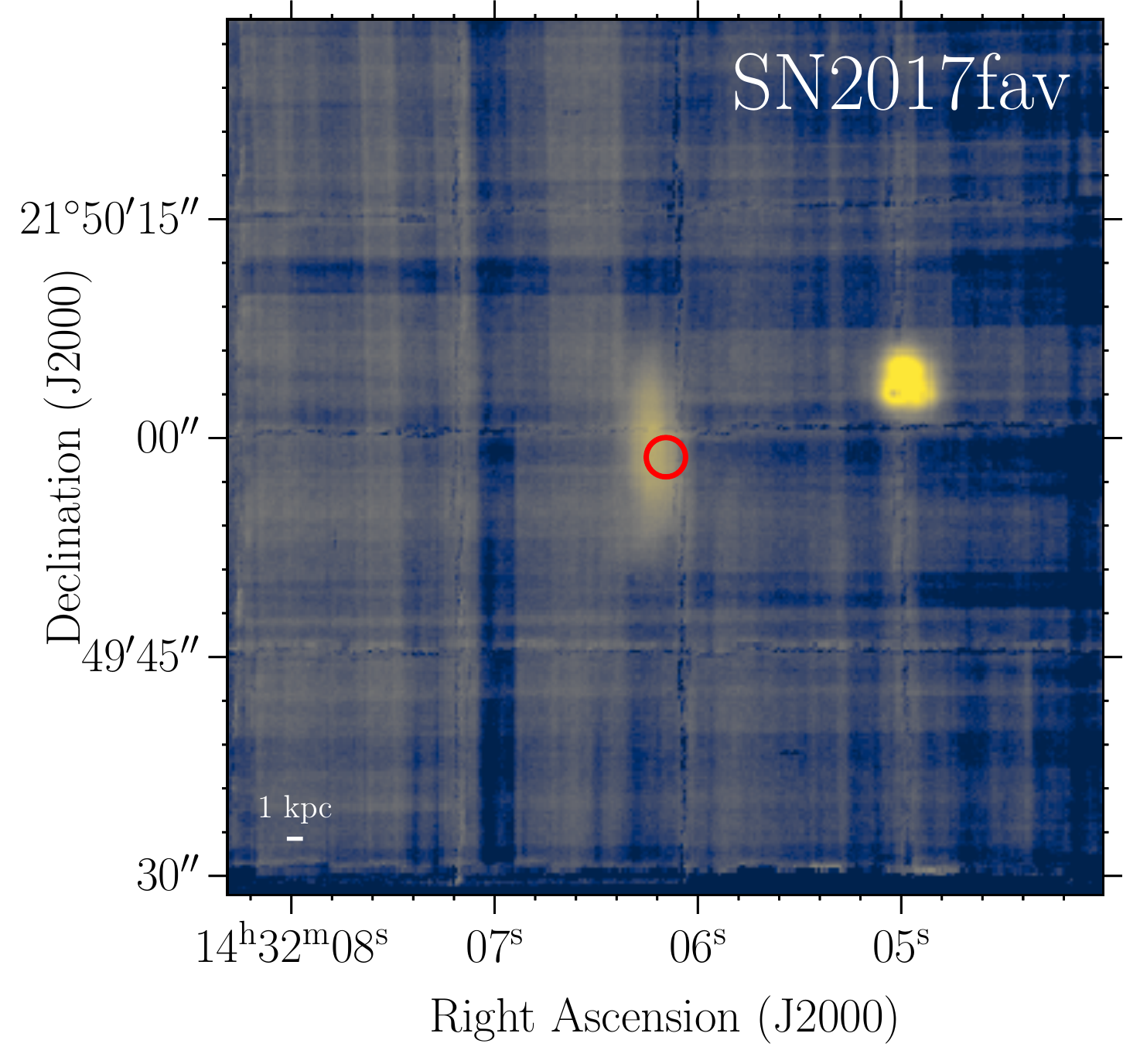}
    \includegraphics[width=0.24\textwidth]{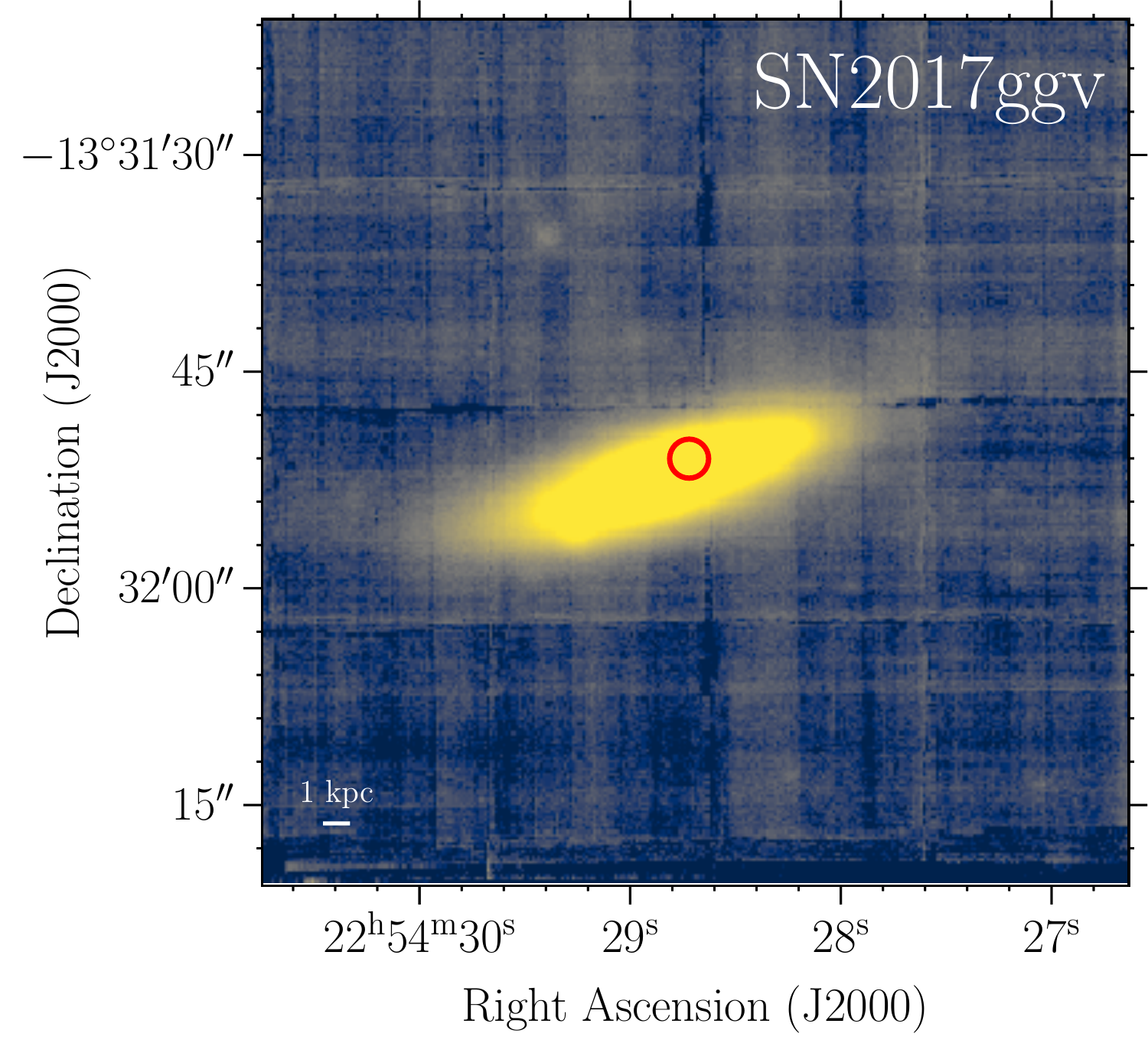}
    \includegraphics[width=0.24\textwidth]{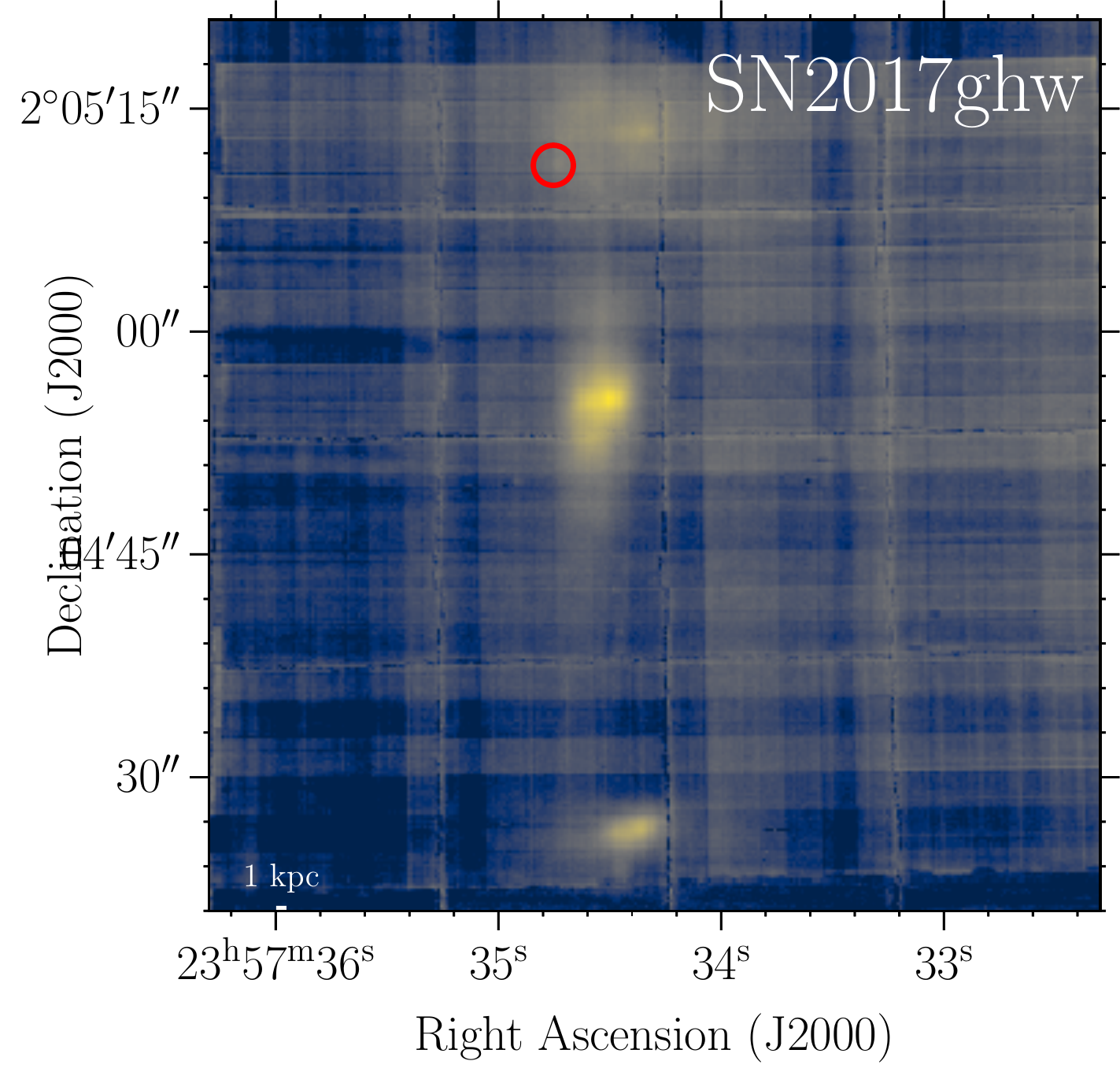}
    \includegraphics[width=0.24\textwidth]{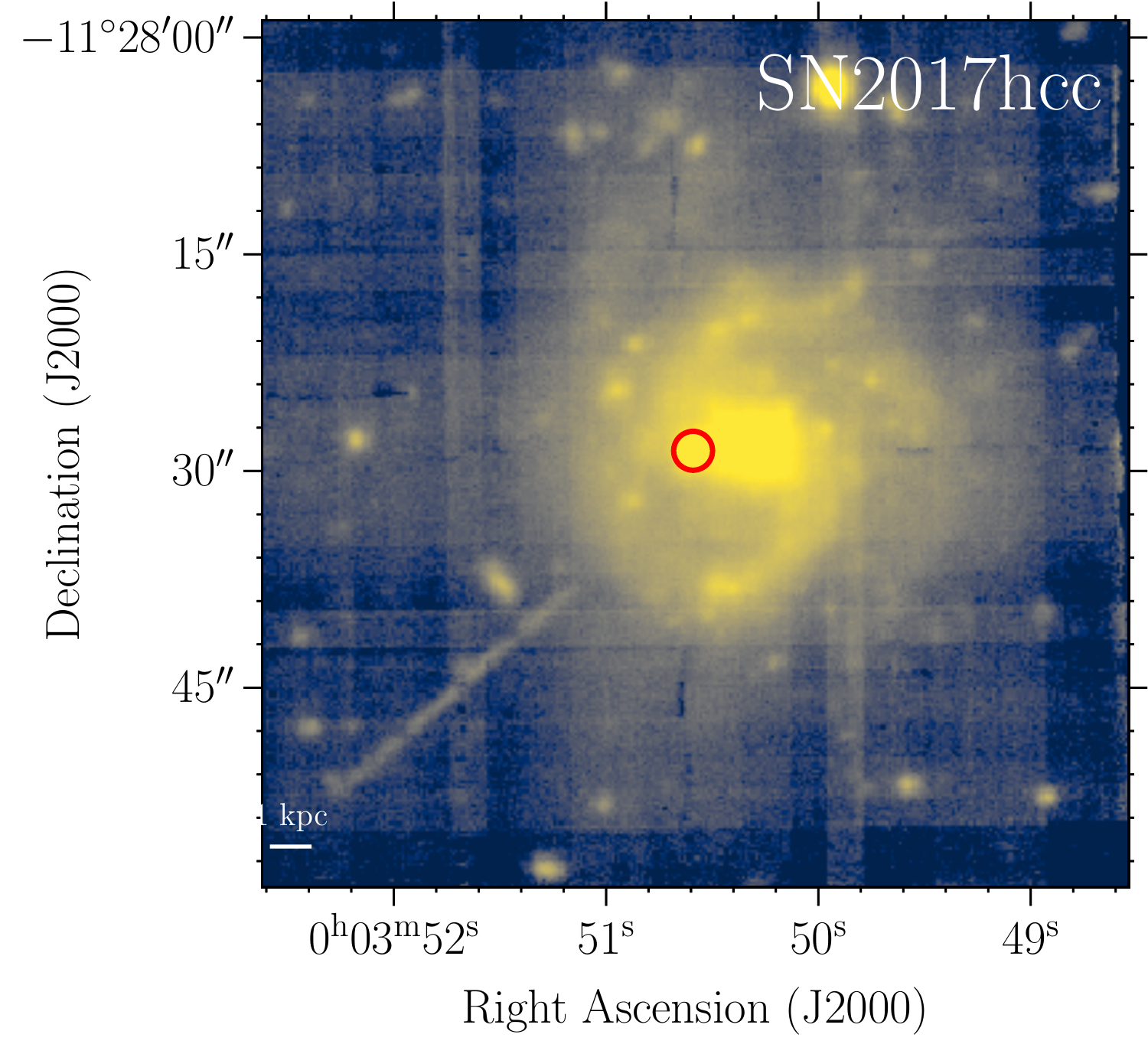}
    \includegraphics[width=0.24\textwidth]{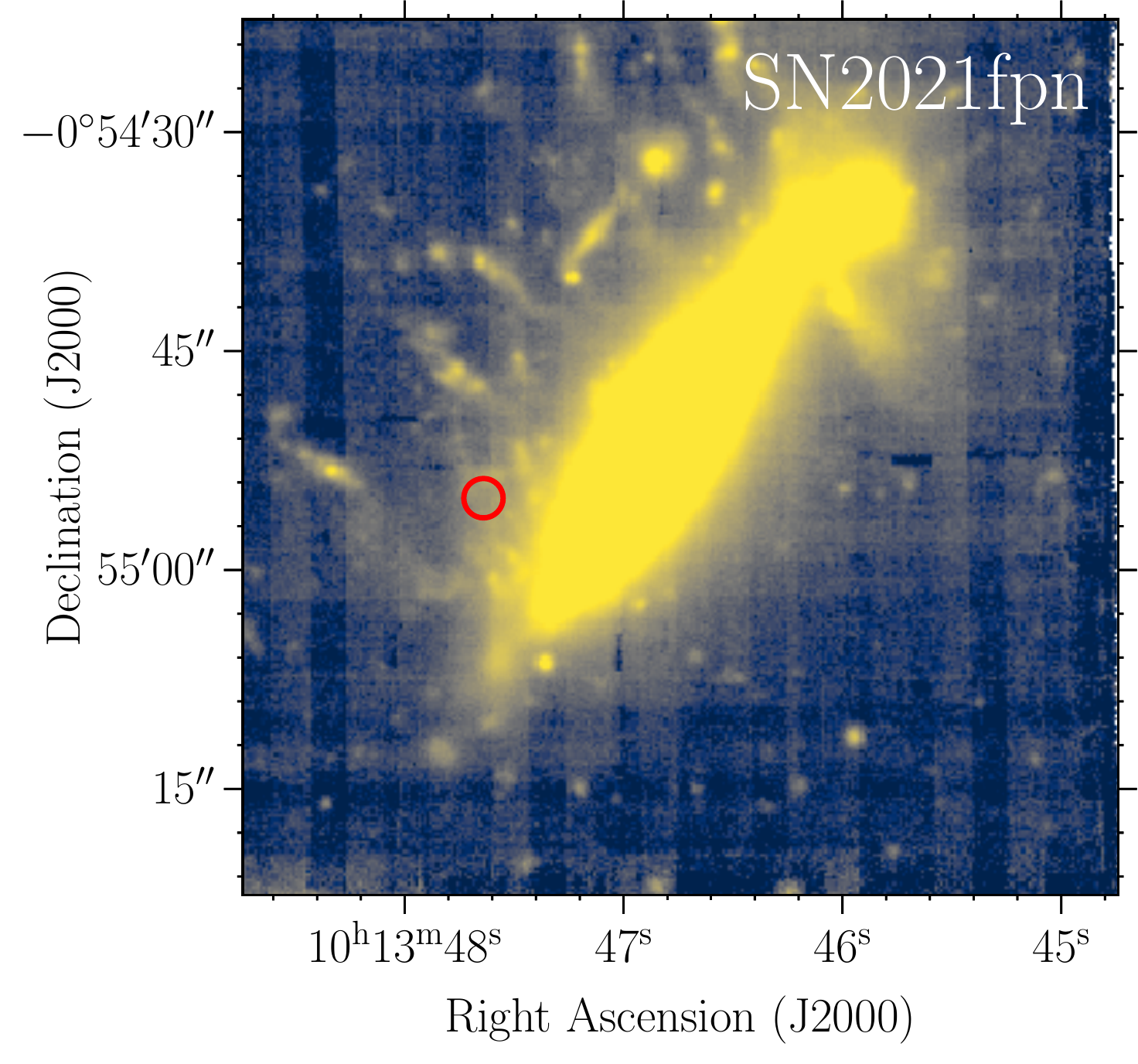}
    \caption{SN~IIn host galaxies in the synthesized \textit{r}~band in our IFS data. The red circle is the SN position with the seeing-sized aperture.} 
    \label{fig:mosaic1}
\end{figure*}

We present supplementary figures presenting each SN environment. Figure~\ref{fig:mosaic1} shows SN host galaxies with SN locations, and Fig.~\ref{fig:envspec1} shows their spectra used for SN environment parameter estimations.

\begin{figure*}
    \includegraphics[width=0.49\textwidth]{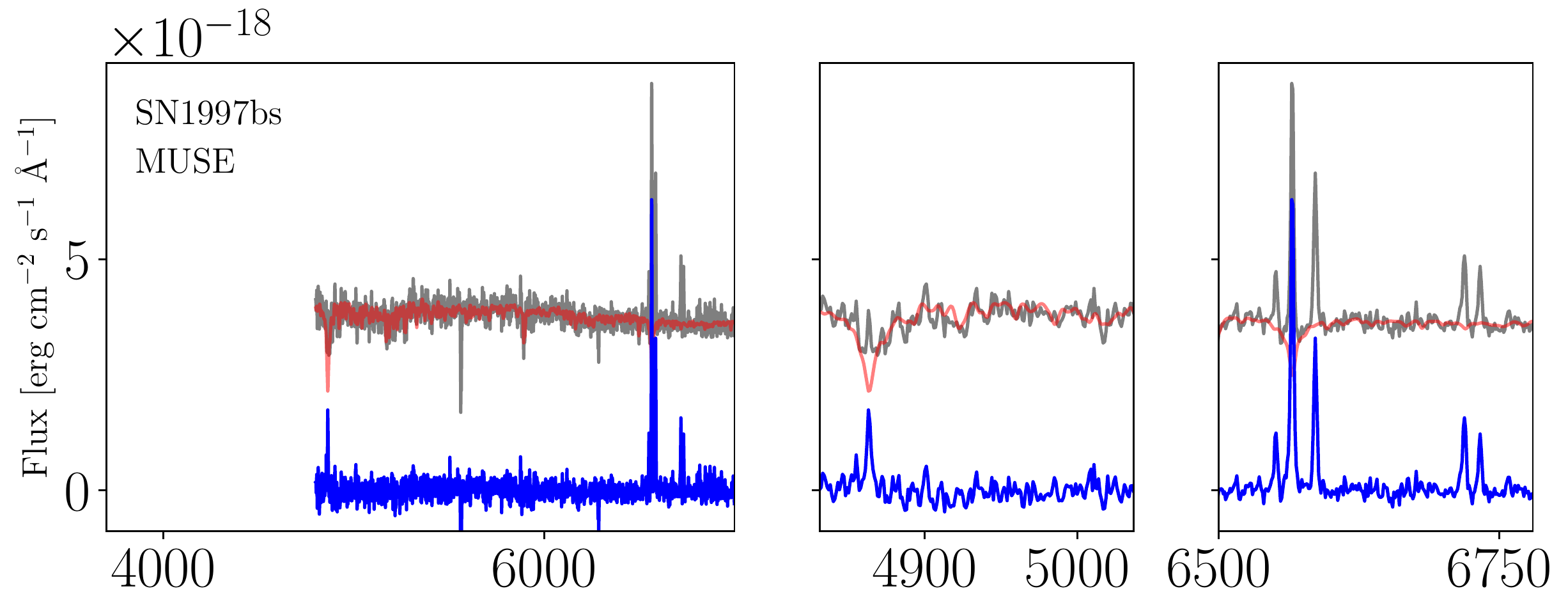}
    \includegraphics[width=0.49\textwidth]{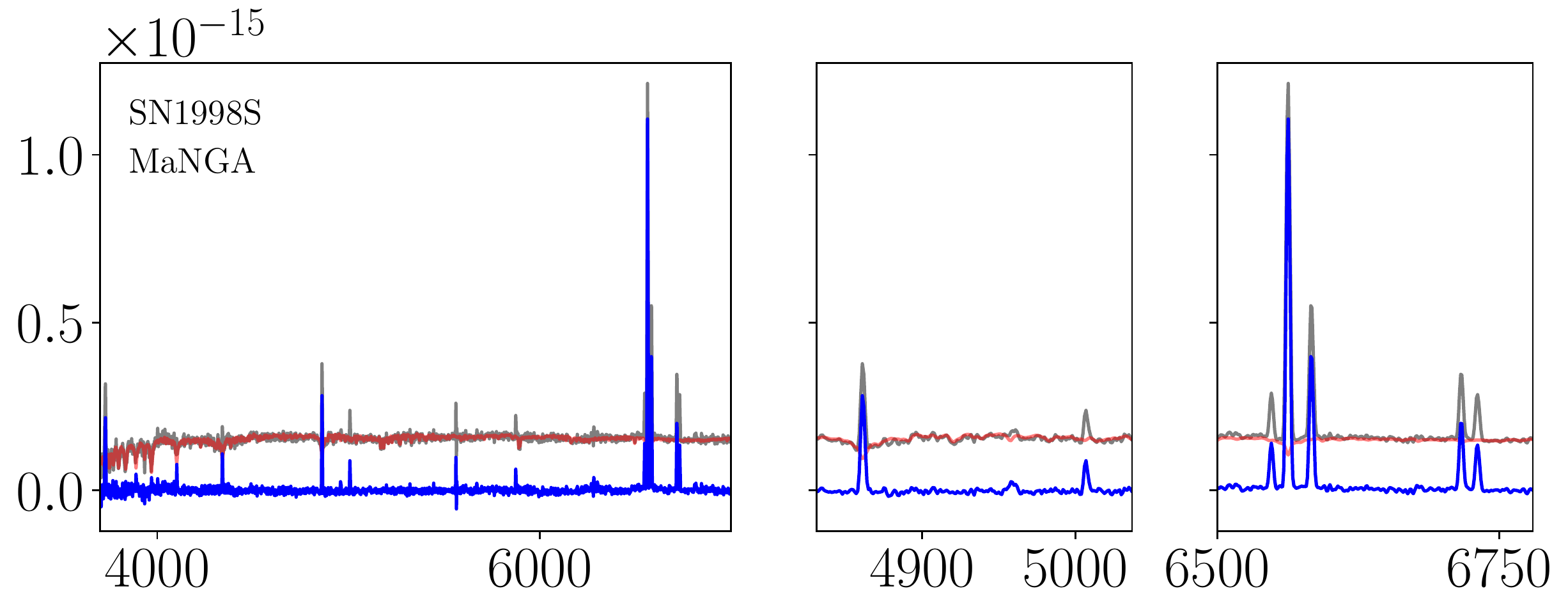}
    \includegraphics[width=0.49\textwidth]{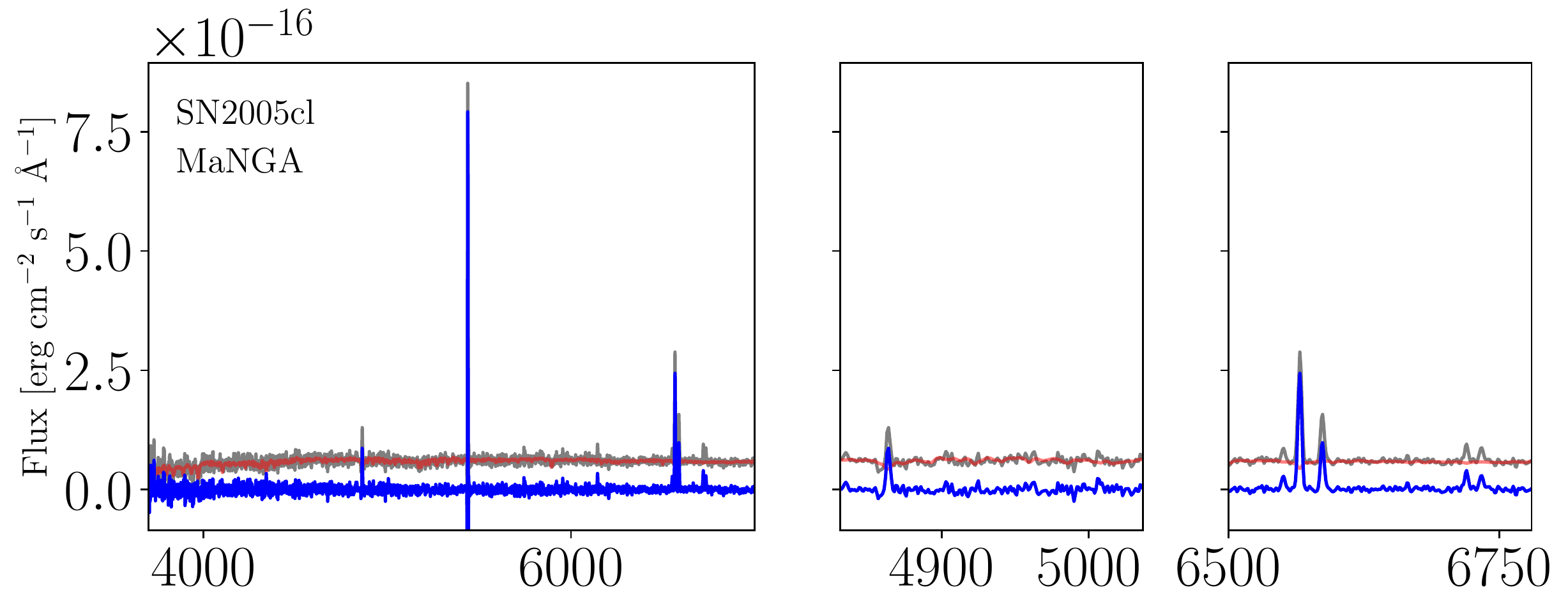}
    \includegraphics[width=0.49\textwidth]{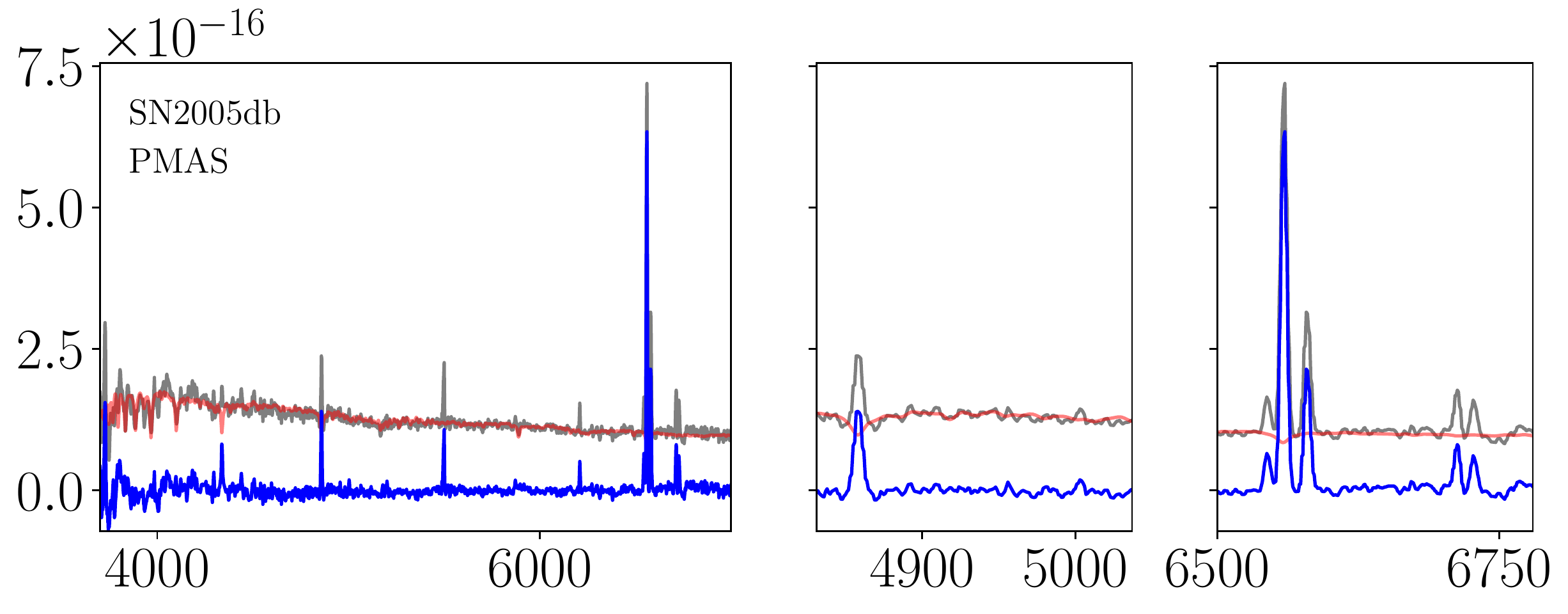}
    \includegraphics[width=0.49\textwidth]{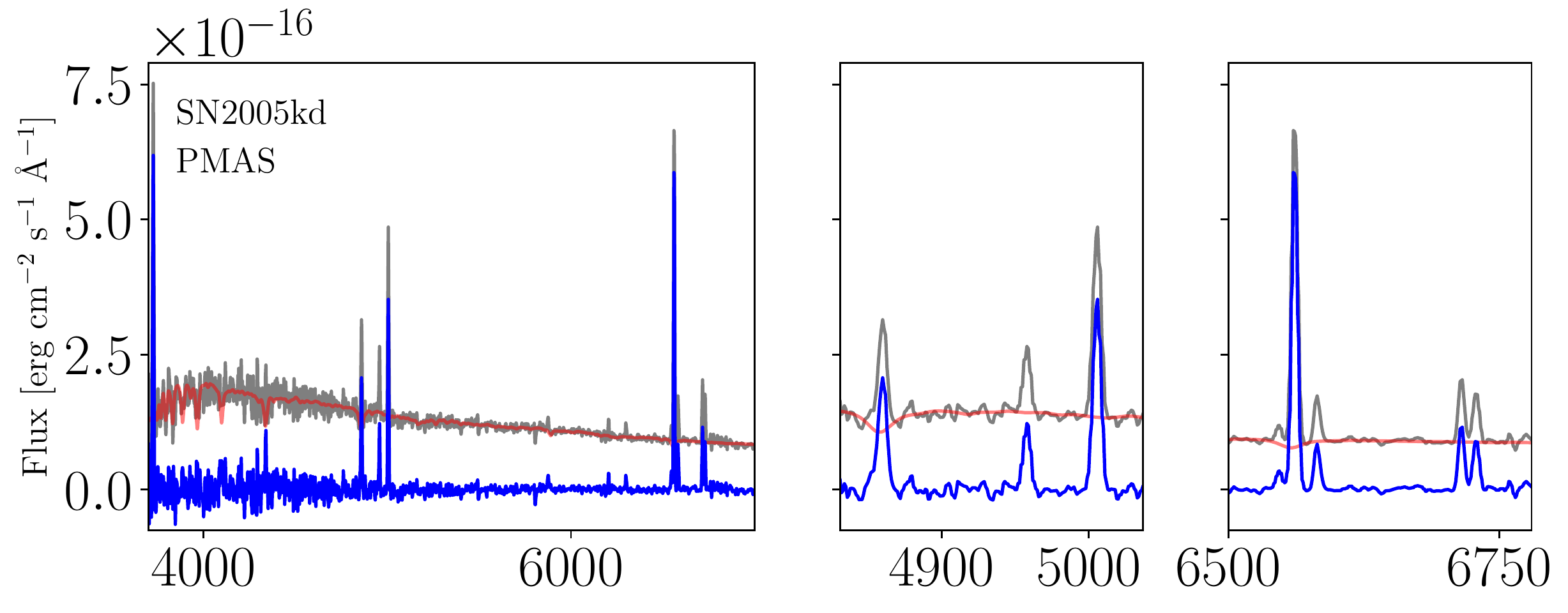}
    \includegraphics[width=0.49\textwidth]{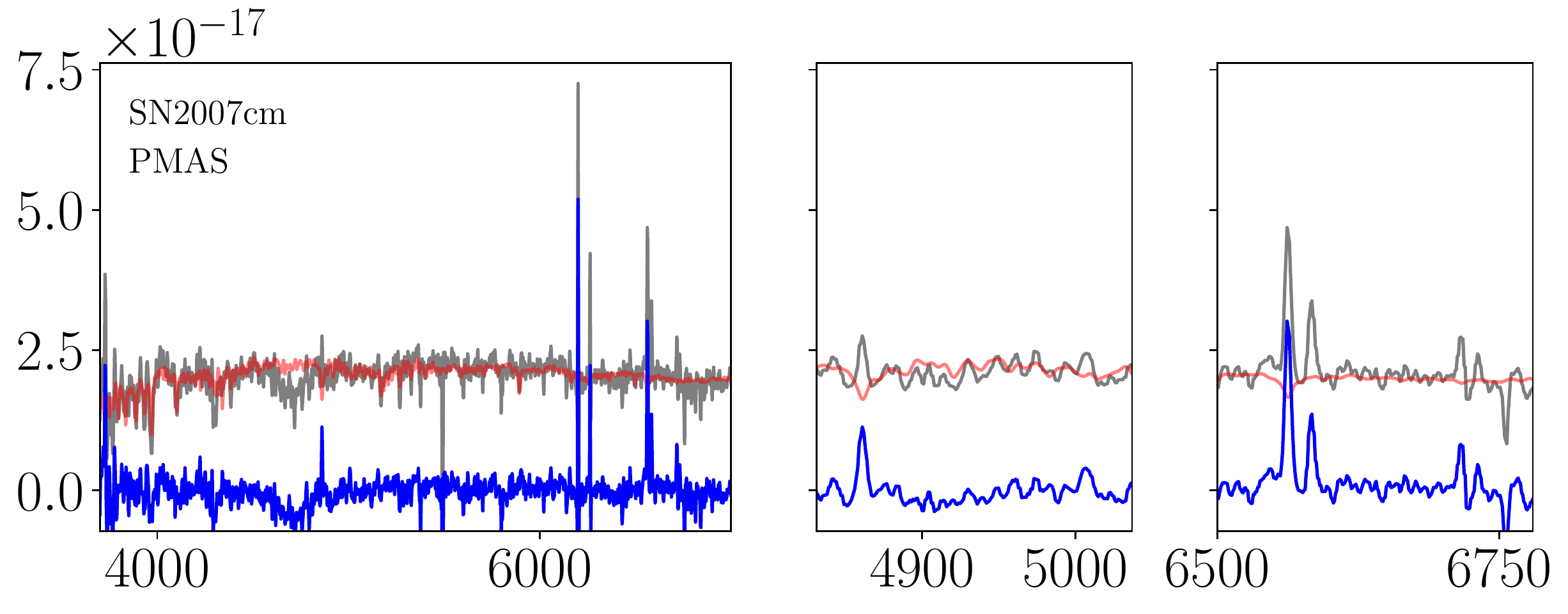}
    \includegraphics[width=0.49\textwidth]{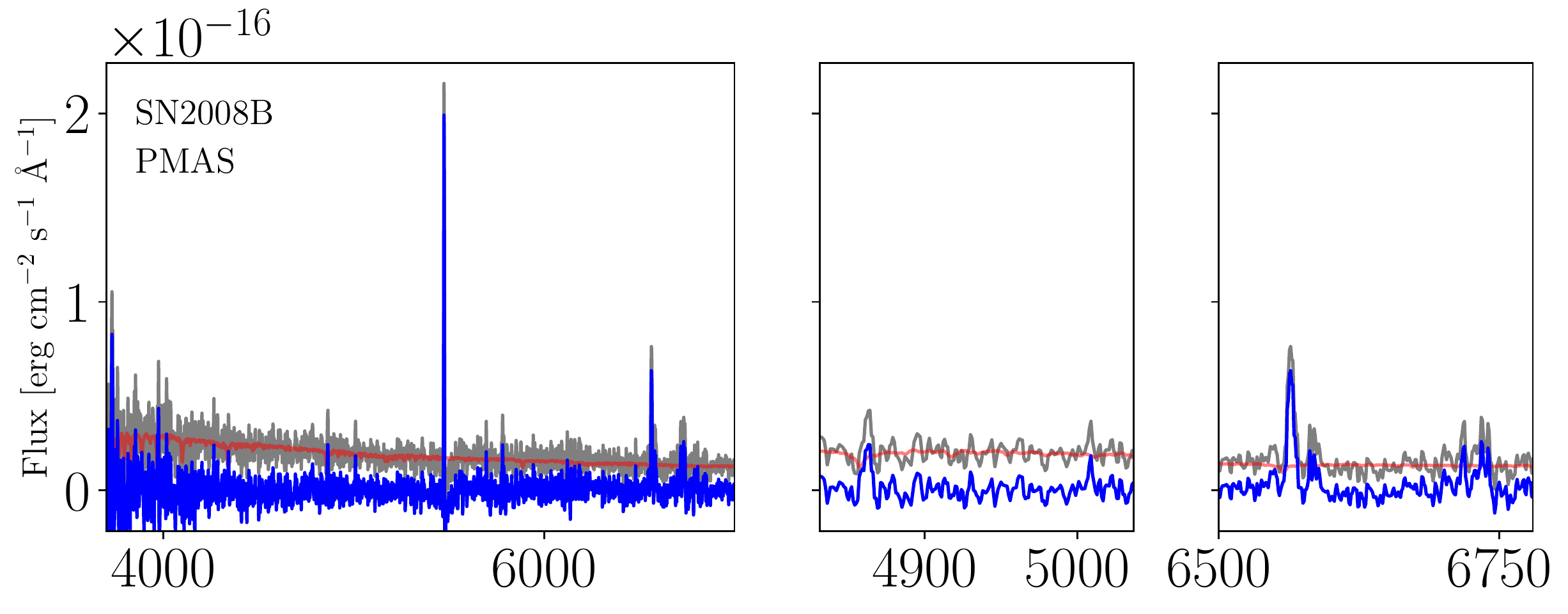}
    \includegraphics[width=0.49\textwidth]{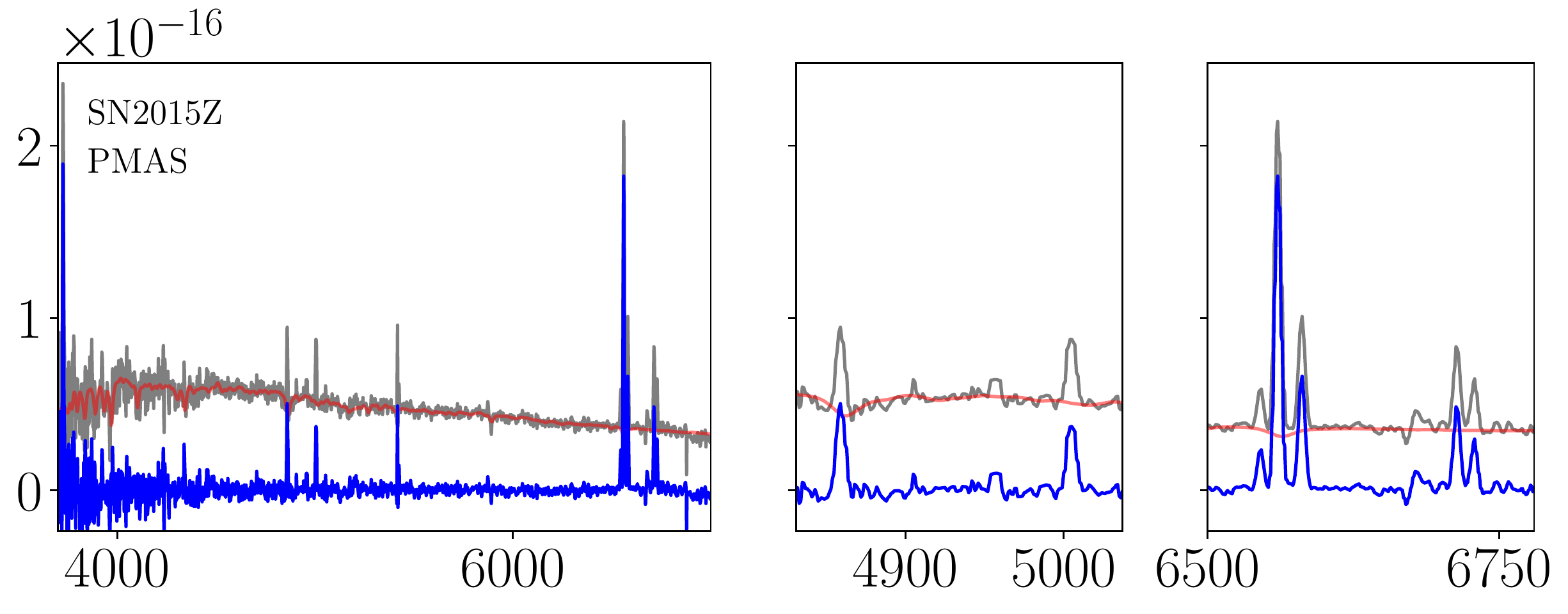}
    \includegraphics[width=0.49\textwidth]{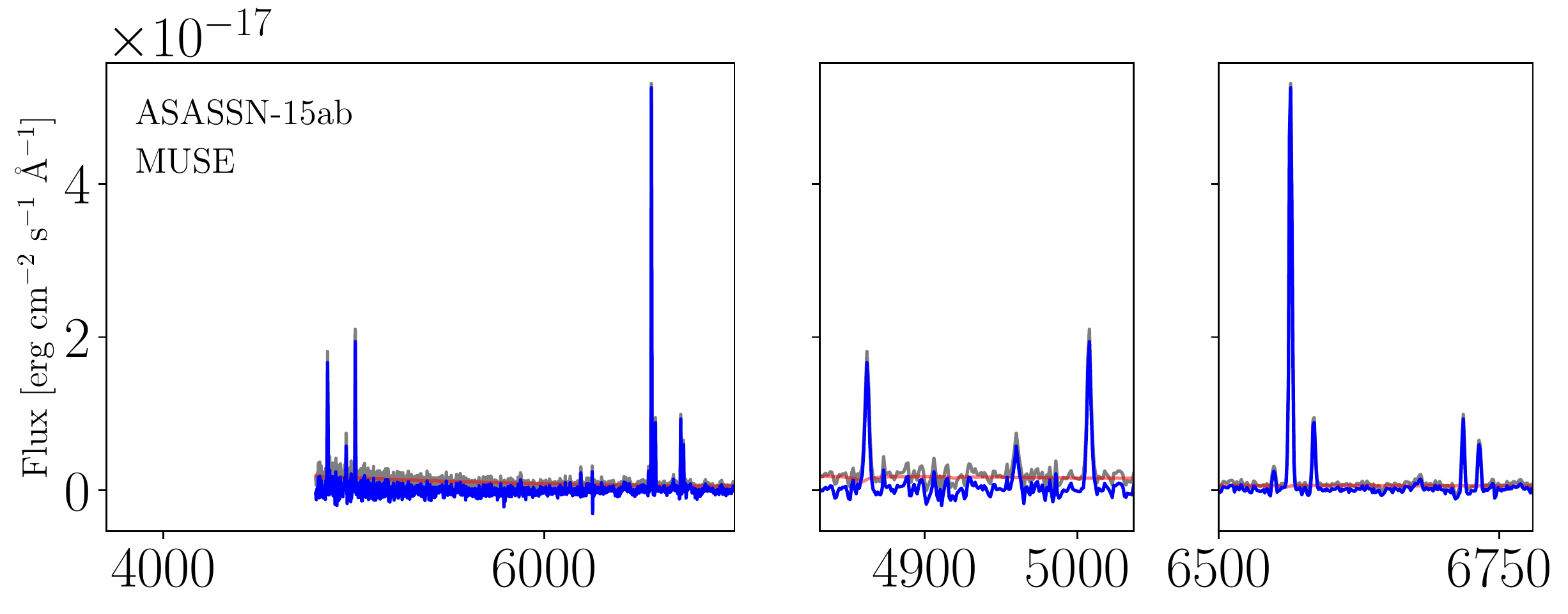}
    \includegraphics[width=0.49\textwidth]{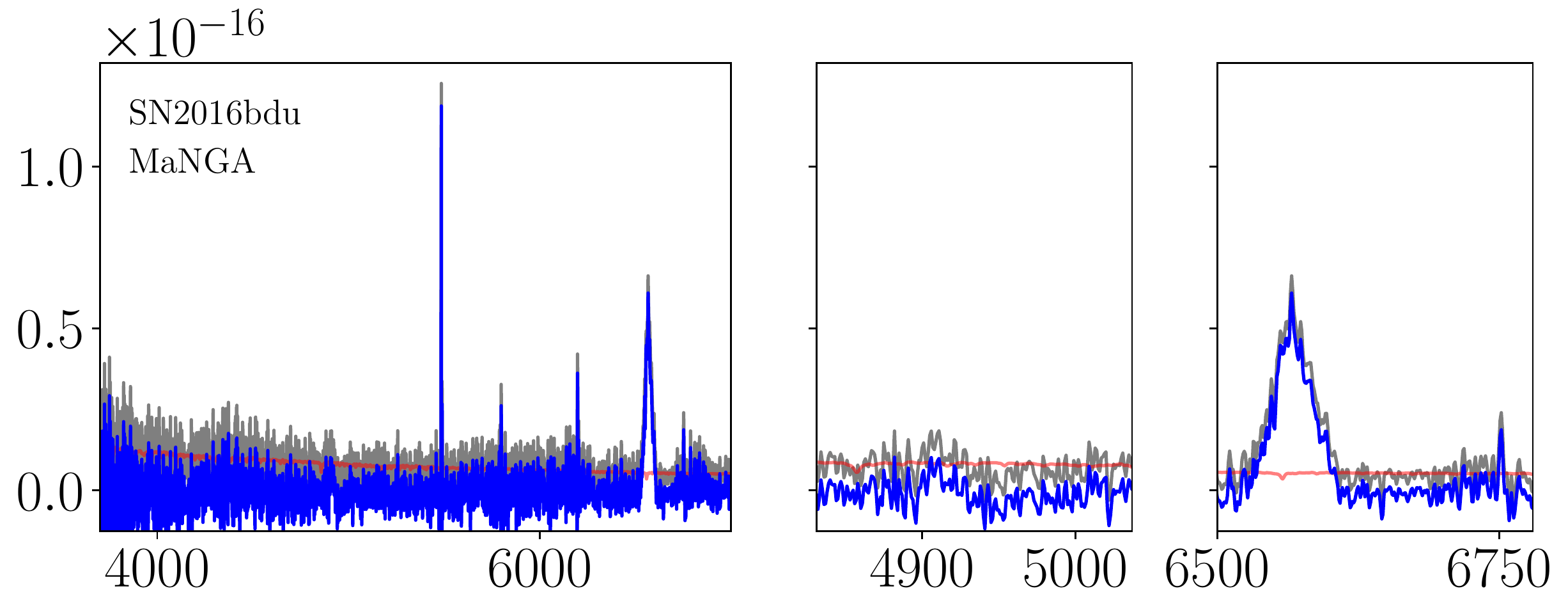}
    \includegraphics[width=0.49\textwidth]{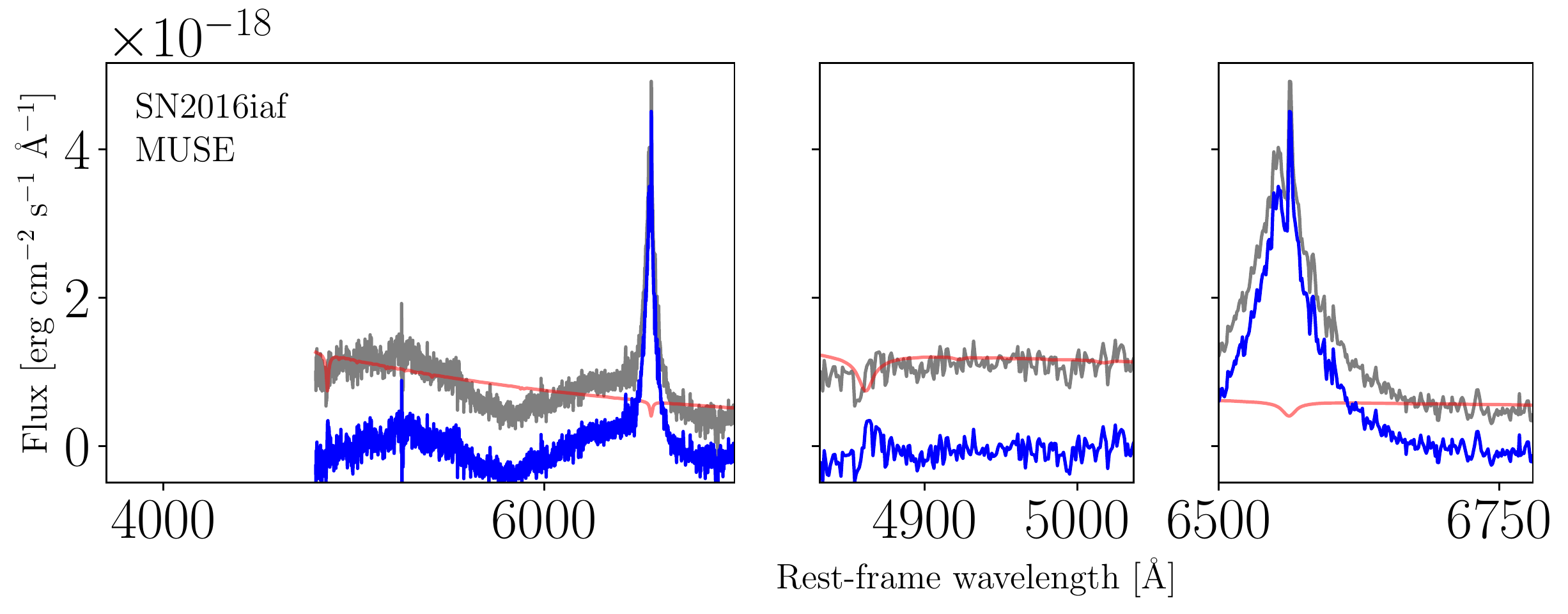}
    \includegraphics[width=0.49\textwidth]{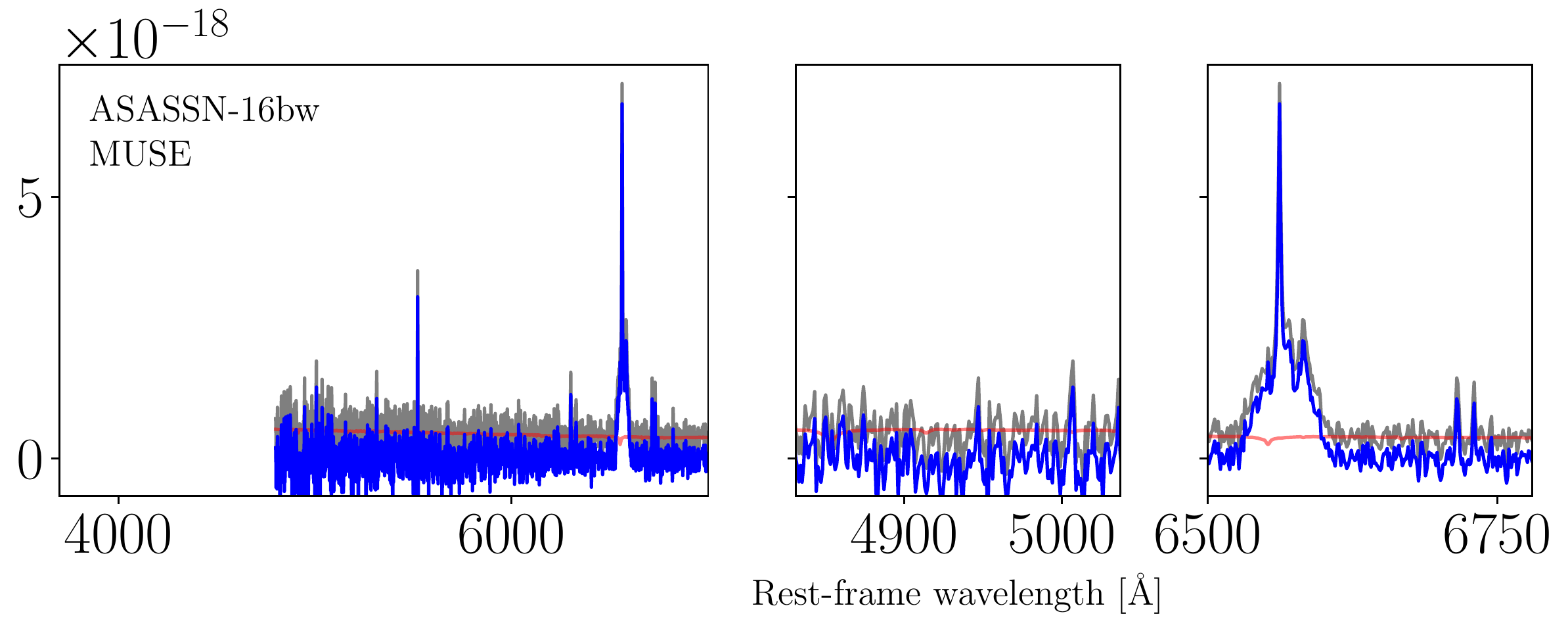}
\caption{Spectra of the SN~IIn environments used for the environmental parameter estimations. Gray, red, and blue spectra are aperture spectra, the best SSP fits, and their resulting gas-pahse emissioin spectra, respectively.}
\label{fig:envspec1}
\end{figure*}
\setcounter{figure}{1}
\begin{figure*}
    \includegraphics[width=0.49\textwidth]{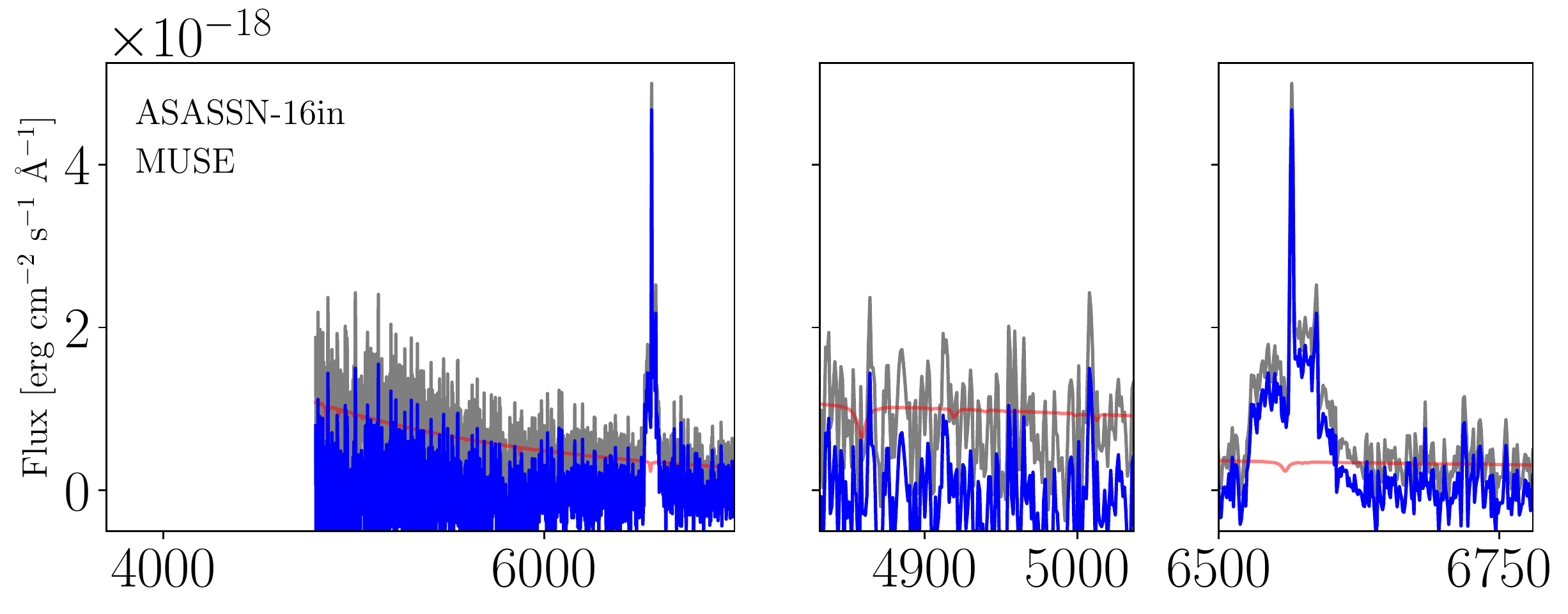}
    \includegraphics[width=0.49\textwidth]{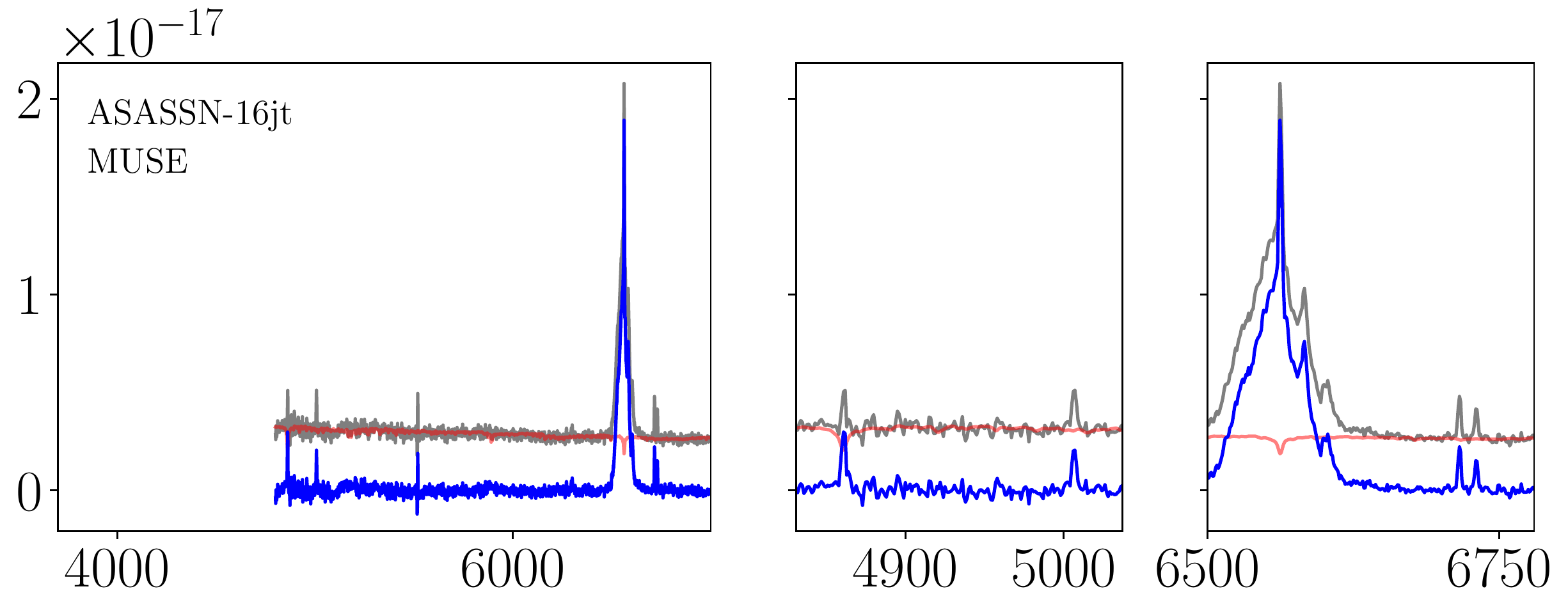}
    \includegraphics[width=0.49\textwidth]{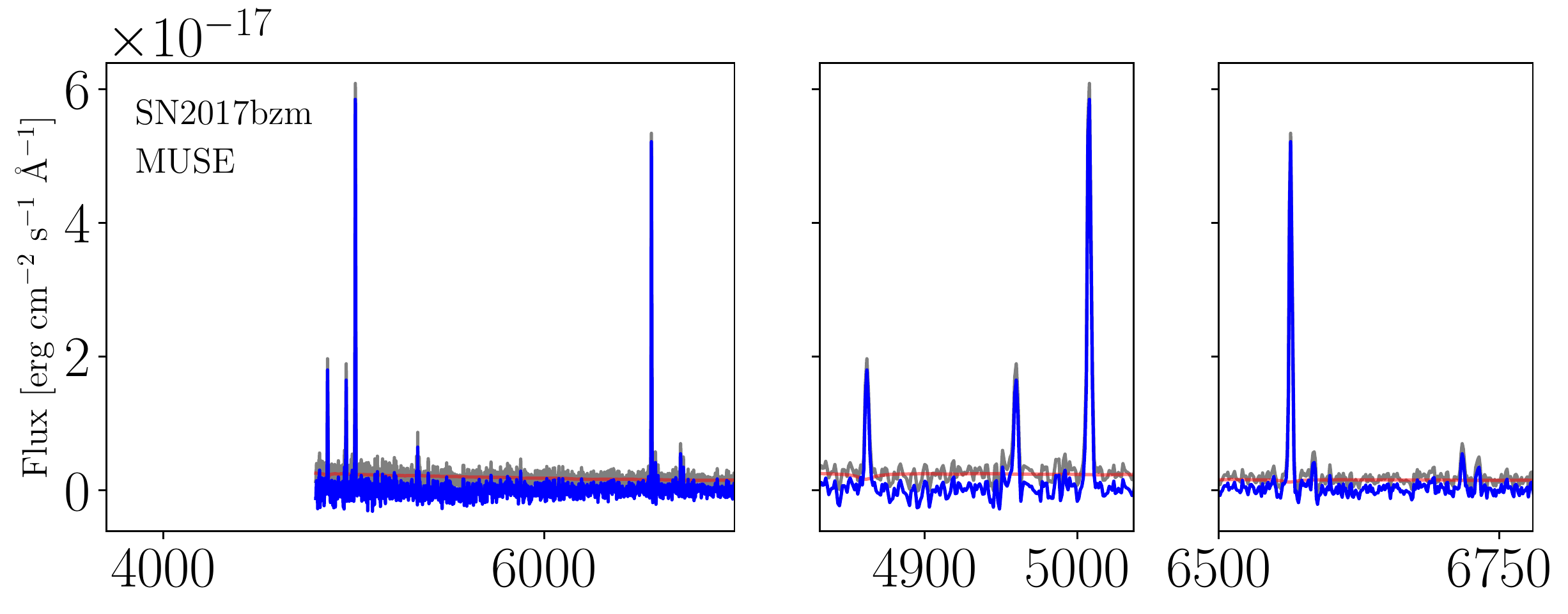}
    \includegraphics[width=0.49\textwidth]{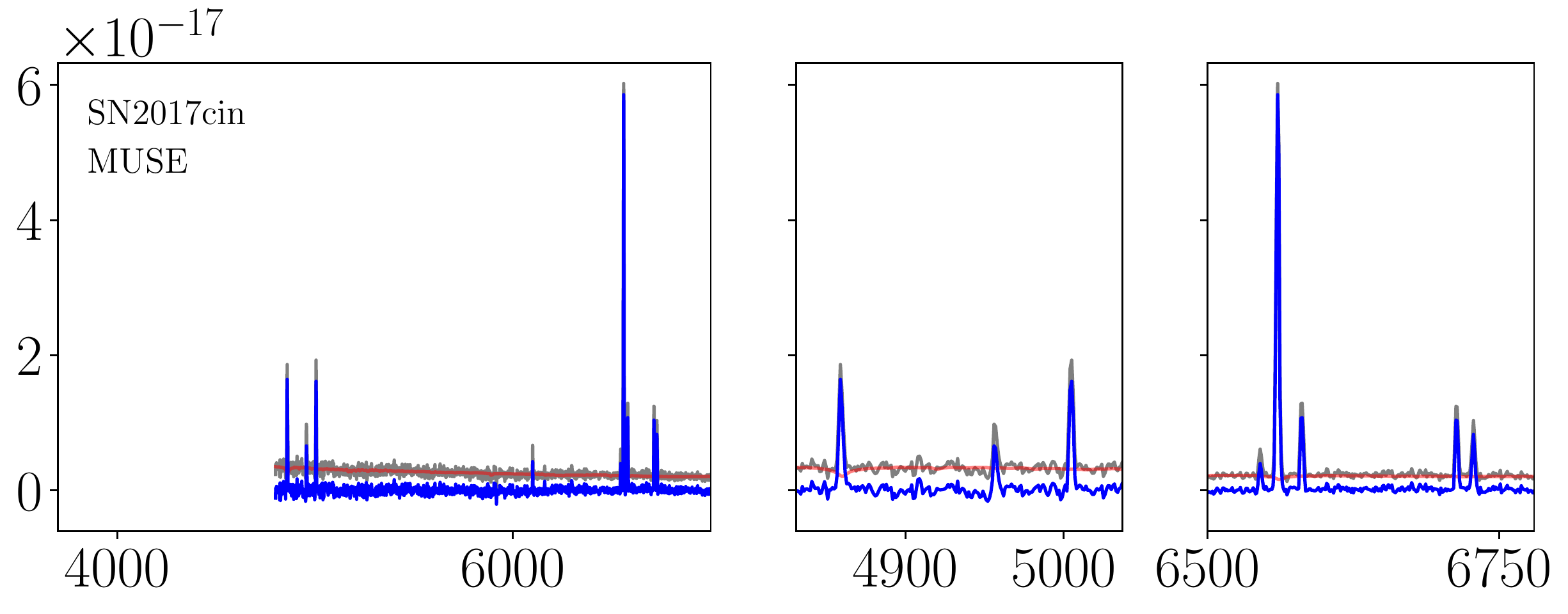}
    \includegraphics[width=0.49\textwidth]{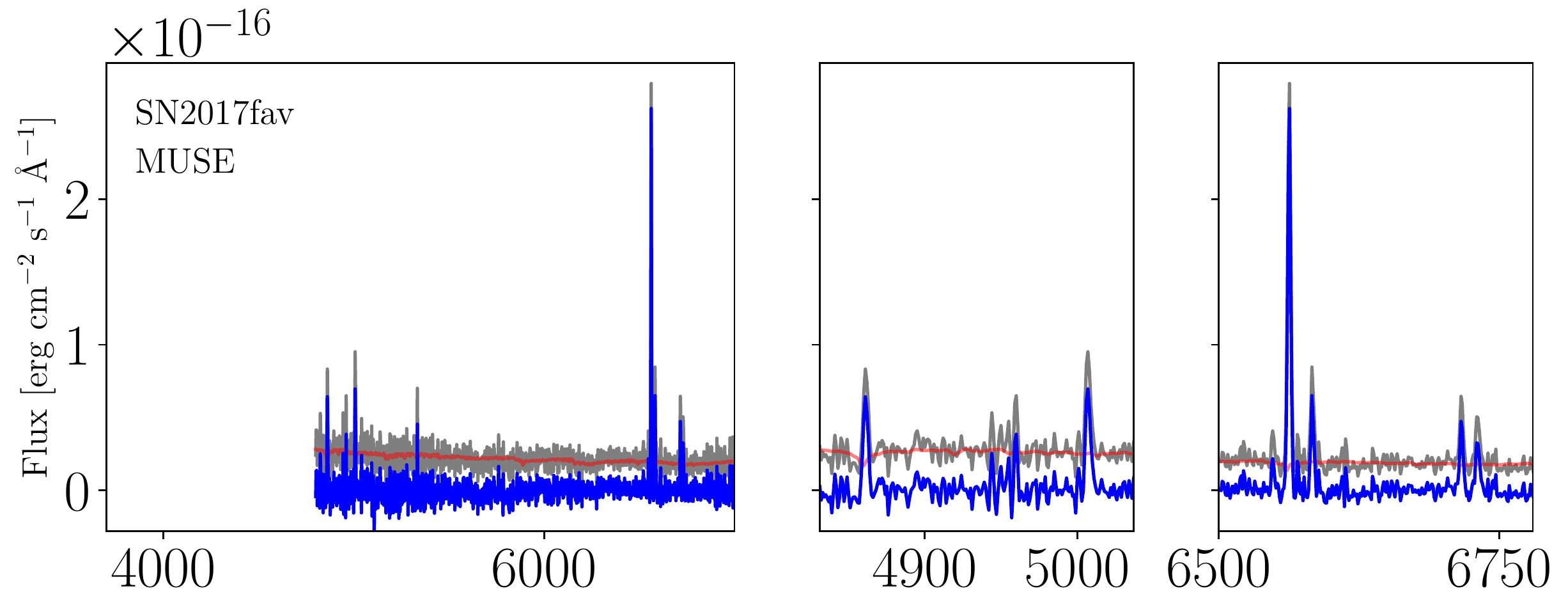}
    \includegraphics[width=0.49\textwidth]{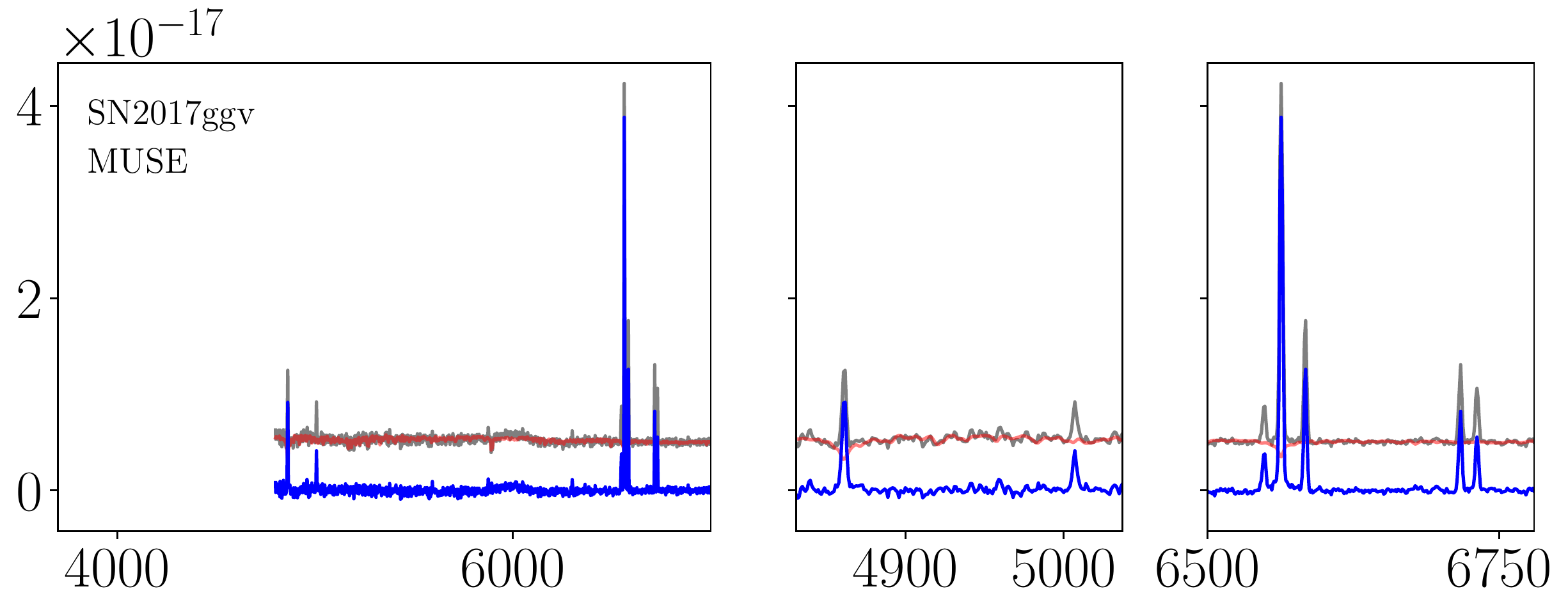}
    \includegraphics[width=0.49\textwidth]{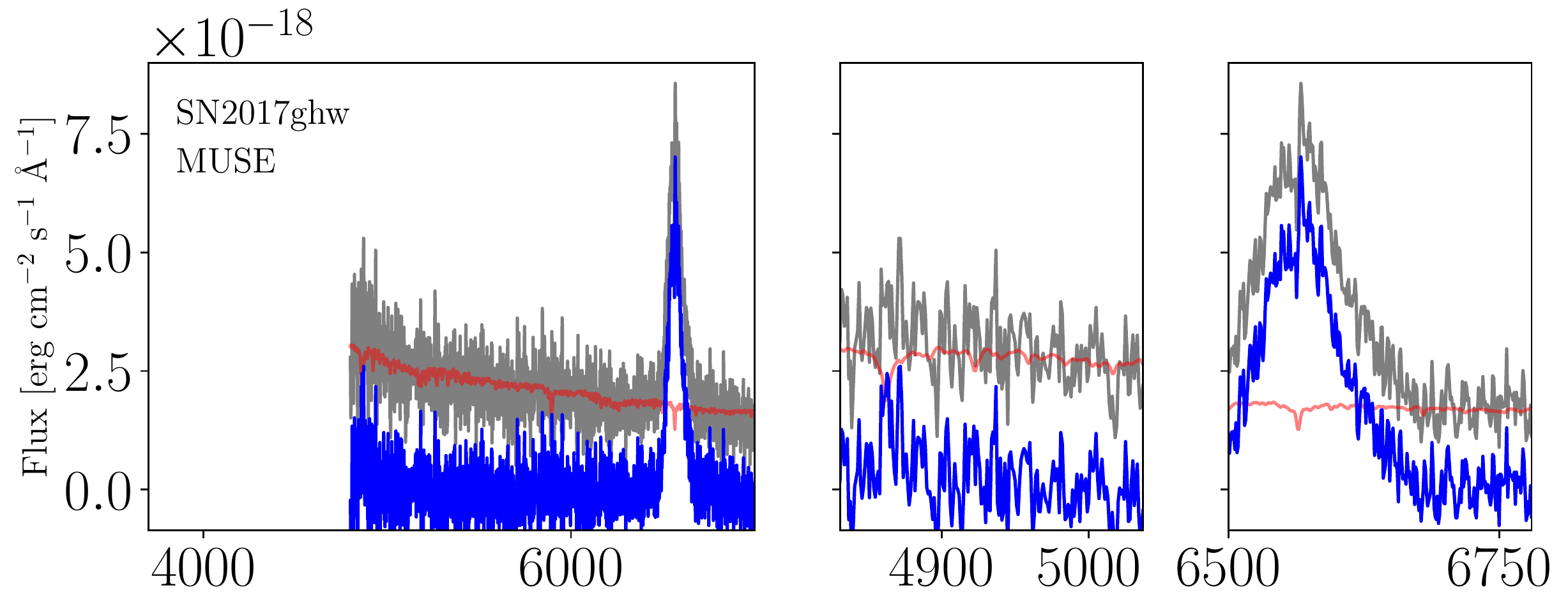}
    \includegraphics[width=0.49\textwidth]{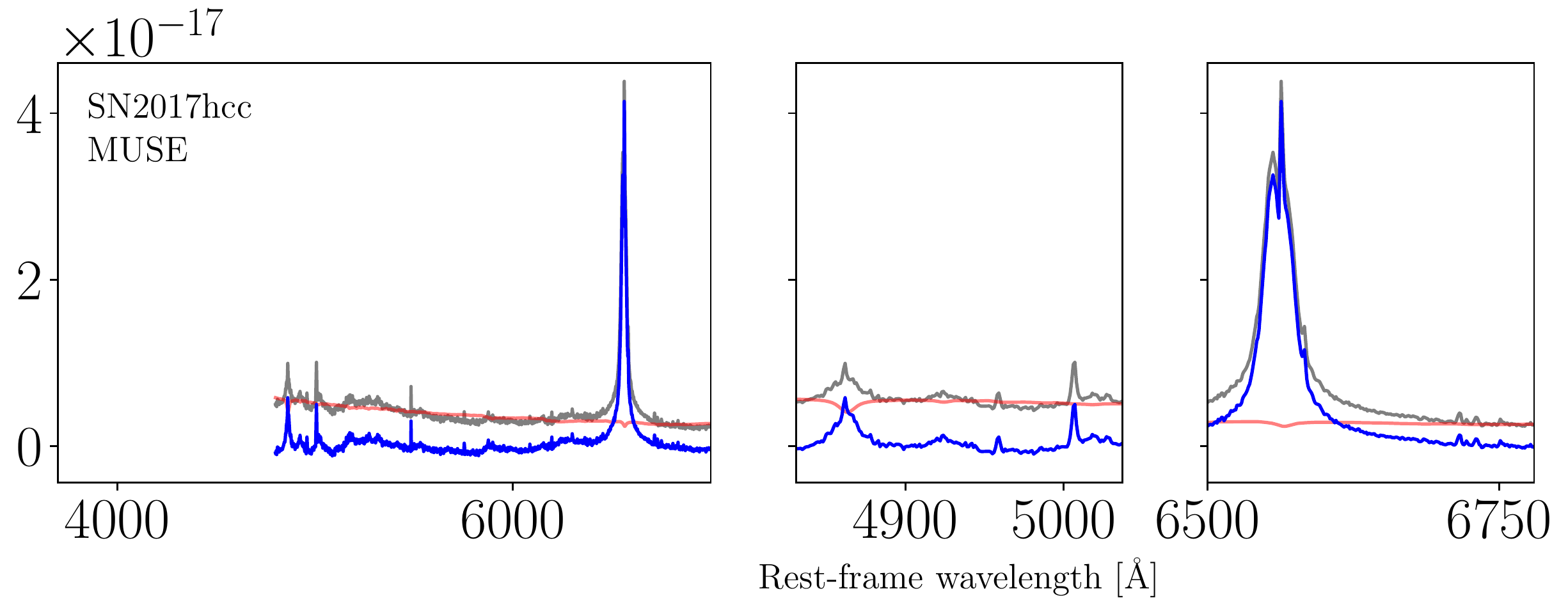}
    \includegraphics[width=0.49\textwidth]{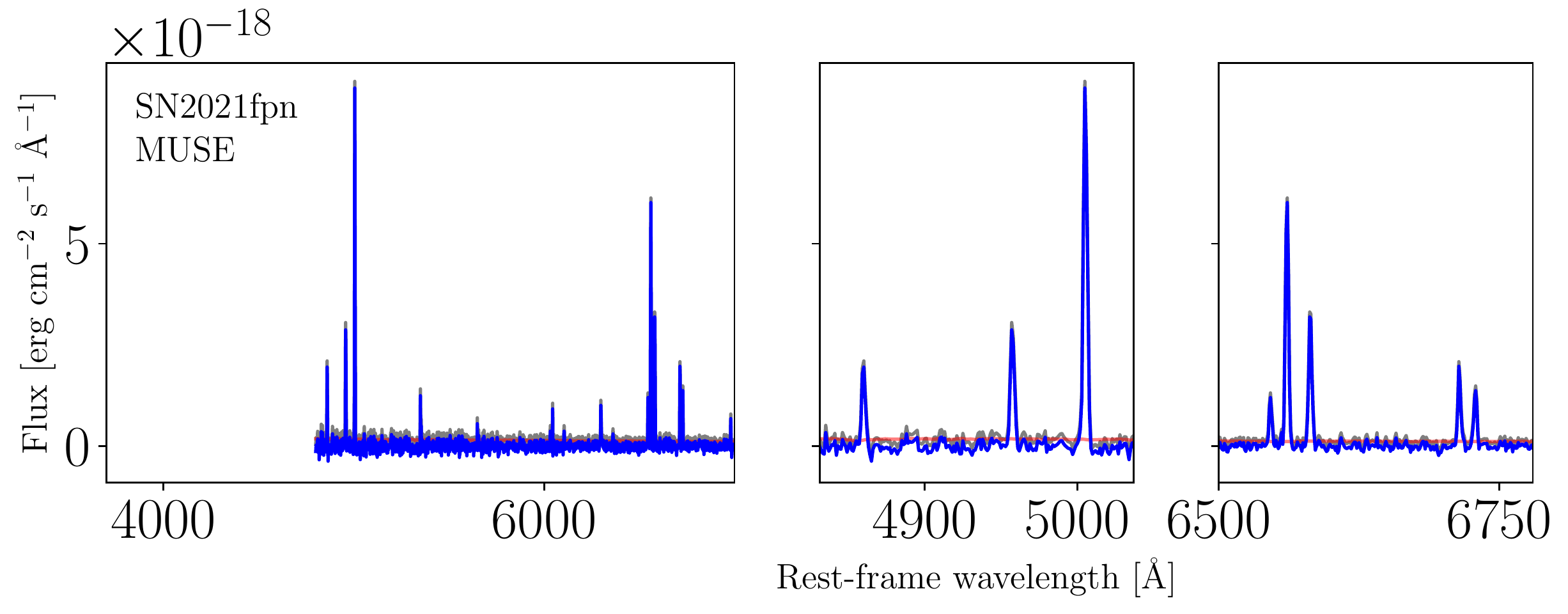}
    \caption{\textit{Continued.}}
\end{figure*}

\section{Light-curve fitting results}
The results of light-curve fitting that are used to estimate rise time and peak luminosity of our SN~IIn sample are presented in Fig.~\ref{fig:lightcurve_fit}. The fitting formula is Eq.~(\ref{eq:fitting}). 

\begin{figure*}
    \centering
    \includegraphics[width=0.31\textwidth]{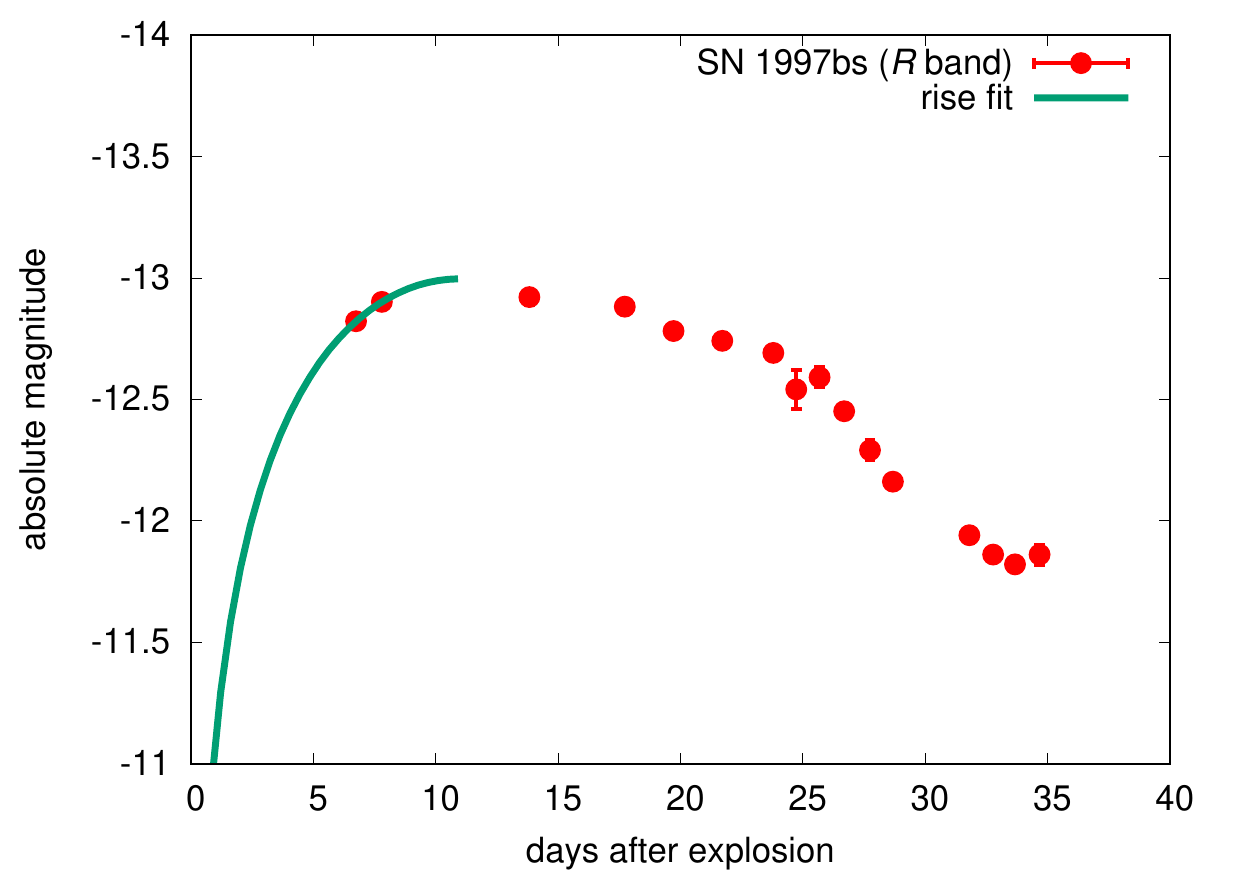} 
    \includegraphics[width=0.31\textwidth]{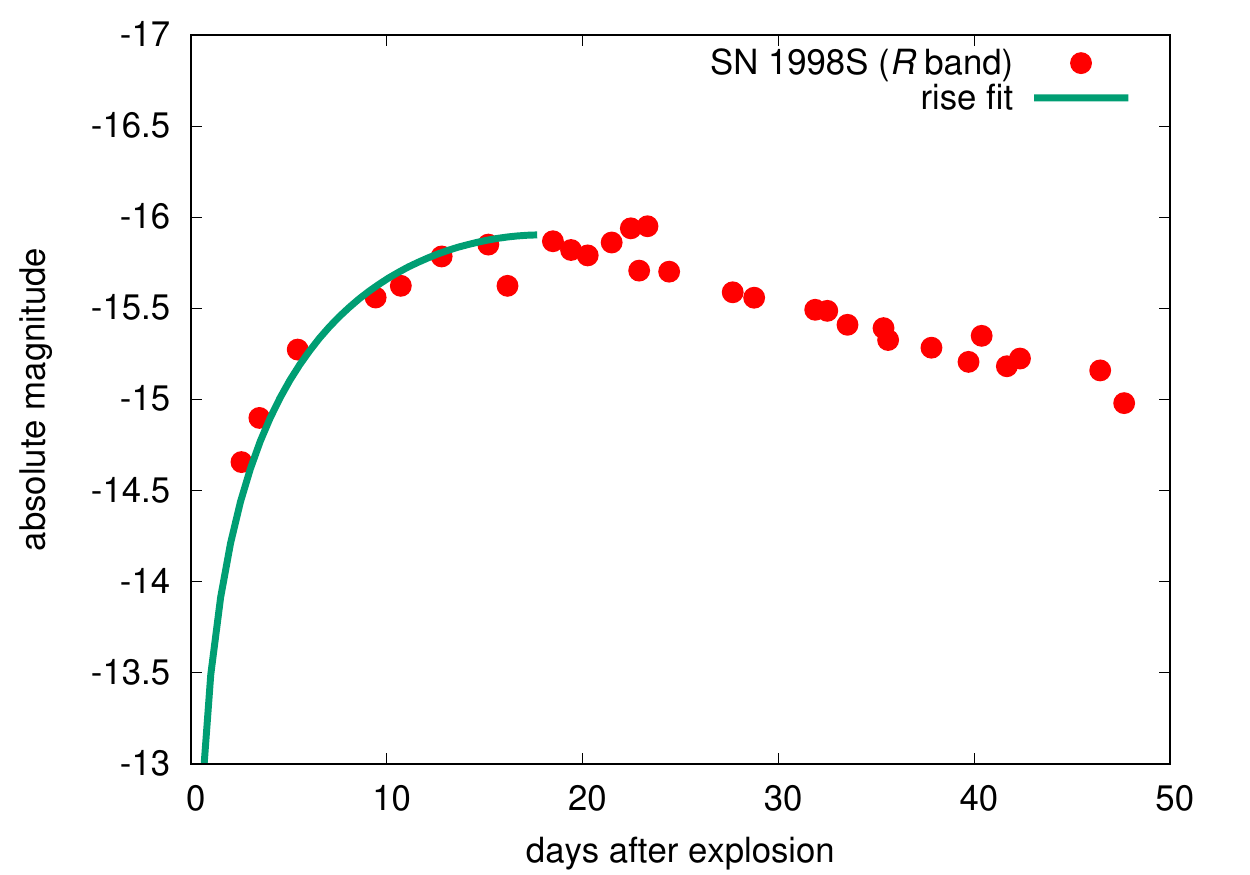} 
    \includegraphics[width=0.31\textwidth]{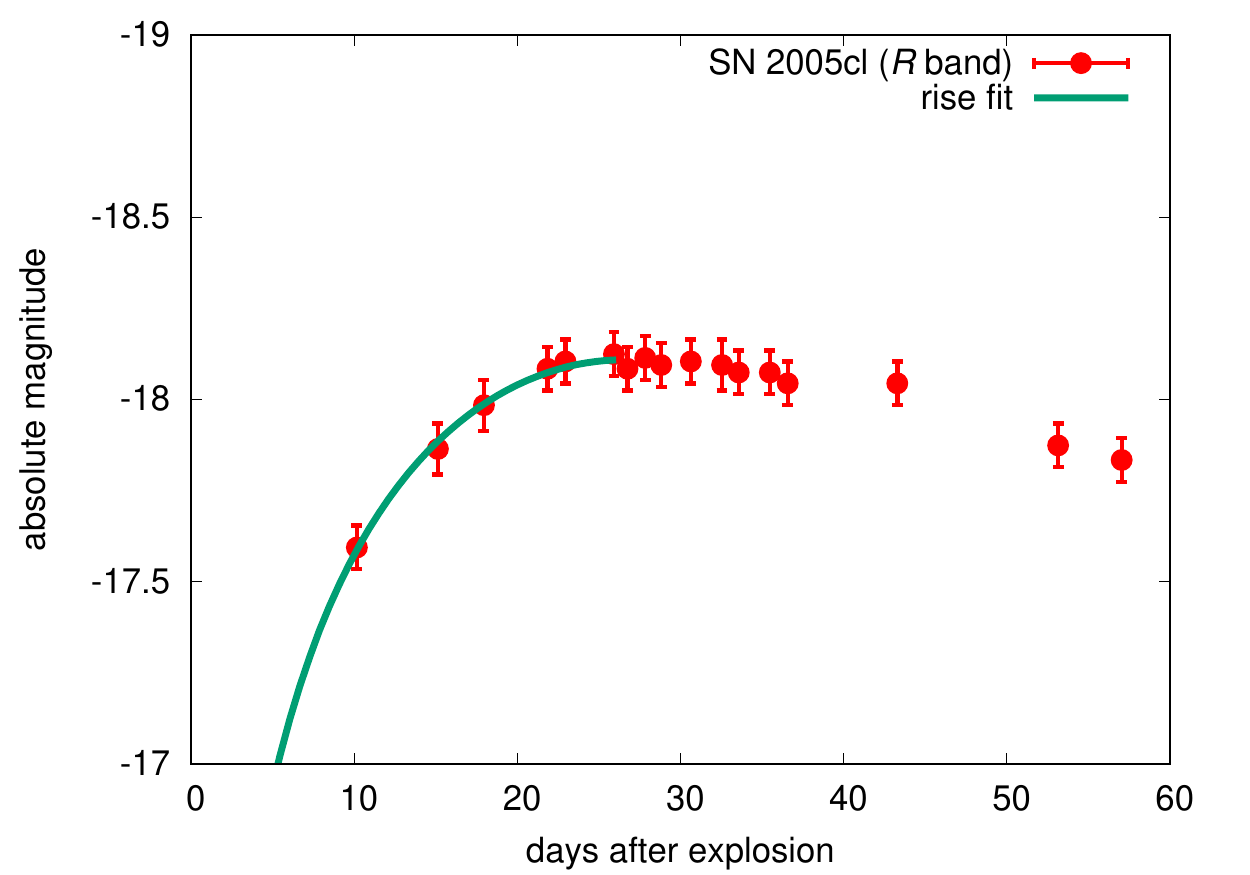} \\    
    \includegraphics[width=0.31\textwidth]{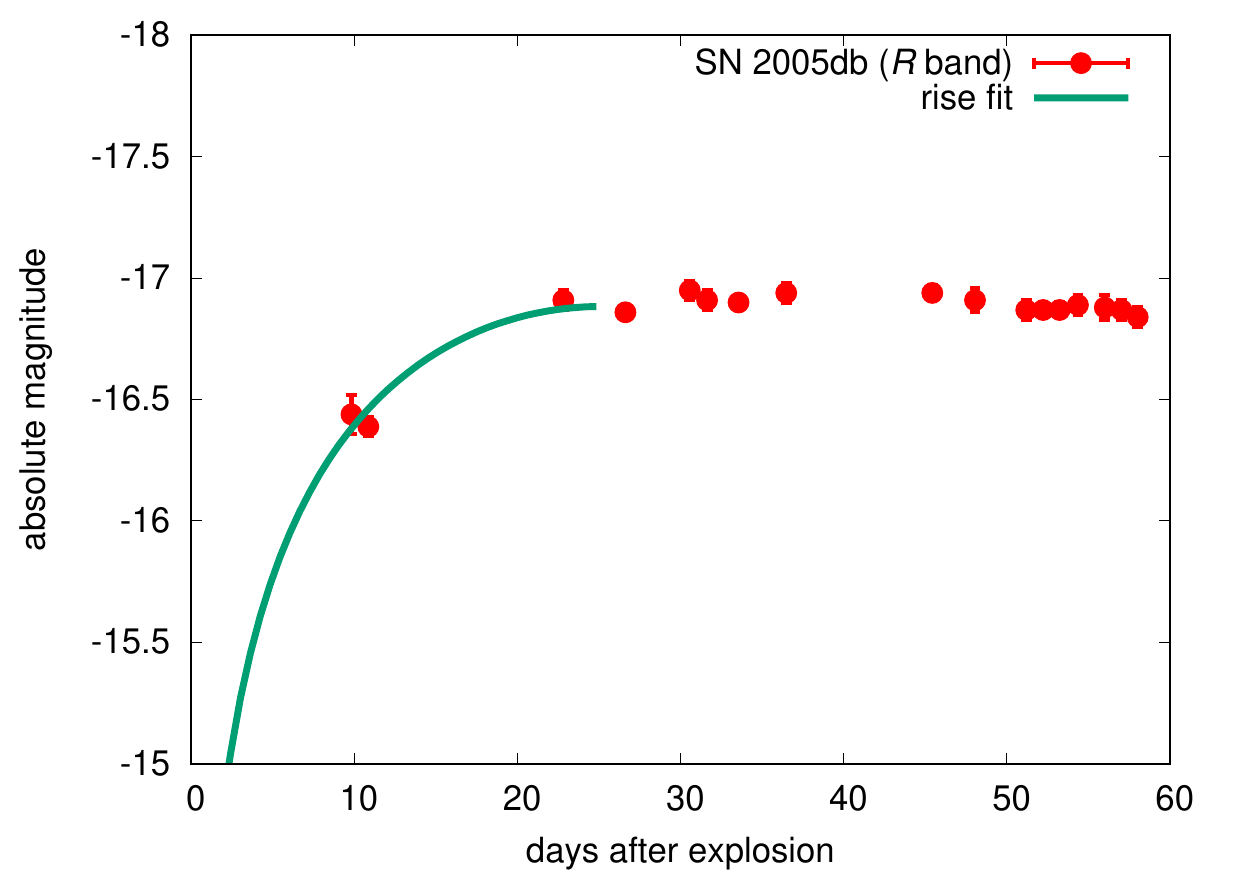}
    \includegraphics[width=0.31\textwidth]{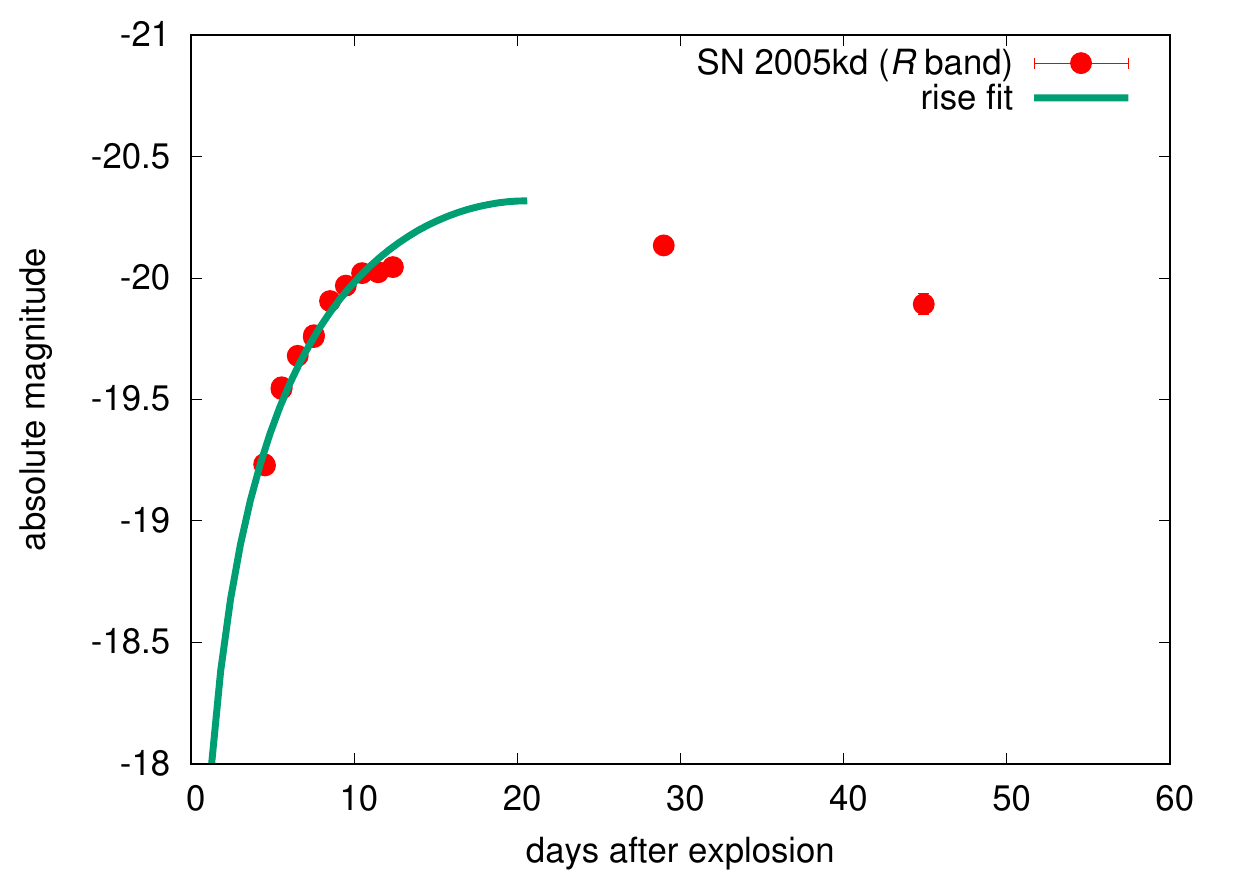}
    \includegraphics[width=0.31\textwidth]{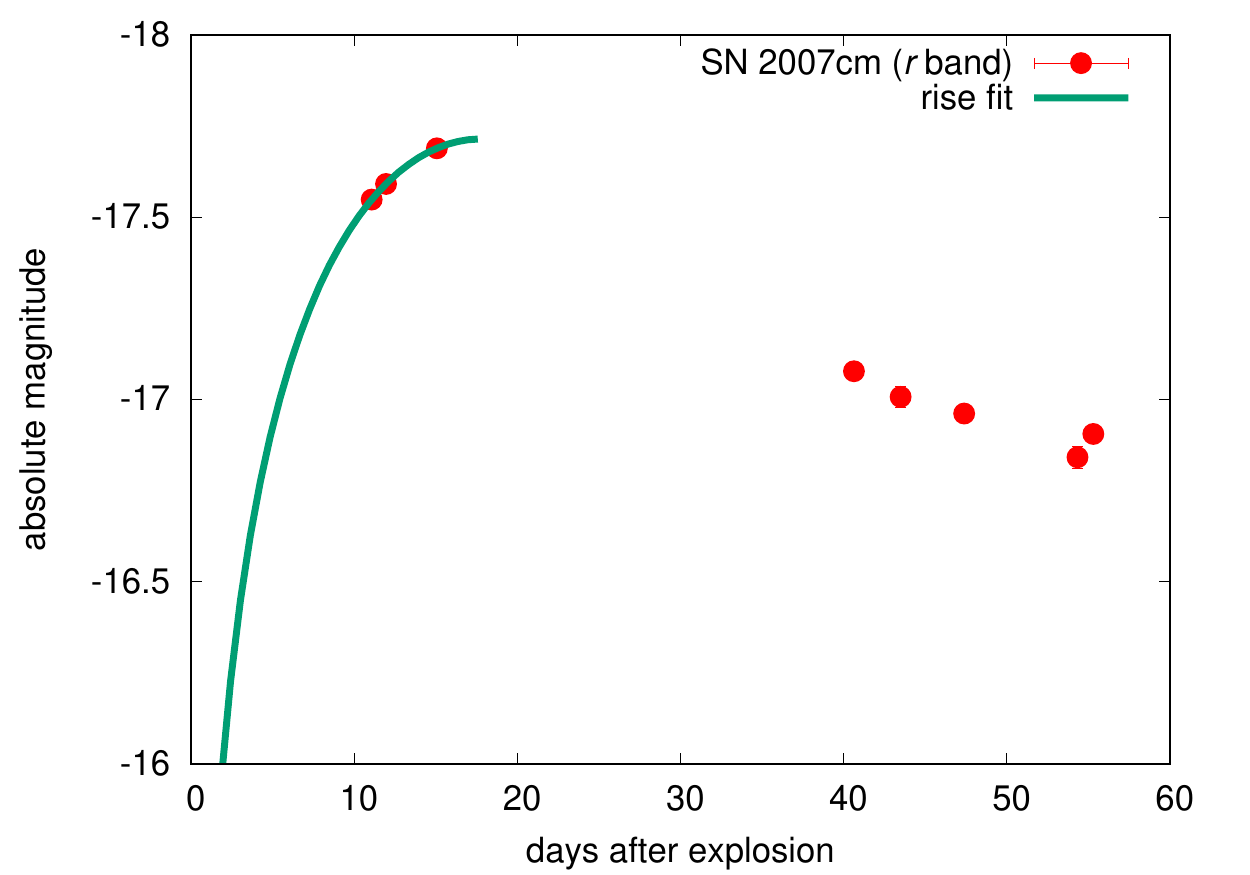} \\
    \includegraphics[width=0.31\textwidth]{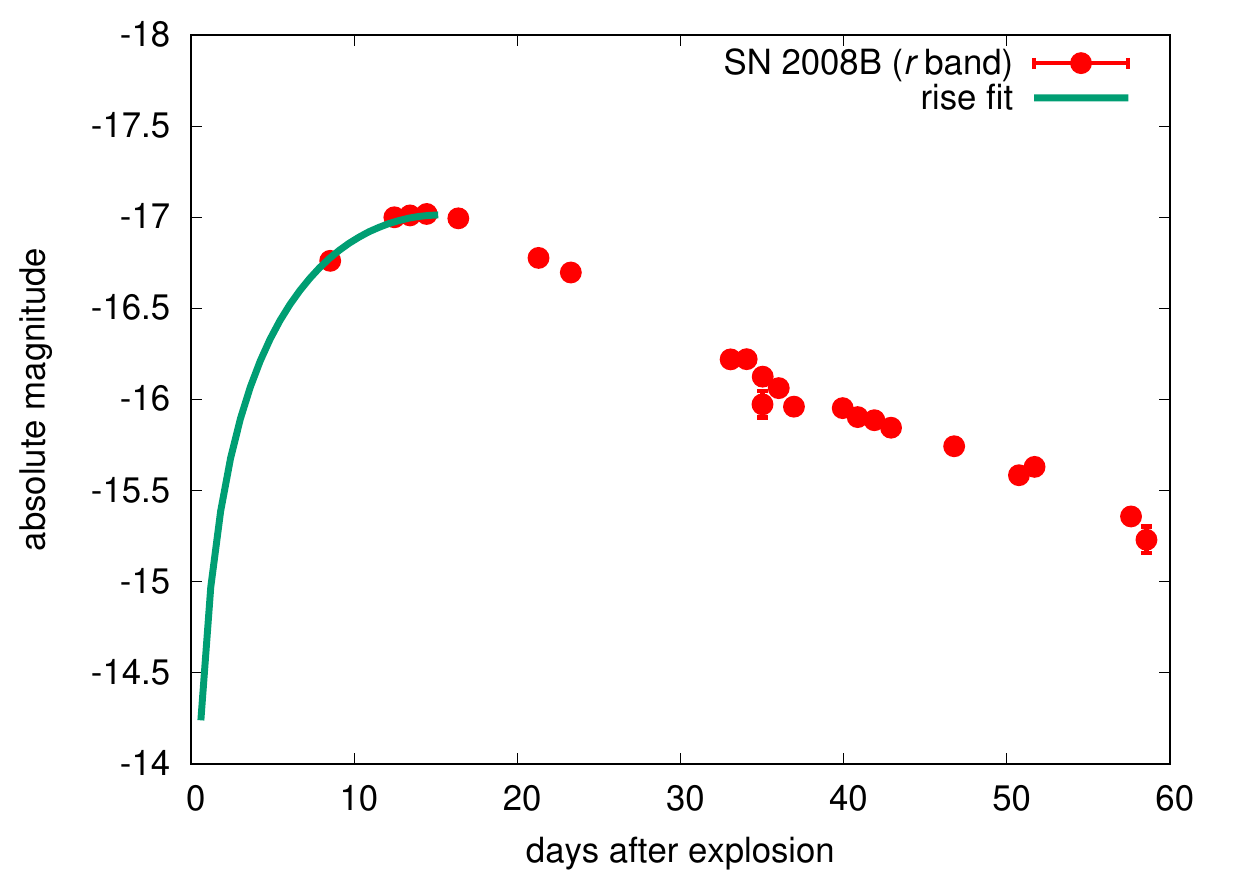}
    \includegraphics[width=0.31\textwidth]{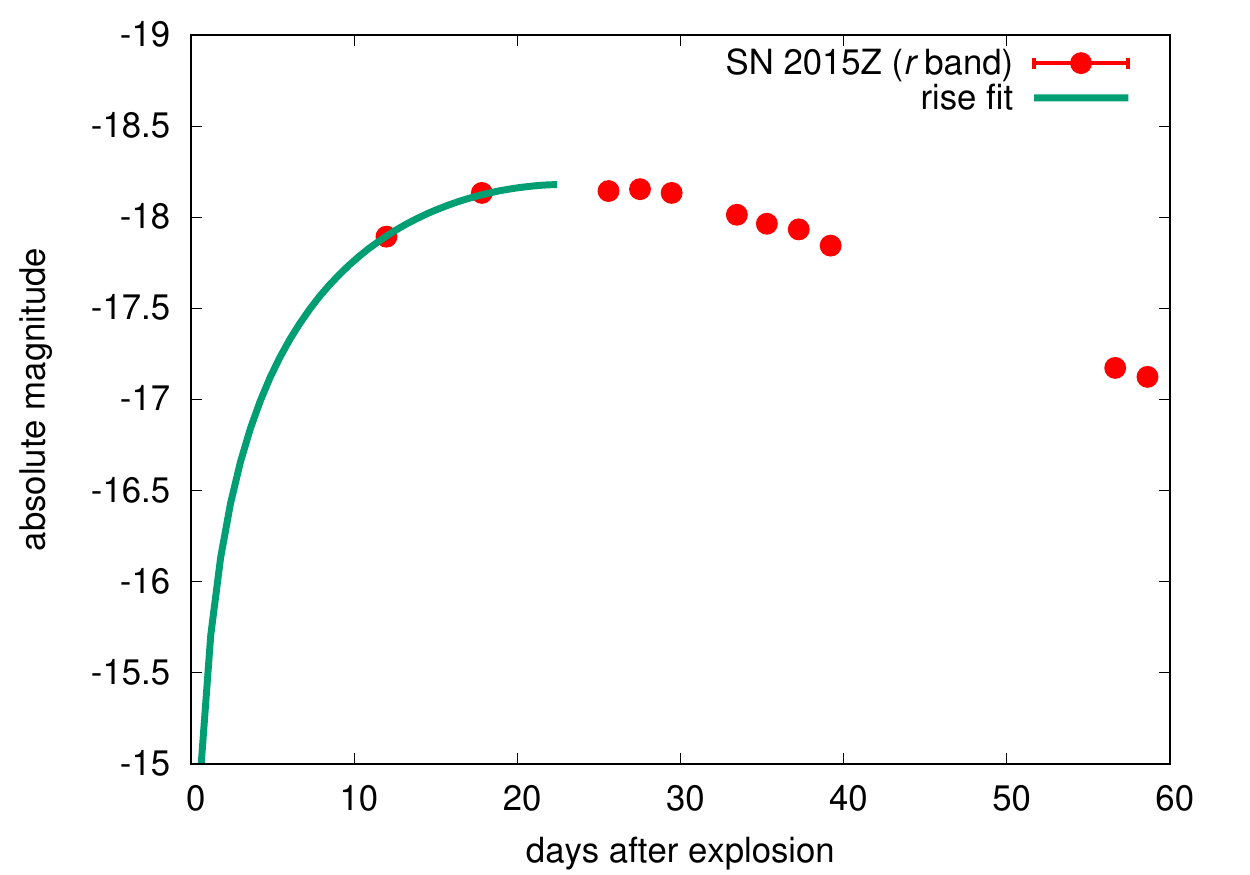}
    \includegraphics[width=0.31\textwidth]{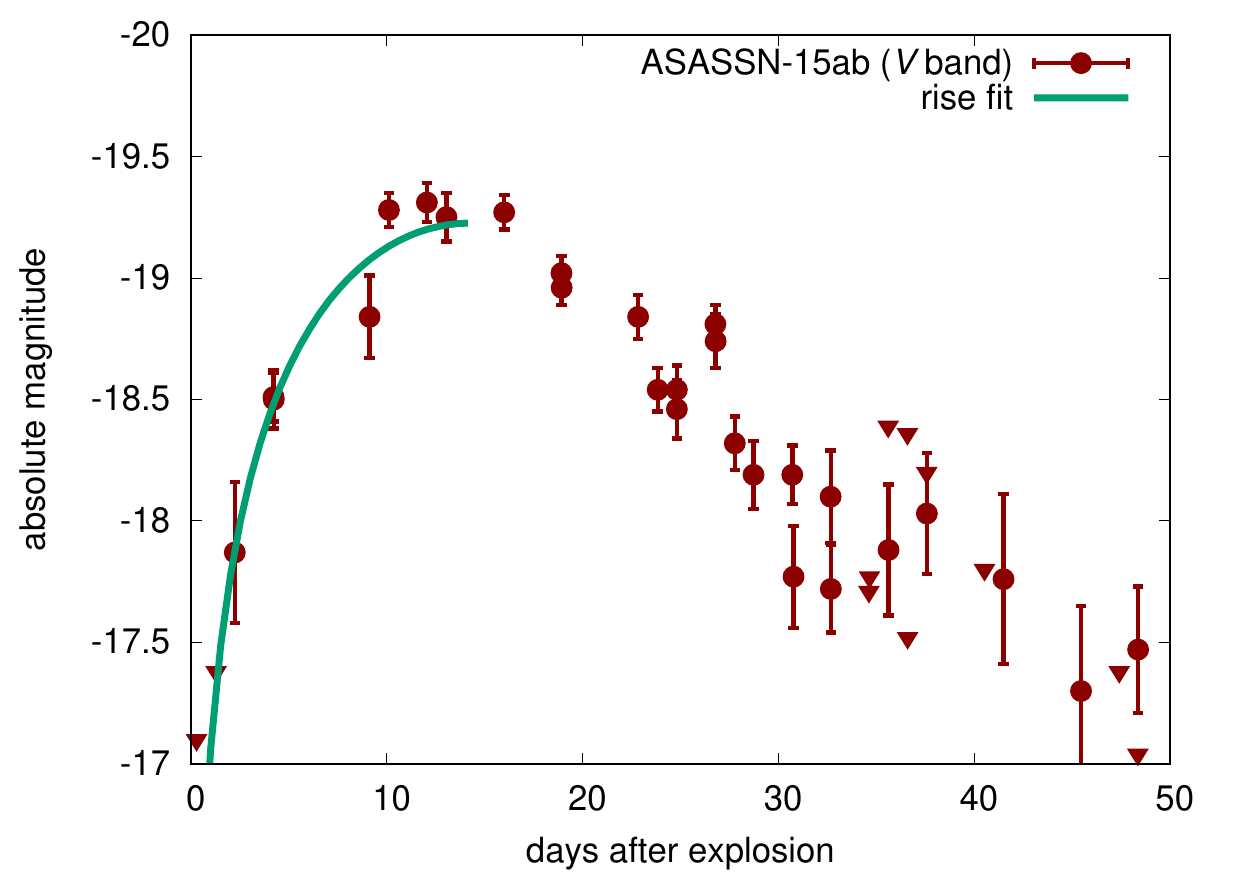} \\
    \includegraphics[width=0.31\textwidth]{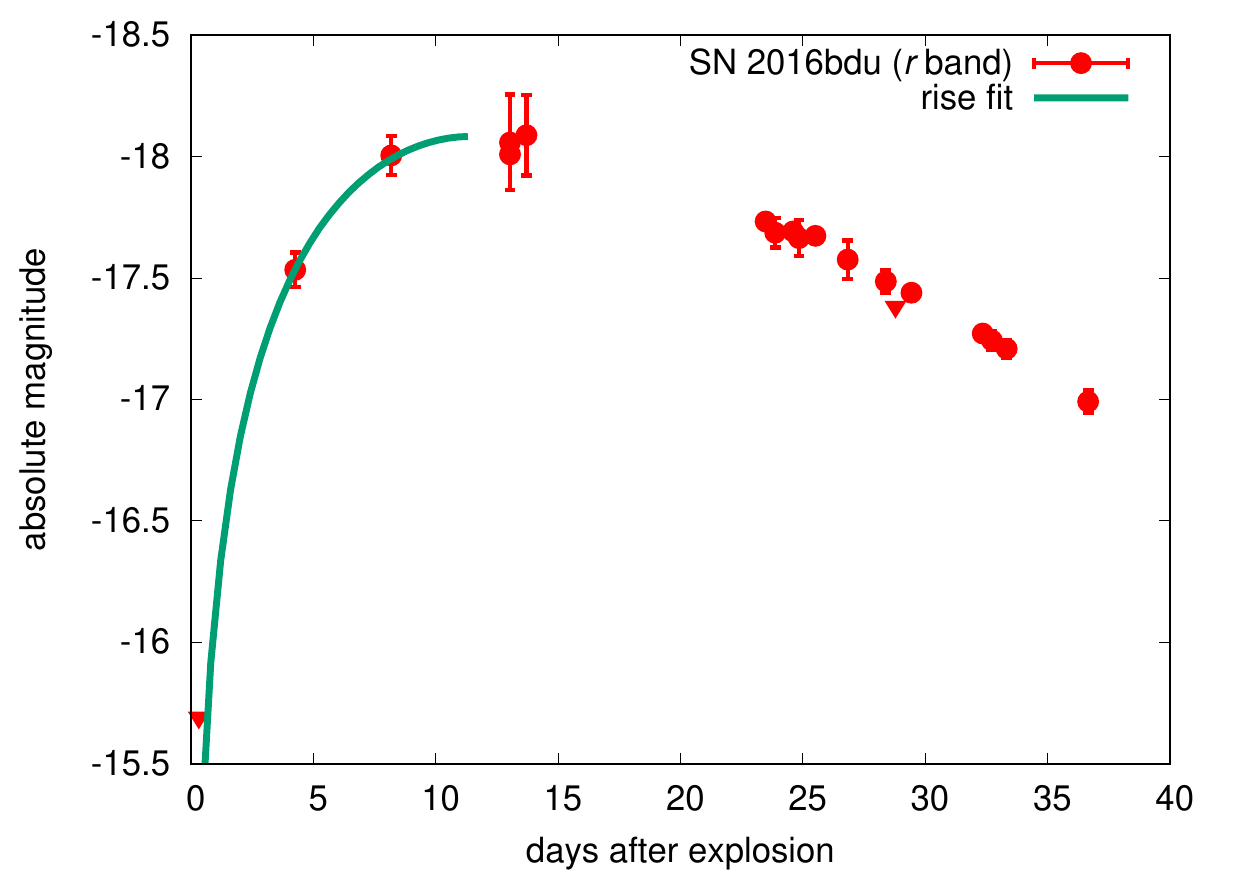} 
    \includegraphics[width=0.31\textwidth]{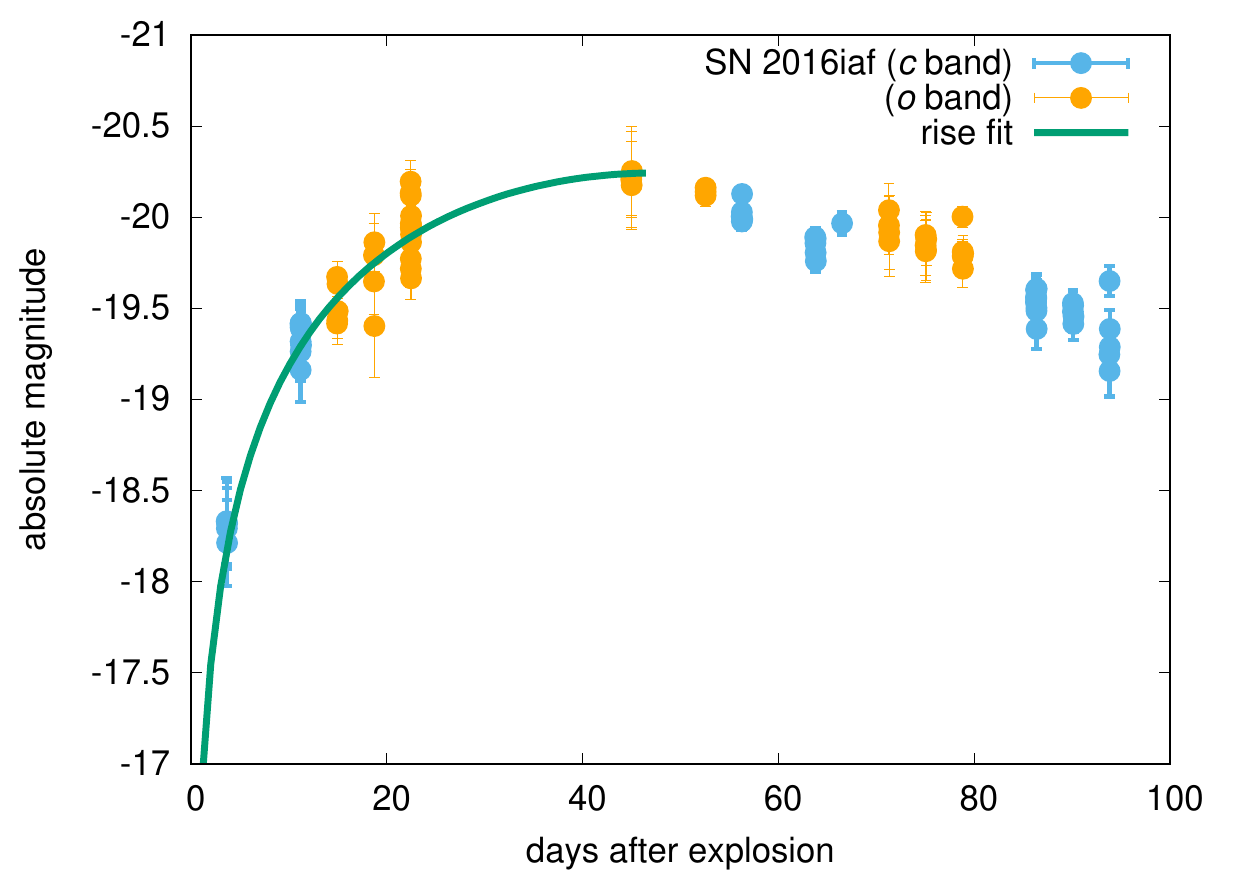}
    \includegraphics[width=0.31\textwidth]{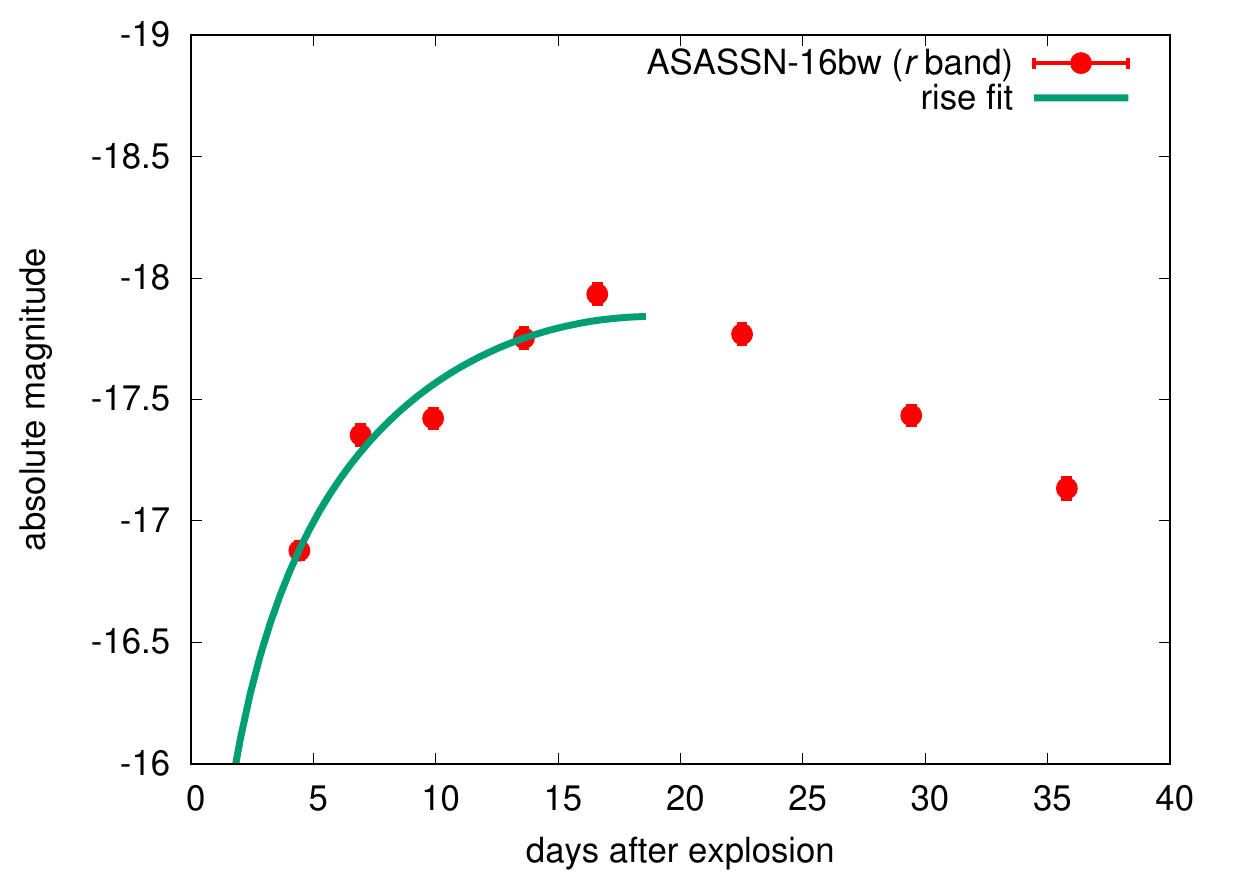} \\
    \caption{
    Light curves and fits for the SNe~IIn used in this work. The light-curve fit to the peak is presented, and the peak is where the fitted line ends.
    }
    \label{fig:lightcurve_fit}
\end{figure*}
\setcounter{figure}{0}
\begin{figure*}
    \centering
    \includegraphics[width=0.31\textwidth]{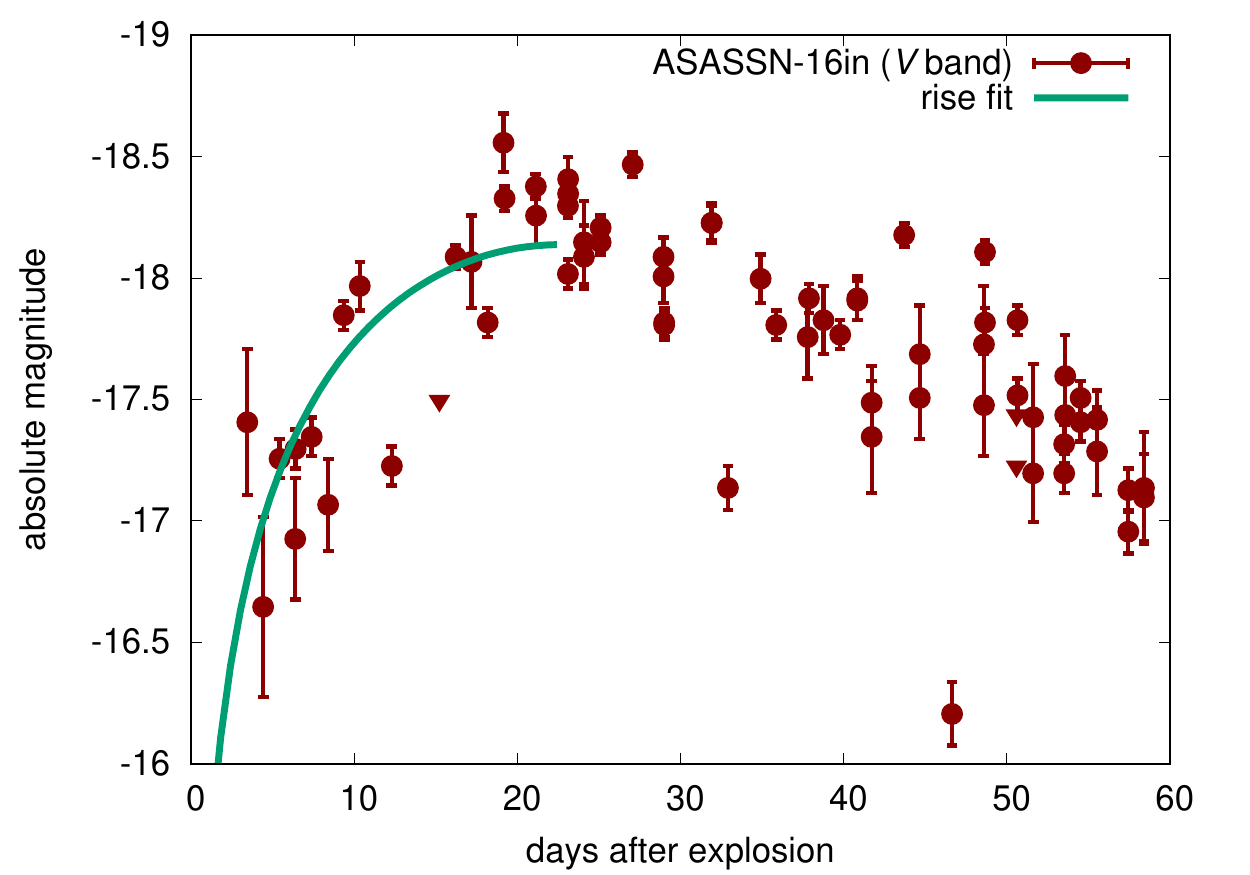}
    \includegraphics[width=0.31\textwidth]{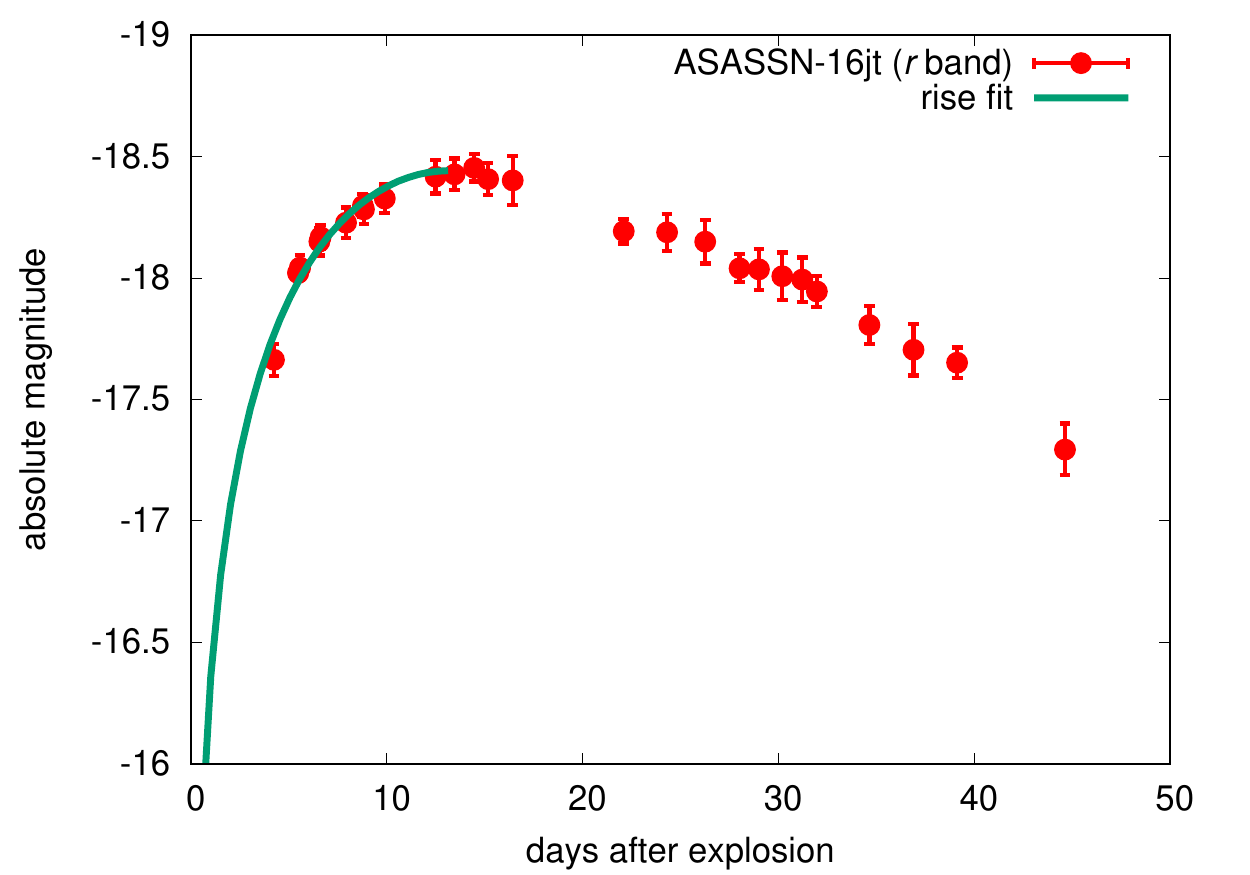}
    \includegraphics[width=0.31\textwidth]{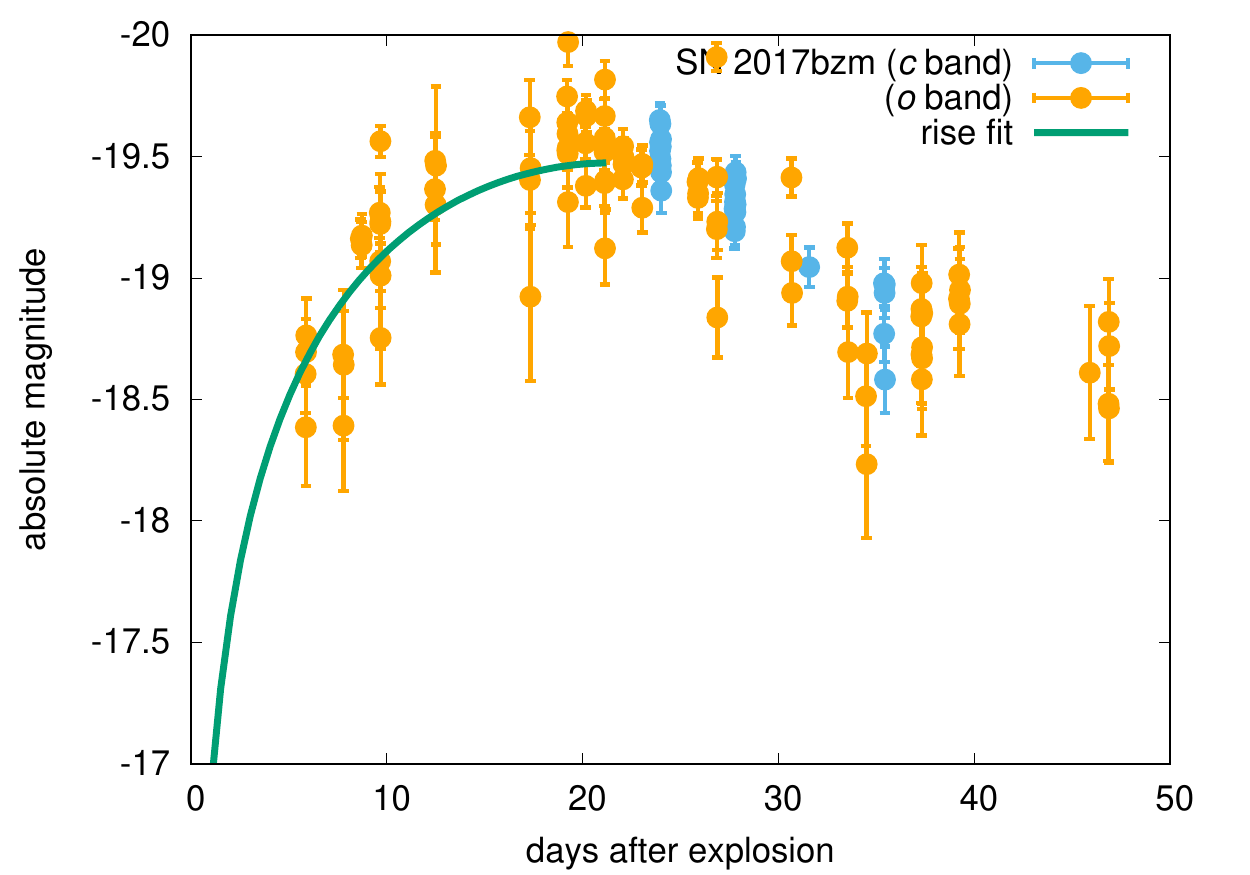} \\
    \includegraphics[width=0.31\textwidth]{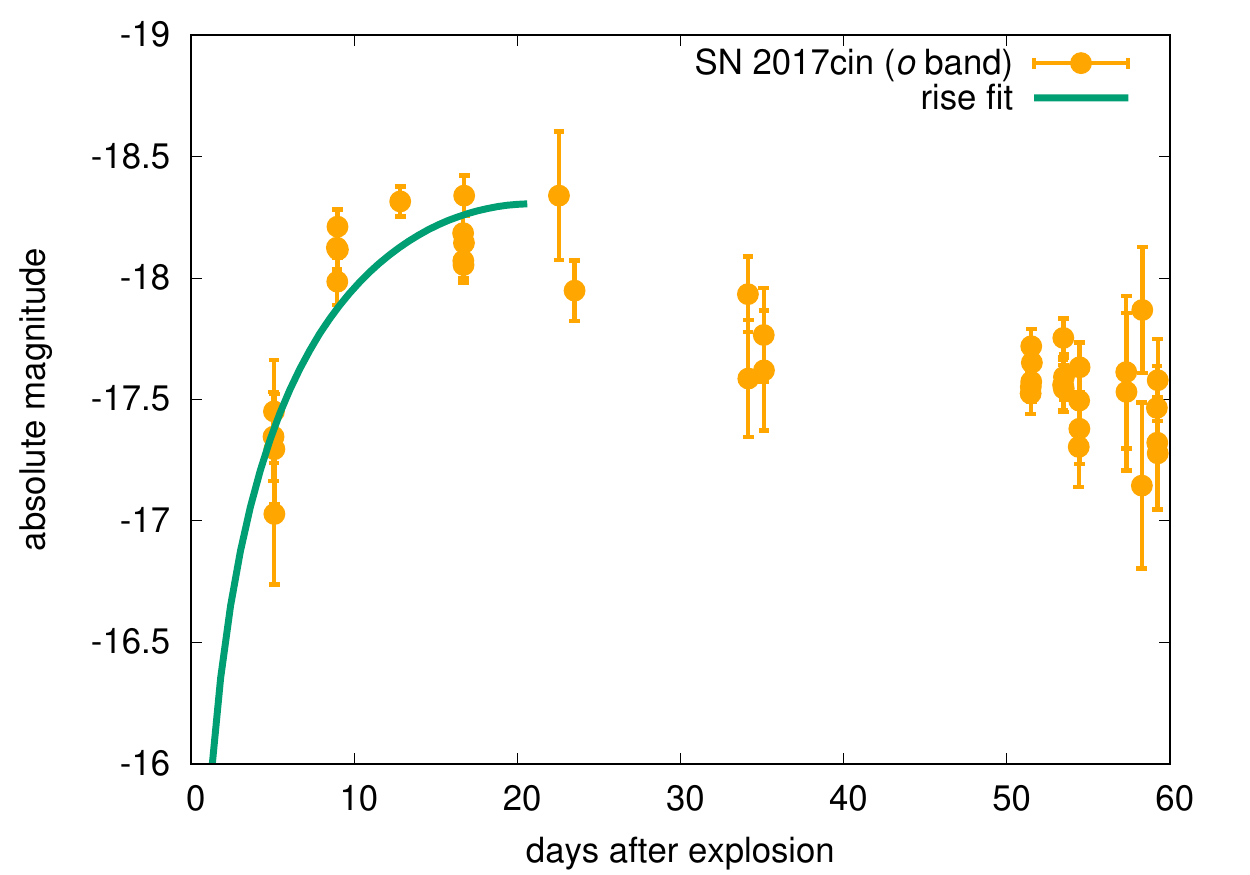} 
    \includegraphics[width=0.31\textwidth]{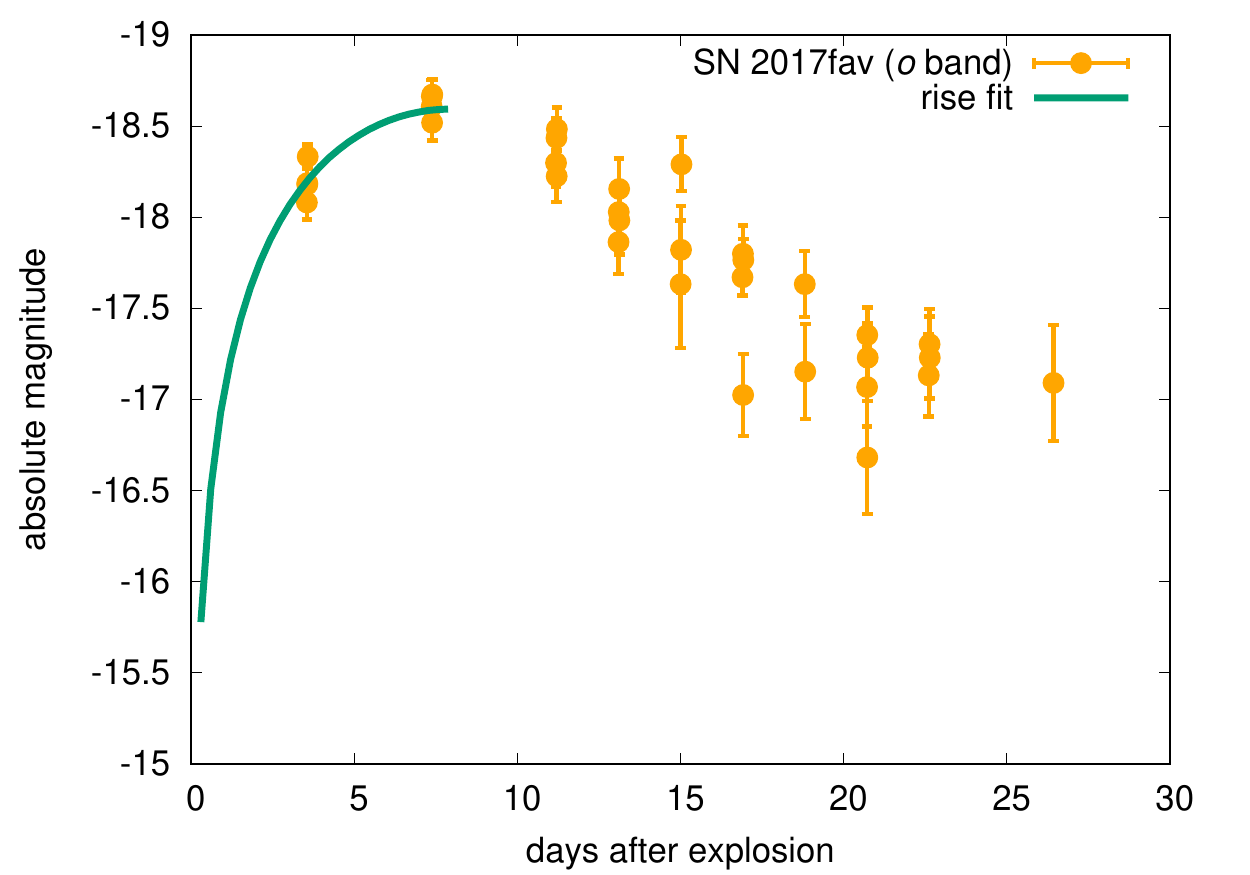}
    \includegraphics[width=0.31\textwidth]{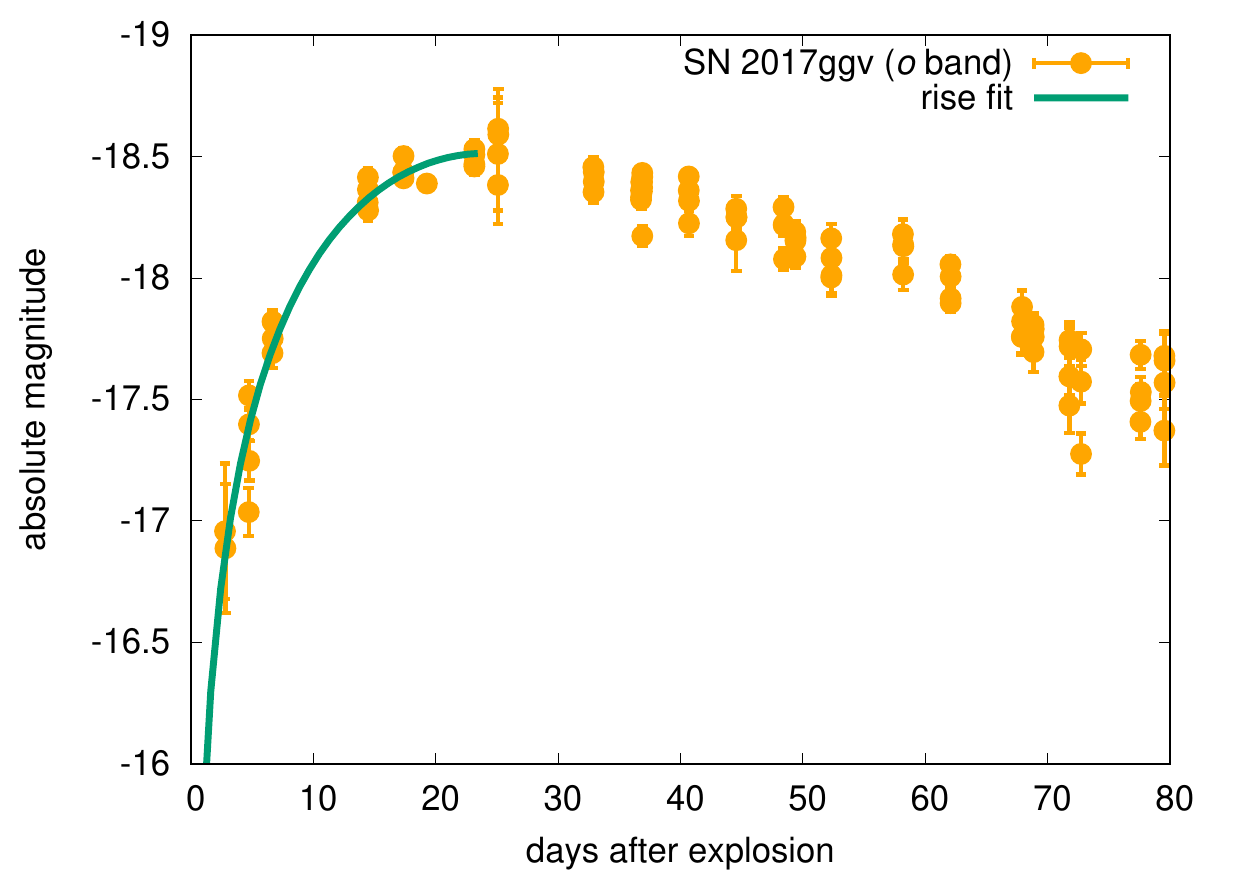} \\
    \includegraphics[width=0.31\textwidth]{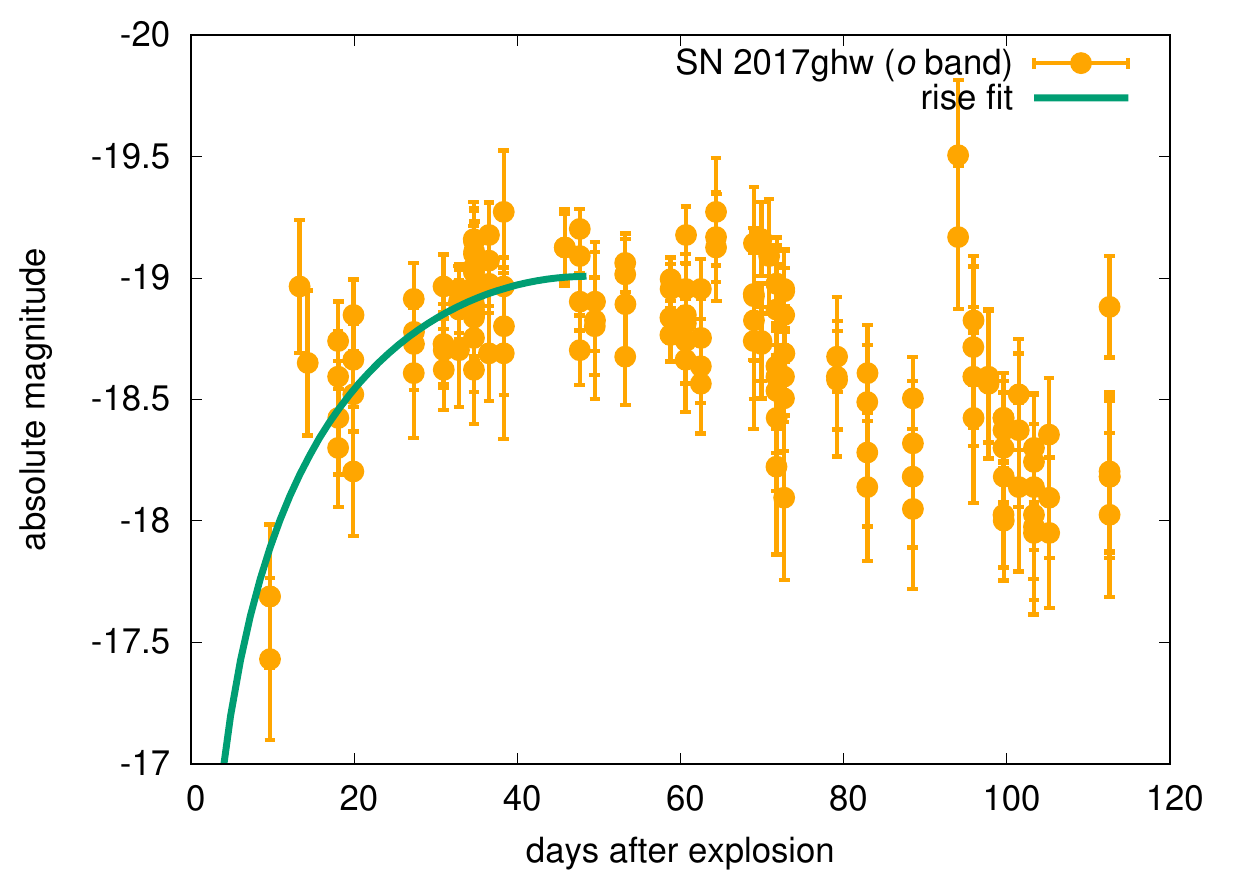} 
    \includegraphics[width=0.31\textwidth]{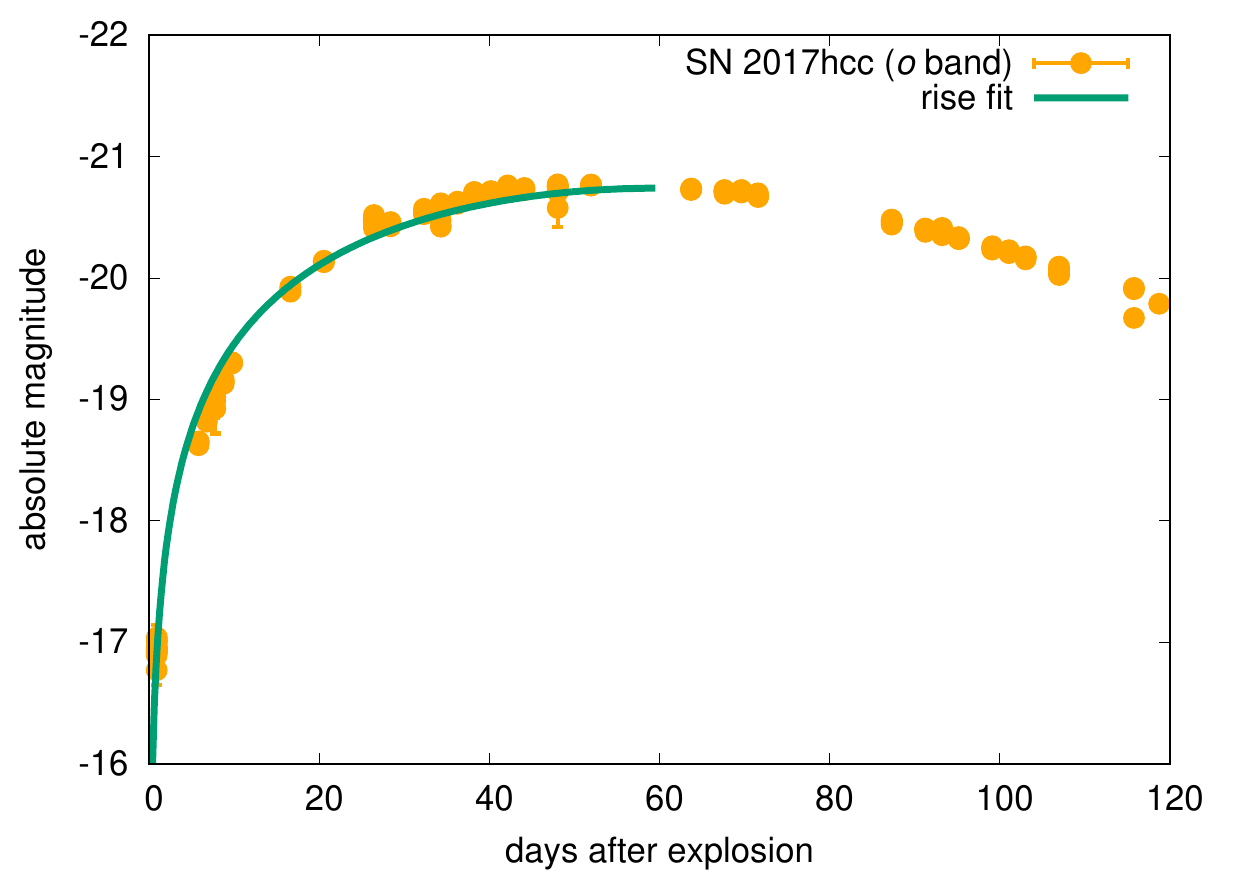} 
    \includegraphics[width=0.31\textwidth]{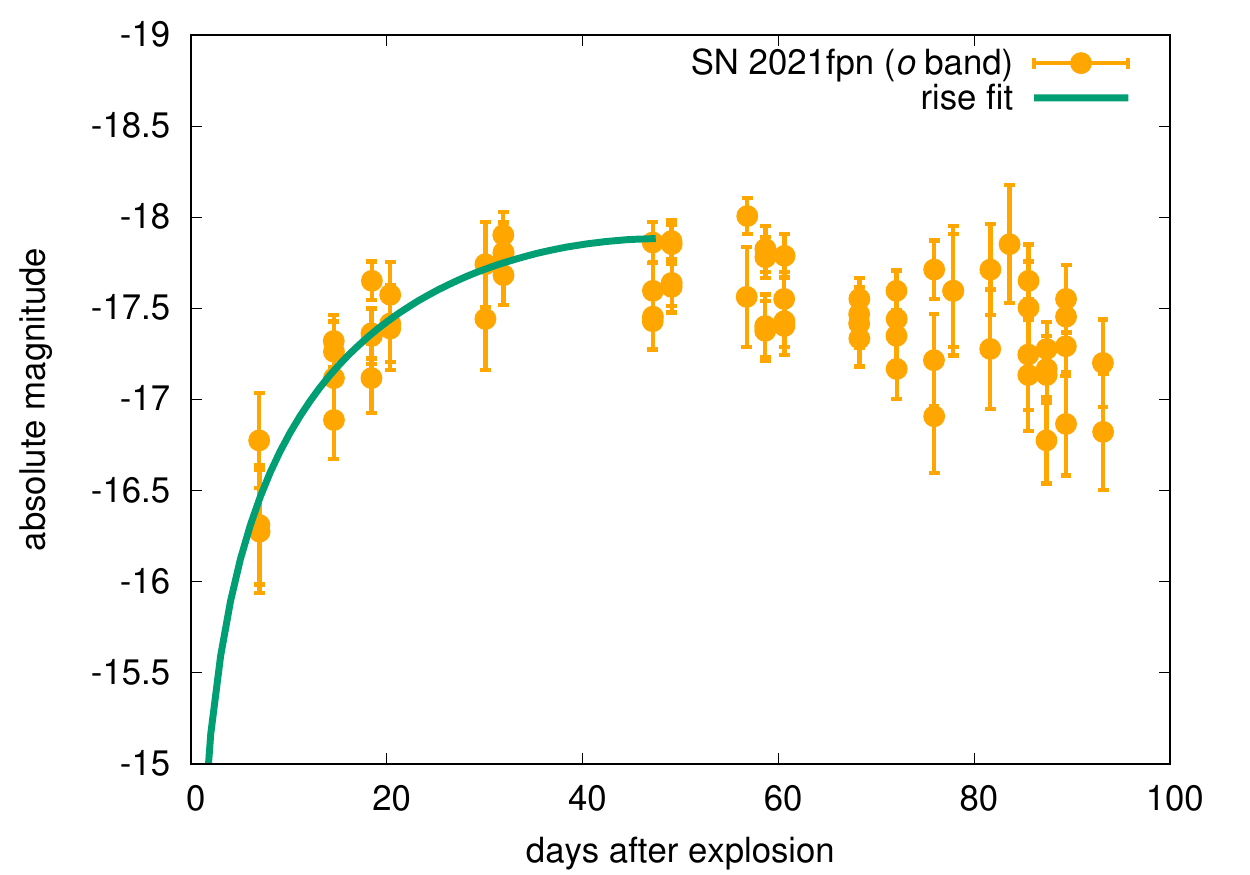}
    \caption{
    \textit{Continued.}
    }
\end{figure*}

\bibliographystyle{aa}
\bibliography{iin} % if your bibtex file is called example.bib

\end{document}